\documentclass[a4paper,preprint]{imsart}

\RequirePackage{amsthm,amsmath,amsfonts,amssymb}
\RequirePackage[authoryear]{natbib}
\RequirePackage[colorlinks,citecolor=blue,urlcolor=blue]{hyperref}
\RequirePackage{graphicx}

\startlocaldefs

\endlocaldefs

\usepackage{booktabs}
\usepackage{tabularx}
\usepackage{multirow}
\usepackage{caption,subcaption}
\usepackage[title]{appendix}
\usepackage[ruled,lined]{algorithm2e}
\usepackage{accents}
\usepackage{mathtools}
\usepackage{luke-macros}

\usepackage[dvipsnames]{xcolor}
\usepackage{tikz}
\usetikzlibrary{arrows,shapes}

\graphicspath{{./figs/}{./analysis-figs/}{./test-figs/}}

\begin{document}

\begin{frontmatter}
\title{Lagged couplings diagnose Markov chain Monte Carlo phylogenetic inference}
\runtitle{Lagged couplings diagnose MCMC phylogenetic inference}
\runauthor{Kelly, Ryder, Clarté}

\begin{aug}
\author[A]{\fnms{Luke J.} \snm{Kelly}\ead[label=e1,mark]{kelly@ceremade.dauphine.fr}},
\author[A]{\fnms{Robin J.} \snm{Ryder}\ead[label=e2,mark]{ryder@ceremade.dauphine.fr}}
\and
\author[A,B]{\fnms{Grégoire} \snm{Clarté}\ead[label=e3]{gregoire.clarte@helsinki.fi}}
\address[A]{CEREMADE, CNRS, UMR 7534, Université Paris-Dauphine, PSL University, \printead{e1,e2}}

\address[B]{Department of Computer Science, University of Helsinki, \printead{e3}}
\end{aug}

\begin{abstract}
    Phylogenetic inference is an intractable statistical problem on a complex space.
    Markov chain Monte Carlo methods are the primary tool for Bayesian phylogenetic inference but it is challenging to construct efficient schemes to explore the associated posterior distribution or assess their performance.
    Existing approaches are unable to diagnose mixing or convergence of Markov schemes jointly across all components of a phylogenetic model.
    Lagged couplings of Markov chain Monte Carlo algorithms have recently been developed on simpler spaces to diagnose convergence and construct unbiased estimators.
    We describe a contractive coupling of Markov chains targeting a posterior distribution over a space of phylogenetic trees with branch lengths, scalar parameters and latent variables.
    We use these couplings to assess mixing and convergence of Markov chains jointly across all components of the phylogenetic model on trees with up to 200 leaves.
    Samples from our coupled chains may also be used to construct unbiased estimators.
\end{abstract}



\end{frontmatter}


\section{Introduction}
\label{sec:introduction}

Phylogenetic inference is the problem of reconstructing the ancestral history of a set of taxa descended from a common ancestor.
The phylogeny is typically represented by a bifurcating tree, where the external leaf nodes correspond to observed taxa and unobserved internal nodes to speciation events.
Phylogenetic inference is a difficult statistical problem.
We attempt to infer a complex, high-dimensional object comprising a discrete tree topology, continuous node ages, and various model parameters and latent variables.
The number of possible topologies grows super-exponentially with the number of taxa, and there are many constraints and dependencies between model components.
Quantifying uncertainty in phylogenetic inference \citep{willis18,willis19,brown19,magee21} and assessing model fit \citep{shepherd18} are difficult tasks as we are fitting a non-standard statistical model on a general state space.
When calculating the likelihood in a phylogenetic model, we attempt to integrate out as many latent variables as possible.
Although the likelihood can often be computed efficiently with a computational cost that grows linearly in the number of taxa, there exist models where it grows exponentially \citep{kelly17}.
There are many software tools for performing model-based Bayesian phylogenetic inference and these methods are routinely applied in various scientific fields to estimate phylogenies of thousands of taxa.

Many modern phylogenetic methods specify a generative model for the data: a branching process on species defines the tree, the species comprise sets or sequences of complex evolutionary traits, and a diversification process acting on the traits represents the evolution of the species along the tree.
Markov chain Monte Carlo (MCMC) is the primary tool for performing Bayesian phylogenetic inference via the Metropolis--Rosenbluth--Teller--Hastings (MH) algorithm \citep{metropolis1953equation,hastings70} and is the focus of our paper.
From initial state $ X_0 \sim \pi_0 $, we construct a Markov chain $ (X_s)_{s \geq 0} $ on the space of phylogenetic trees and model parameters $ \cX $ whose equilibrium distribution $ \pi $ is the posterior distribution under our model.
For a function of interest $ f $, we approximate $ \EE_{\pi}[f(X)] $ by the asymptotically exact estimator $ (S + 1)^{-1} \sum_{s = 0}^S f(X_s) $.
As we initialise the chain at some distribution $ \pi_0 \neq \pi $ and iteratively draw a finite number of dependent samples, the marginal distribution of iterates may never reach $ \pi $ exactly in practice.
However, we can still perform valid inference provided that the error in approximating $ \pi $ by a finite collection of dependent samples is negligible.

The quality of a finite MCMC sample depends on its speed of convergence; that is, how quickly the distribution of samples approaches $ \pi $.
The bias due to the initialisation in Monte Carlo estimators is generally considered to decay faster than the standard error \citep{geyer11}, so in a typical MCMC analysis we discard initial samples as burn-in and run our chains for sufficiently many iterations that Monte Carlo estimators have a desired level of accuracy.
Removing burn-in reduces the unknown initialisation bias in MCMC estimators but does not eliminate it, so we cannot simply combine samples or average estimators across multiple chains.
Using the same samples to estimate burn-in and perform inference may bias inference \citep{cowles99possible}.
Constructing efficient MCMC sampling schemes for phylogenetic models and confidently assessing how well finite chains approximate their target posterior distribution are extremely difficult in practice.

Phylogenetic posterior distributions are often multimodal \citep{beiko06}, even when all but one parameter are fixed \citep{dinh17shape}, and topologies with high posterior support are frequently isolated from each other by multiple rearrangement operations \citep{whidden15}.
Constructing a chain which efficiently explores a broad class of phylogenetic posterior distributions is an active topic of research.
\texttt{MrBayes} \citep{ronquist12} uses parallel tempering, whereby chains at different temperatures interact through swap moves, to increase exploration between modes.
\texttt{Blang} \citep{bouchard21} implements non-reversible parallel tempering \citep{syed22} to further increase the speed of exploration.
Many leading Bayesian phylogenetic software packages implement adaptive proposal schemes to sample model components more efficiently.
\citet{baele17} implement adaptive methods for phylogenetic model parameters in \texttt{BEAST} \citep{suchard18}.
\citet{douglas21} develop adaptive proposal schemes in \texttt{BEAST 2} \citep{bouckaert19}.
\citet{meyer21} constructs adaptive proposals for sampling topologies.
\citet{hohna12} describe guided proposals on topologies using Metropolised Gibbs steps.
In certain classes of problems, we can use Hamiltonian Monte Carlo (HMC) \citep{zhao16,dinh17,bastide21}, piecewise-deterministic Markov processes \citep{zhang21,koskela22}, sequential Monte Carlo \citep{wang19} or variational approximations \citep{zhang19} to sample elements of the posterior more efficiently than in standard MH approaches.
\citet{whidden20} use systematic search and optimisation to construct a set of high-likelihood trees.
In any case, we lack methods to properly quantify convergence or mixing of Markov schemes on the space of trees and model parameters when running one or more independent chains, so we struggle to separate modelling and fitting errors \citep{fourment19}.

Phylogenetic inference is computationally expensive so we desire to identify convergence shortly after it occurs so as to avoid wasting output by discarding it as burn-in.
As data sets increase in size, practitioners resort to running ever longer chains from different initial configurations and checking whether various summary statistics converge to similar values or distributions at similar rates \citep{cowles96markov,roberts04general}.
As components of the model are not independent, these approaches are unlikely to be sufficient for diagnosing convergence jointly across the entire model and are dominated by the slowest mixing summaries \citep{vats19multivariate}.
Furthermore, we lack tools to remove potential burn-in bias in estimators which would allow us to safely combine estimates from multiple independent chains.

A coupling of distributions $ p $ and $ q $ on $ \cX $ is any distribution on $ \cX \times \cX $ such that $ (X, Y) $ drawn from the coupling has $ X \sim p $ and $ Y \sim q $.
Couplings have been used to derive theoretical bounds on convergence of MCMC algorithms \citep[and references therein]{roberts04general} and develop sampling schemes \citep{propp96exact} but are often difficult to apply in practice \citep{johnson98}.
In a recent series of important papers, Pierre Jacob and collaborators have developed techniques using lagged couplings of Markov chains to debias MCMC estimators and estimate convergence bounds on general state spaces under mild conditions.
Our description follows \citet{biswas19} who build on the framework developed by \citet{jacob20}.

Let $ (X_s)_{s \geq 0} $ and $ (Y_s)_{s \geq 0} $ be Markov chains on a space $ \cX $ with common initial distribution $ \pi_0 $, stationary distribution $ \pi $ and transition kernel $ P $.
We construct a coupled Markov chain $ (X_s, Y_s)_{s \geq 0} $ which evolves according to a coupling $ \bar{P} $ of the marginal transition kernels such that the lag-$ l $ staggered chains meet at a random, finite time $ \tau^{(l)} = \inf\{s \geq l : X_s = Y_{s - l}\} $ and remain coupled thereafter:
\begin{itemize}
    \item sample $ X_0 \sim \pi_0 $ and $ Y_0 \sim \pi_0 $;
    \item for $ 1 \leq s \leq l $, draw $ X_s \sim P(X_{s - 1}, \cdot) $;
    \item for $ s > l $, draw $ (X_s, Y_{s - l}) \sim \bar{P}((X_{s - 1}, Y_{s - l - 1}), \cdot) $.
\end{itemize}
We assume throughout this paper that $ \PP(\tau^{(l)} > s) $ decays geometrically in $ s $, so the coupled kernel $ \bar{P} $ must be carefully constructed in order to achieve this.
\citet{middleton20} relax the tail assumption on $ \tau^{(l)} $ to polynomial decay.
The coupling preserves the marginal properties of $ (X_s)_{s \geq 0} $ and $ (Y_s)_{s \geq 0} $, so by this construction the marginal distributions are equal at each iteration $ s \geq 0 $; that is, $ \pi_s(A) = \PP(X_s \in A) = \PP(Y_s \in A) $ for any measurable $ A \subset \cX $.
Using this framework, \citet{jacob20} construct unbiased MCMC estimators of $ {\EE}_{\pi}[f(X)] $ which can be averaged across independent pairs of chains to reduce their variance.

\citet{biswas19} derive the following bound on the total variation (TV) distance between the marginal distribution of a chain and its target:
\begin{equation}
    \label{eq:tv-bound}
    d_{\mathrm{TV}}(\pi_s, \pi)
        \leq \EE\left[0 \vee \ceil*{\frac{\tau^{(l)} - l - s}{l}}\right],
\end{equation}
where $ a \vee b = \max(a, b) $ and $ \ceil{\cdot} $ is the ceiling function.
This generalises and sharpens the bound for lag $ l = 1 $ described by \citet{jacob20}.
In addition to $ \pi_s $ and $ \pi $, the bound also depends on the quality of the coupling: tighter couplings will produce earlier meeting times and a smaller bound.
We can sample exactly from the distribution of $ \tau^{(l)} $ under our coupling so construct a Monte Carlo estimate of the TV bound \eqref{eq:tv-bound} from the meeting times of pairs of coupled chains.
\citet{craiu20} use control variates to derive a tighter lagged coupling TV bound but we do not pursue their approach here.

The lag $ l $ is a free parameter with a simple interpretation and guidelines for choosing it.
Slow mixing between modes of the target distribution manifests as plateaus in the bound: a pair $ (X_l, Y_0) $ in the same mode when we start sampling from the coupled kernel will typically produce earlier meeting times than those in separate modes.
To account for multimodality, we initialise chains far apart and increase the lag so that chains are more likely to explore the posterior before meeting.
Informally, if the lag $ l $ is such that $ X_l \sim \pi $, then further increasing $ l $ will not change the distribution of $ \tau^{(l)} - l $ as it depends entirely on $ \bar{P} $.
Increasing $ l $ produces a sharper bound but the returns eventually diminish as the bound is not tight and the coupling is not optimal.
In practice, we run multiple pairs of chains at an increasing sequence of lags until the estimated bounds stabilise.
\citet{biswas19} discuss the choice of lag in more detail.

The TV distance uniformly bounds the difference in probabilities assigned by the two distributions and is particularly attractive for phylogenetic inference problems as it does not require a metric on the space $ \cX $ of trees, parameters and other components.
In order to successfully implement these methods in practice, we require a coupling which produces a positive probability that the chains will meet on at least some region of the state space.
Previous efforts at constructing couplings have focused on situations where the state space has a straightforward geometry, typically a subset of $ \R^d $ \citep{heng19,biswas20,bou20}, or problems such as the Ising model \citep{jacob20}.
In this paper, we describe techniques to couple Markov chains exploring a posterior distribution over a space of phylogenetic trees and model parameters so that we can produce a useful, qualitative bound on the TV distance for diagnosing convergence.
This is a vast improvement over existing convergence diagnostics and allows us to have greater confidence in any subsequent analyses.
We can run multiple pairs of coupled chains independently in parallel so the computational cost is not prohibitive.
In attempting to couple pairs of chains, we can identify moves to add to our proposal kernel from those which fail to meet.
That pairs of chains meet and do not separate is a strong validation of our software implementation and complements other tools for assessing software consistency, such as the joint distribution testing methodology of \citet{geweke04getting} with an appropriate choice of test functions \citep{wang2015bayesian}.
In addition to diagnosing convergence, samples from our lag-$ l $ coupled chains may also be used to construct unbiased estimators following the framework developed by \citet{jacob20}.
Pairs of chains must have the same transition kernel at each iteration, otherwise they could meet and later separate, so we do not attempt to couple adaptive MCMC algorithms.
Note that this framework is not what has also been termed \emph{coupled MCMC for phylogenetic inference} in other works such as \citet{muller20} who implement an adaptive parallel tempering scheme for phylogenetic inference.

The extension of this coupling approach to phylogenetic problems is not straightforward for a number of reasons.
MCMC proposal distributions which update the entire state of the chain or large portions of it are computationally intractable.
There is a trade-off between making computationally cheap proposals which only update small components of the state and larger updates which increase the speed of exploration but with an increased computational cost.
For example, an HMC proposal simultaneously on all of the branch length parameters \citep{zhao16} can explore the posterior more efficiently per MCMC iteration than a random walk update but requires numerical integration and an appropriate choice of tuning parameters.
The MCMC transition kernel we consider is a mixture of several kernels, each proposing local modifications to some aspect of the state, and may be extended to accommodate new models and parameter configurations.
In attempting to couple chains, the region of space where we can make an identical proposal for a component of both states is extremely small and requires the states to already be similar in many respects.
In order to obtain a successful coupling from arbitrary initial configurations, we must make proposals to both chains which make the states increasingly similar until a componentwise meeting of chains becomes possible.

As a running example throughout this paper, we assume that species diversify according to the Stochastic Dollo (SD) model \citep{nicholls08,ryder11,kelly17}, which posits a birth-death process of evolutionary traits along a rooted, clock-like tree.
Any MCMC on a space of phylogenetic trees requires a kernel which is, to a certain extent, model-specific, and our SD example is no exception.
Nonetheless, the SD model has the advantage of exhibiting many features present in other evolutionary models: in addition to the tree topology and node ages, the parameters to sample include several correlated scalar parameters (the various rates), one discrete latent parameter per edge (the number of catastrophes), and a varying number of scalar latent parameters (the positions of catastrophes along branches).
We believe the techniques that we describe to couple the topology, node ages, scalar parameters and latent variables therefore cover a wide enough variety of cases to be transferable to a large class of phylogenetic models, including unrooted phylogenetic trees and networks or models with branch-specific parameters.


The remainder of this paper is arranged as follows.
In \secref{sec:bayesian-phylogenetic-inference}, we introduce the problem of Bayesian phylogenetic inference via MCMC and describe existing methods for assessing convergence in this setting.
\secref{sec:couplings-generic} provides an overview of methods to couple generic MCMC transition kernels.
\secref{sec:coupled-mcmc-phylogenetics} describes our couplings for phylogenetic proposal kernels, and
\secref{sec:experiments} illustrates our approach on a variety of data sets.
We defer much of the technical description of our couplings to \appref{app:coupling-proposals-phylogenetic}. Scripts to
All the code used for this paper is available online at \url{https://github.com/lukejkelly/CoupledPhylogenetics}.


\section{Bayesian phylogenetic inference}
\label{sec:bayesian-phylogenetic-inference}

\subsection{Phylogenetic trees and models}
\label{sec:phylogenies}

\begin{figure}[tb]
	\centering
	\includegraphics[width=\textwidth]{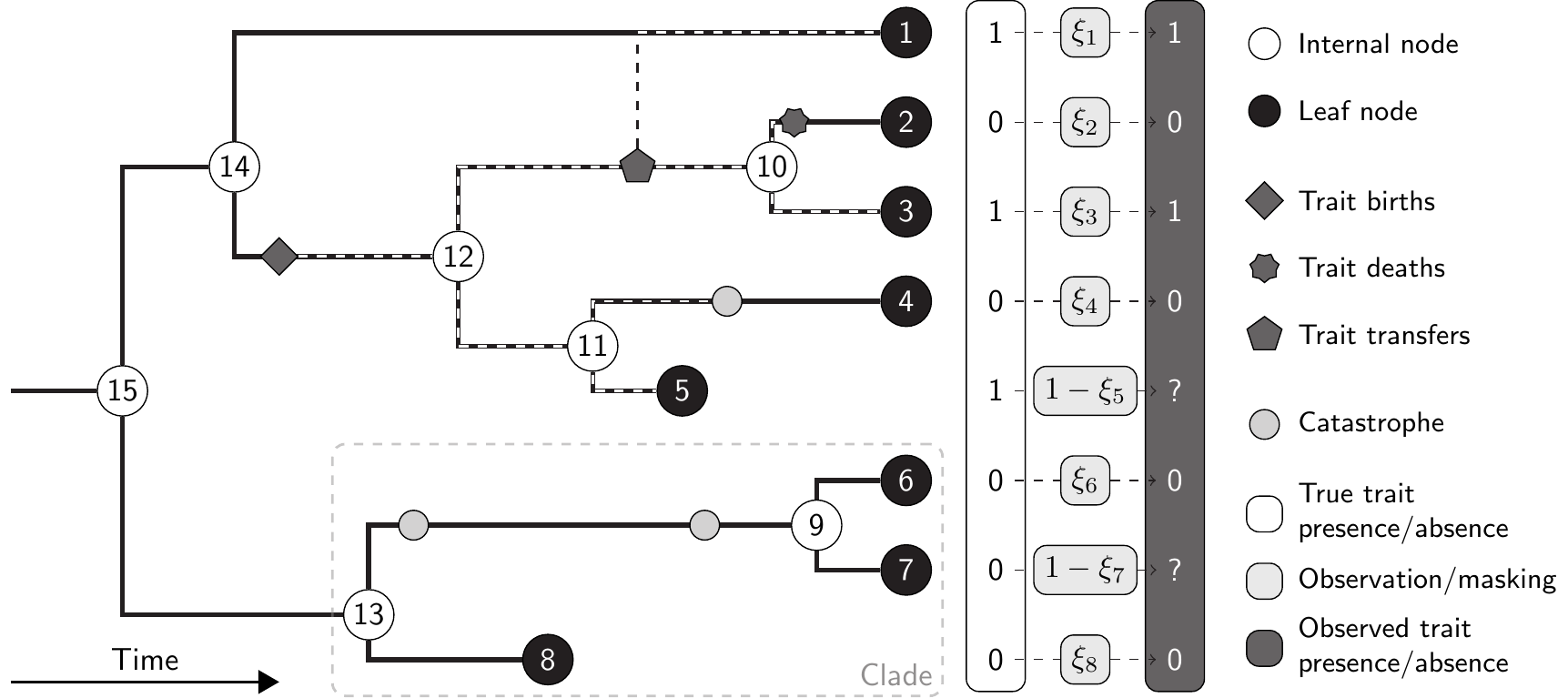}
    \caption{
        A rooted phylogenetic tree on observed taxa $ L = \{1, \dotsc, 8\} $.
        Internal node labels are arbitrary, the root node index is $ 15 $ here.
        Catastrophes represent instantaneous bursts of evolutionary activity relative to the other branches.
        Superimposed on the phylogeny is the history of a trait drawn from the Stochastic Dollo model: the trait was born on branch $ 12 $ and survived to be present in leaves $ 1 $, $ 3 $ and $ 5 $ but a missing-at-random masking process obscured the true status of the trait in taxa $ 5 $ and $ 7 $.
    }
    \label{fig:sd-example}
\end{figure}

We assume that the set of observed taxa $ L $ are the terminal observations of a branching stochastic process.
We represent the history of this process by a dated phylogenetic tree $ g = (V, E, T) $ with vertex set $ V $, edge set $ E $ and node ages $ T $.
The observed taxa are the leaves of the tree, the unobserved internal nodes represent the most recent common ancestors of their descendant leaves.
Edges in the tree correspond to evolving species and their lengths represent elapsed evolutionary time.
We focus on rooted, bifurcating trees such as in \figref{fig:sd-example}.
The root node represents the most recent common ancestor of all of the leaves, time runs forward from the root to the leaves.
For mathematical convenience, we assume that the edge leading into the root is of infinite length and do not include its parent node in $ V $.
Each node $ i \in V $ has an associated time $ t_i \in T $ and, with the exception of the root, a parent $ \pa(i) $ and sibling $ \sib(i) $.
We refer to branches by their offspring node index.
A clade is a set of leaves which group together on the tree.

In this paper, we focus on the Stochastic Dollo (SD) model for binary trait presence/absence data \citep{nicholls08} and its extensions for missing data and rate heterogeneity \citep{ryder11} and lateral transfer \citep{kelly17}.
We briefly describe the generative model here and provide a full description in \appref{app:sd-model}.
New traits arise at rate $ \lambda $ along branches of the tree and are copied into offspring branches at speciation events.
Each trait on a branch dies at per capita rate $ \mu $ and attempts to transfer a copy of itself into other branches at rate $ \beta $.
Catastrophes occur at rate $ \rho $ along branches of the tree, each causing an instantaneous burst of evolutionary activity with strength $ \kappa $.
The set $ C $ indexes the catastrophes on the tree, where each catastrophe comprises a branch index and relative location along it.
We record the presence or absence of each trait at the leaves.
Data are missing-at-random in each taxon, $ \xi_i $ is the probability of observing the true state of a trait at leaf $ i $ and $ \Xi = \{\xi_i : i \in L\} $.
\figref{fig:sd-example} displays the history of a trait drawn from the SD model.

The likelihood calculation numerically integrates over unobserved trait events on the tree under the model.
We analytically integrate $ \lambda $ and $ \rho $ out of the posterior under our choice of priors.
The target of our inference is the posterior distribution on $ (g, \mu, \beta, \Xi, C, \kappa) $ when lateral trait transfer, missing data and catastrophes are included in the model.
We fix leaves at their sampling times, if known, and place an upper bound on the root age.
We incorporate prior knowledge of the tree through calibration constraints which restrict the space of admissible topologies to those with subtrees on specified clades and may also bound some ancestral node ages.
In certain settings, additional constraints may be required so that components of the model are identifiable and we discuss these issues further in \secref{sec:experiments}.

\subsection{Bayesian inference in phylogenetic models}
\label{sec:bayesian-approach}

We focus on MCMC approaches to inference so construct an ergodic Markov chain $ (X_s)_{s \geq 0} $ on $ \cX $ with transition kernel $ P $ that leaves the posterior distribution $ \pi $ invariant.
At each iteration $ s $, we draw $ X_s \sim P(X_{s - 1}, \cdot) $ via the MH algorithm.
We denote $ Q $ our proposal kernel and define the Hastings ratio $ h(X, X') = \pi(X') Q(X', X) / [\pi(X) Q(X, X')] $ for any pair of states $ X $ and $ X' $.
From initial state $ X_0 \sim \pi_0 $, the MH algorithm proceeds at each iteration $ s \geq 1 $ as follows:
\begin{enumerate}
    \item sample proposal $ X' \sim Q(X_{s - 1}, \cdot) $ and $ U \sim \Unif{0, 1} $;
    \item if $ U \leq h(X_{s - 1}, X') $, then $ X_s \leftarrow X' $, otherwise $ X_s \leftarrow X_{s - 1} $.
\end{enumerate}
As is standard in phylogenetic inference, we use a mixture of local proposal kernels $ (Q_m)_m $ with weights $ (\epsilon_m)_m $, where each $ Q_m $ proposes a different type of modification to the current state and $ \sum_m \epsilon_m = 1 $.
At each iteration of the MH algorithm, we sample a kernel $ Q \sim \sum_m \epsilon_m Q_m $ then use it to make a proposal.
In practice, we opt for proposals which make local rearrangements of the current state.
As described in \secref{sec:introduction}, both chains must have the same transition kernel $ P $ at each iteration so the proposal kernels and weights remain fixed throughout.

\begin{figure}[tb]
	\centering
     \begin{subfigure}[t]{0.30\textwidth}
         \centering
         \includegraphics[height=0.10\textheight]{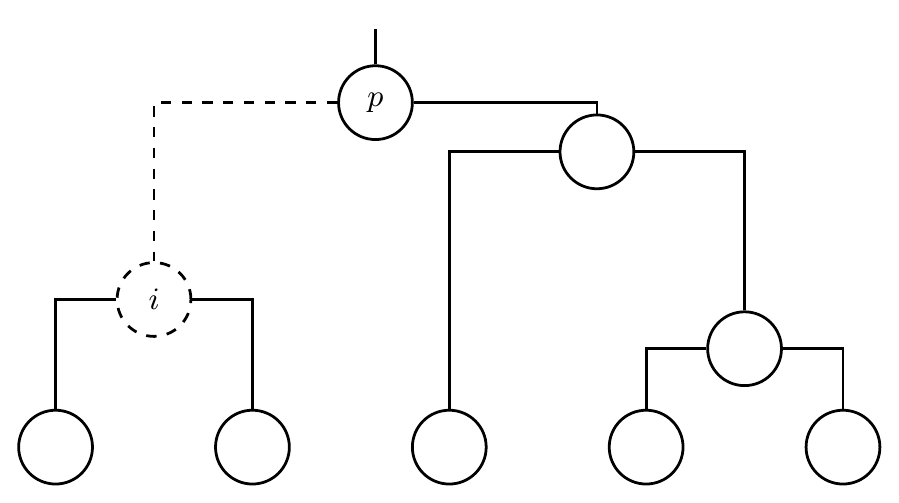}
         \caption{Select a subtree root index $ i $ from the nodes beneath the root, let $ p = \pa(i) $}
         \label{fig:spr-i}
     \end{subfigure}
     \hfill
     \begin{subfigure}[t]{0.33\textwidth}
         \centering
         \includegraphics[height=0.10\textheight]{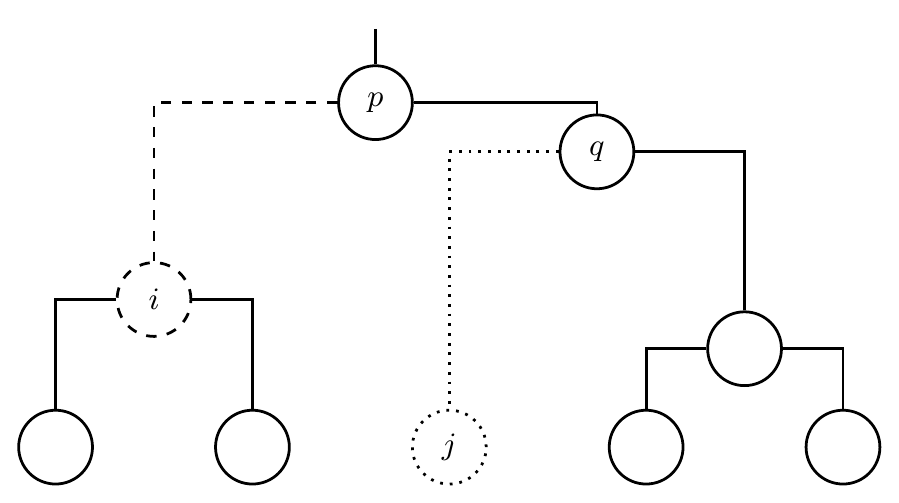}
         \caption{Select a destination branch index $ j \neq i $ with parent $ q =  \pa(j) $ such that $ t_q > t_i $}
         \label{fig:spr-j}
     \end{subfigure}
     \hfill
     \begin{subfigure}[t]{0.30\textwidth}
         \centering
         \includegraphics[height=0.10\textheight]{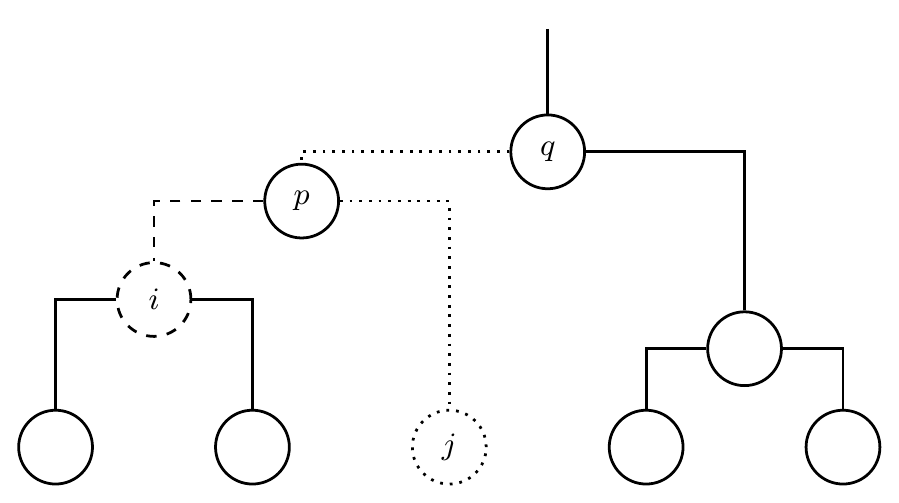}
         \caption{Detach $ p $ and reattach it at a randomly chosen time $ t_p' > t_i $ along $ j $}
         \label{fig:spr-t}
     \end{subfigure}
    \caption{
        Example of a rooted subtree prune-and-regraft proposal.
        For the reverse move, we detach $ p $ and reattach it along the branch leading into $ q $.
    }
    \label{fig:spr}
\end{figure}

The rooted subtree prune-and-regraft (SPR) proposal, also known as a Wilson--Balding move \citep{drummond02}, is one of the primary methods for exploring the space of clock-constrained trees.
In an SPR proposal, a randomly chosen subtree is moved to a new location while respecting the time-ordering of nodes.
Given the current state $ X $ with root index $ r $, we can sample an SPR proposal as follows:
\begin{enumerate}
    \item sample a subtree root $ i \sim \Unif{V \setminus \{r\}} $;
    \item select a destination branch $ j \sim \Unif{\{j' \in V : j' \neq i, t_{\pa(j')} > t_i\}} $, we choose at random from the branches where we could reattach $ \pa(i) $ at some time $ t_{\pa(i)}' > t_i $;
    \item draw a new time for $ \pa(i) $ along branch $ j $:
    \begin{itemize}
        \item if $ j \neq r $, then $ t_{\pa(i)}' \sim \Unif{t_i \vee t_j, t_{\pa(j)}} $;
        \item if $ j = r $, then draw $ \delta \sim \Exp{\theta} $ and set $ t_{\pa(i)}' \leftarrow t_j + \delta $, where $ \theta $ is a constant;
    \end{itemize}
    \item detach $ \pa(i) $ from its location in $ X $ and reattach it at time $ t_{\pa(i)}' $ along $ j $ to form the proposed state $ X' $.
\end{enumerate}
\figref{fig:spr} illustrates this proposal.
If clade constraints are imposed, then we select a destination branch from those under the same restrictions as $ i $.
If $ j = \pa(i) $ or $ j = \sib(i) $, we are only proposing to change $  t_{\pa(i)} $ and not the topology, then the move fails and $ X' \leftarrow X $.
We use a separate proposal to change the age of an individual node, so rather than account for the mixture of proposal densities for the two move types in the Hastings ratio, the SPR move fails in this setting.
We could modify the SPR move to always propose a new topology but have not done so.

The 19 local kernels we use are listed in \tabref{tab:proposals} of \appref{app:coupling-proposals-phylogenetic} and described in detail therein.
Broadly speaking, the following classes of proposals are of interest for this work.
\begin{itemize}
    \item We explore the space of rooted topologies through SPR moves and by swapping randomly chosen pairs of subtrees.
    \item We propose to change a single node time $ t_i \in T $ by sampling on the interval between its eldest child and parent, and groups of nodes times through rescaling by a common factor.
    \item We propose updates to the catastrophe set $ C $ through addition or deletion of single catastrophes, changing a catastrophe's position within a branch or moving it to a neighbouring branch, or resampling the entire catastrophe set on a branch with a draw from the prior.
    \item We propose updates to the scalar parameters $ \mu $, $ \beta $, $ \kappa $ and $ \Xi $ by rescaling.
\end{itemize}
If a proposal violates a model constraint then it is rejected.
These are just a small subset of the proposals in use in Bayesian phylogenetic inference but are sufficient to construct ergodic chains \citep{drummond02}.


\subsection{Diagnosing mixing and convergence of phylogenetic MCMC samplers}
\label{sec:convergence}

Theoretical analyses of Markov chains on the space of phylogenies are restricted to relatively simple settings.
For a Markov chain with transition kernel $ P $ on finite space $ \cX $, let $ \tau_{\mathrm{mix}} = \min\{s \geq 0 : \max_{x \in \cX} d_{\mathrm{TV}}(P^s(x, \cdot), \pi) \leq 1/4\} $, its mixing time, and $ \tau_{\mathrm{rel}} = 1 / (\text{spectral gap of $ P $}) $, its relaxation time.
\citet{aldous00} constructs a random walk on the space of unrooted tree topologies and derives an upper bound on $ \tau_{\mathrm{mix}} $ and a lower bound on $ \tau_{\mathrm{rel}} $, both functions of the number of leaves.
Generally, $ \tau_{\mathrm{rel}} = \cO(\tau_{\mathrm{mix}}) $ so the lower bound on $ \tau_{\mathrm{rel}} $ also applies to $ \tau_{\mathrm{mix}} $.
\citet{spade14} upper bound $ \tau_{\mathrm{rel}} $ for random walks on rooted trees via nearest-neighbour interchange and SPR proposals.
\citet{mossel06} show that $ \tau_{\mathrm{mix}} $ can grow exponentially with the number of traits when data are drawn from a mixture on two trees.

Practitioners typically assess MCMC mixing and convergence in phylogenetic problems through the comparison of several runs of the same algorithm from different initial states under the principle that graphical and numerical summaries of samples should exhibit similar behaviour as chains converge \citep{nascimento17,bromham18}.
Even if the asymptotic values of these summaries are well defined, their finite-sample behaviour is generally unknown so their use as convergence diagnostics is typically subjective or based on guidelines developed from a small collection of test data sets and models.
The summaries used are often marginal on each parameter or use low-dimensional projections from the space of trees, so their power to quantify convergence jointly across all components of the model is unknown, particularly as models grow in size and complexity.

\texttt{Tracer} \citep{rambaut18} is a popular tool for exploring the output of one or more chains through graphical and numerical summaries of parameter samples: the graphical summaries include marginal and joint traceplots, histograms and contour plots; the numerical summaries include the mean, standard deviation and effective sample size (ESS).
The ESS of an MCMC sample estimates the equivalent number of independent draws from the target posterior.
\citet{magee21} develop a number of ESS measures for tree topologies and assess their ability to quantify Monte Carlo error in estimating posterior summaries.
A minimum ESS for each parameter is a frequent stopping criteria in phylogenetic analyses but recommended thresholds vary: \citet{mrbayes} suggest at least $ 100 $; \texttt{Tracer} flags values below $ 100 $ or $ 200 $; \citet{drummond06relaxed}, \citet{dellicour21} and \citet{hoffman21} recommend $ 200 $; \citet{fabreti22} advocate $ 625 $ under the assumption that marginal posterior distributions are Gaussian; \citet{nascimento17} propose $ 10^3 $ or $ 10^4 $ but remark that $ 200 $ is common.
\texttt{MrBayes} \citep{ronquist12} uses the Potential Scale Reduction Factor (PSRF) \citep{gelman92} as a convergence diagnostic for parameters, branch lengths and overall tree length.
The PSRF approaches $ 1 $ as chains converge but the thresholds used to diagnose convergence vary: \citet{mrbayes} recommend between $ 1 $ and $ 1.2 $ for all components; \citet{vats21revisiting} draw a connection between PSRF and ESS and develop a principled method to choose a convergence threshold which is typically much lower than those used in practice.

Rather than diagnose convergence directly on the high dimensional space of tree topologies, many approaches first project samples to the lower dimensional space of splits.
In an unrooted tree, which we can form from a rooted tree by deleting the root node and merging its offspring edges, each edge corresponds to a bipartite split of the taxa.
We can recover a tree topology from a set of compatible splits \citep{bryant99} and splits form the basis of many convergence diagnostics.
\citet{beiko06} compare the posterior distributions on split frequencies from an extremely long chain to multiple shorter chains and conclude that likelihood traceplots often stabilise well before the corresponding split distributions.
\citet{nylander08} propose to diagnose convergence from trace plots of split frequencies across multiple independent chains, as well as cumulative split frequencies, split presence/absence and tree distances within and across chains.
\citet{ali17} assess convergence on a pair of chains by conducting Mann--Whitney U-tests on parameters and $ \chi^2 $ tests on split distributions.
\citet{fabreti22} propose a Kolmogorov--Smirnov test on split frequency distributions with an ESS greater than $ 625 $ for each split.
\citet{meyer21} uses multiple long independent runs to estimate a reference posterior distribution on splits, then for subsequent experiments diagnoses convergence when the Euclidean distance between the sampled and reference split distributions falls below $ 0.02 $.

The Average Standard Deviation of Split Frequencies (ASDSF) \citep{lakner08} is widely used as a convergence diagnostic.
Suppose we have $ M $ independent chains and they visited $ K $ unique splits, let $ f_m^{(k)} $ denote the proportion of sampled topologies in chain $ m $ which imply split $ k $ and $ \bar{f}^{(k)} = M^{-1} \sum_{m = 1}^M f_m^{(k)} $.
\texttt{MrBayes} \citep{ronquist12,mrbayes} defines the ASDSF as
\begin{equation}
    \label{eq:asdsf}
    \mathrm{ASDSF} = \frac{1}{K} \sum_{k = 1}^K \sqrt{\frac{1}{M - 1} \sum_{m = 1}^M (f_m^{(k)} - \bar{f}^{(k)})^2}.
\end{equation}
By default, \texttt{MrBayes} runs two independent quartets of parallel-tempered chains and in computing the ASDSF ignores any split $ k $ where $ f_m^{(k)} \leq 0.1 $ in all $ M $ chains.
\texttt{RWTY} \citep{warren17} computes the ASDSF on cumulative disjoint sliding windows.
Although the ASDSF will decay to 0 as chains converge and the window size increases, the accuracy of estimates depend on a variety of factors and its behaviour has only been studied empirically.
\citet{fourment19} remark that ``Typically an ASDSF below $ 0.01 $ is taken to be evidence that two MCMC analyses are sampling the same distribution,'' but instead use the root mean squared deviation (RMSD) of split frequencies in their study and propose cut-offs for \emph{good agreement}, \emph{acceptable agreement} and \emph{substantial disagreement} with ground-truth estimates.
They also compare split frequency distributions using the Kullback--Leibler divergence and obtain broadly similar conclusions.
As the error in estimating frequencies of splits with low posterior support is relatively high, \citet{fabreti22} propose a modification of ASDSF to account for the true frequency of each split.

The ASDSF can be used as a stopping rule in \texttt{MrBayes}, where the default setting is to stop sampling once the ASDSF calculated on the most recent 75\% of samples decreases below $ 0.05 $.
For a variety of real data sets, \citet{whidden15} use \texttt{MrBayes} to draw samples from a varying number of independent chains (two to eight) until their ASDSF falls below $ 0.01 $ or they reach $ 10^8 $ iterations.
They compute the RMSD between the split frequency distributions and ground truth estimates from chains of length $ 10^9 $ iterations, discarding the first 25\% of samples in each case, and find that the ASDSF from two independent single-chain or parallel-tempered ensembles is often insufficient to diagnose non-convergence.
Although increasing the number of chains reduces the error in estimating ASDSF, thus providing a more stringent convergence diagnostic, eight chains were insufficient to diagnose convergence in some of their experiments with multimodal distributions on topologies.

To estimate the speed of convergence and sampling efficiency of various tree proposal kernels, \citet{hohna08clock} compute the maximum discrepancy in posterior support for clades against estimates from extremely long chains, and diagnose convergence once it has decreased below a fixed threshold.
\citet{whidden15} propose a number of graphical tools to visualise tree space and identify bottlenecks in phylogenetic posterior distributions.
\citet{whidden15} also propose to diagnose convergence with a topological variation of the PSRF using SPR distances and which empirically exhibits similar behaviour to the ASDSF.
The \texttt{TopologyTracer} tool within \texttt{BEAST} \citep{suchard18} computes the distance between a reference tree and each tree visited by the MCMC, the output can be loaded into \texttt{Tracer} for further analysis.
\citet{lanfear16} propose to assess convergence on topologies by plotting a distance between pairs of sampled trees against the number of MCMC transitions between them.
\citet{brown19} analyse properties of Fr\'{e}chet means and variances of trees in the treespace of \citet{billera01}, and suggest that variances computed on sliding windows of trees may be used to assess convergence.
\citet{kim20} define an $ L_2 $-medoid with respect to their tree metrics then assess convergence of their MCMC samplers by plotting the distance between a running $ L_2 $-medoid and one computed using all of the samples.
\citet{magee21} propose to compare split probabilities across independent chains with error bars computed using their tree ESS measures as a convergence diagnostic.
\citet{smith21robust} analyses a number of commonly used metrics on trees and concludes that many of them are unsuited for phylogenetic inference problems.

\citet{harrington20} perform a comprehensive empirical study of 18,588 phylogenetic analyses comparing various popular convergence diagnostics.
The authors find similar behaviour between many diagnostic tools but also some incongruence between them.
The ESS for every non-topological parameter was above $ 200 $ in 98.8\% of their analyses while the corresponding PSRFs were all below $ 1.02 $ in 98\%.
The approximate topological ESS of \citet{lanfear16} was above $ 200 $ in 97.3\% of their analyses but the ASDSF was below $ 0.01 $ only 37.5\% of the time.
\citet{whidden15} observed that an ASDSF below $ 0.01 $ coincided with other convergence diagnostics being satisfied, such as the PSRF for branch lengths and ESS for the tree length.

The behaviour of a phylogenetic MCMC algorithm is often specific to the model and data set so we would like to be able to quantify mixing and convergence for a given problem without relying on ad hoc guidelines developed on other problems.
Although many of the methods we describe above are asymptotically consistent, in that they converge once the chain has reached stationarity, their finite-sample behaviour, power to detect convergence and the effect on inference of using them as stopping times are not well understood.
In contrast, couplings allow us to estimate an upper bound on convergence in total variation distance and diagnose convergence jointly across all components of the model.
This approach does not require arbitrary convergence thresholds like ASDSF or PSRF, running longer chains beyond their meeting times will not change our estimate of convergence on a given problem, and we can use the same samples for diagnosing convergence and performing unbiased inference.


\section{Coupling generic transition kernels}
\label{sec:couplings-generic}

A coupling of $ X \sim p $ and $ Y \sim q $ on the same general space $ \cX $ is any joint distribution on $ \cX \times \cX $ whose marginals are $ p $ and $ q $.
For $ (X, Y) $ drawn from a coupling of $ p $ and $ q $, the coupling inequality states that $ \PP(X \neq Y) \geq d_{\mathrm{TV}}(p, q) $, with equality achieved by a maximal coupling.
Maximal couplings are typically not unique, we use a maximal coupling with independent residuals which samples $ X $ and $ Y $ independently when $ X \neq Y $.
For the most part, we use the following general approach to sample $ (X,  Y) $ from such a coupling of $ p $ and $ q $, which requires that we can evaluate their densities (also denoted $ p $ and $ q $) at every point in $ \cX $ \citep{lindvall2002lectures,jacob20}:
\begin{enumerate}
    \item sample $ X \sim p $ and $ U \sim \Unif{0, 1} $;
    \item if $ U \leq q(X) / p(X) $, then return $ Y \leftarrow X $;
    \item otherwise, draw $ Y' \sim q $ and $ U' \sim \Unif{0, 1} $ until $ U' > p(Y') / q(Y') $, then return $ Y \leftarrow Y' $.
\end{enumerate}
The expected computational cost of the above algorithm is two units, where each unit is one sample and two density evaluations, but its variance increases without bound as $ d_{\mathrm{TV}}(p, q) \rightarrow 0 $ \citep{jacob20}.
In certain settings, such as the following example, we can sample directly from a maximal coupling.
A maximal coupling of $ X \sim \Bern{p} $ and $ Y \sim \Bern{q} $ has $ \PP(X = 0, Y = 0) = (1 - p) \wedge (1 - q) $ and $ \PP(X = 1, Y = 1) = p \wedge q $, where $ a \wedge b = \min(a, b) $.
We can sample from this coupling by drawing $ U \sim \Unif{0, 1} $ and setting $ (X, Y) \leftarrow (\ind{U \leq p}, \ind{U \leq q}) $.
\appref{app:sample-maximal-couplings} discusses in more detail the algorithms we use to sample from maximal couplings.

\secref{sec:introduction} describes a lag-$ l $ coupling of Markov chains $ (X_s)_{s \geq 0} $ and $ (Y_s)_{s \geq 0} $ on space $ \cX $ with common initialisation $ \pi_0 $, target $ \pi $, Markov transition kernel $ P $ and MH proposal kernel $ Q $.
At for each iteration $ s \geq l $, we sample $ (X_s, Y_{s - l}) \sim \bar{P}((X_{s - 1}, Y_{s - l - 1}), \cdot) $, a coupling of the marginal transition kernels $ P(X_{s - 1}, \cdot) $ and $ P(Y_{s - l - 1}, \cdot) $.
We are free to choose the coupling $ \bar{P} $ provided that it produces a satisfactory distribution on meeting times $ \tau^{(l)} $ and that chains do not separate after meeting.
Although not a maximal coupling of the marginal transition kernels, \citet{jacob20} demonstrate that the following widely applicable approach due to \citet{johnson98} produces meeting times $ \tau^{(l)} $ with appropriate tails in a variety of problems.
At iteration $ s > l $:
\begin{enumerate}
    \item sample proposals $ (X', Y') \sim \bar{Q}((X_{s - 1}, Y_{s - l -1}), \cdot) $, a maximal coupling of $ Q(X_{s - 1}, \cdot) $ and $ Q(Y_{s - l - 1}, \cdot) $;
    \item accept or reject the proposals by sampling from a maximal coupling:
    \begin{itemize}
        \item draw $ U \sim \Unif{0, 1} $;
        \item if $ U \leq h(X_{s - 1}, X') $, then $ X_s \leftarrow X' $, otherwise $ X_s \leftarrow X_{s - 1} $;
        \item if $ U \leq h(Y_{s - l - 1}, Y') $, then $ Y_{s - l} \leftarrow Y' $, otherwise $ Y_{s - l} \leftarrow Y_{s - l - 1} $.
    \end{itemize}
\end{enumerate}
As we sample from a maximal coupling at both the proposal and acceptance steps, chains which meet will stay together by design.
The closer $ \bar{P} $ is to being maximal, the earlier chains will meet and the tighter the coupling TV bound \eqref{eq:tv-bound} will be.
\citet{wang21} describe techniques to sample from a maximal coupling of transition kernels.

\section{Coupled MCMC for phylogenetic inference}
\label{sec:coupled-mcmc-phylogenetics}

We follow the approach in \secref{sec:couplings-generic} and construct our coupling of phylogenetic transition kernels by separately sampling from couplings of the proposal distributions and accept/reject steps of the MH algorithm.
Similar to the Gibbs sampling example of \citet{jacob20} and the mixture of HMC and random walk kernels of \citet{heng19}, our coupling of proposal kernels is a mixture of coupled local kernels.
We first draw a coupled local kernel $ \bar{Q} \sim \sum_m \epsilon_m \bar{Q}_m $, where each $ \bar{Q}_m $ is a coupling of $ Q_m $ with itself, then sample a pair of proposals from $ \bar{Q} $.
We cannot construct a maximal coupling of many of our local proposal kernels as we cannot tractably evaluate their densities at every point in $ \cX $.
As each local kernel is a composition of simple operations involving draws from standard probability distributions, we instead sample from a maximal coupling at each step of a proposal.

In order for the local kernels to overlap and coupled proposals to be accepted or rejected together, we require the other elements of the states to be close.
We introduce a housekeeping operation, described in \secref{sec:housekeeping}, to identify common components of trees so that we can make coupled proposals which have similar or identical effects in both states.
Through our construction, we obtain a non-trivial coupling of the transition kernels and distributions on meeting times which decay at most geometrically in practice.
\algoref{alg:coupled-mcmc} describes our implementation of the coupled MCMC algorithm of \citet{biswas19} to sample a meeting time $ \tau^{(l)} $.
If a minimum of $ S $ samples are desired for inference with unbiased estimators, then we would run \algoref{alg:coupled-mcmc} for $ S \vee \tau^{(l)} $ iterations in total, sampling from the marginal kernel after the chains meet, and return $ (X_0, X_1, \dotsc, X_{S \vee \tau^{(l)}}) $ and $ (Y_0, Y_1, \dotsc, Y_{\tau^{(l)} - l - 1}) $; see \citet{jacob20} for further details.

\begin{algorithm}[tb]
    \DontPrintSemicolon
    \SetKwBlock{Begin}{}{}
    $ X_0 \sim \pi_0 $ \;
    \For{$ s = 1 $ \KwTo $ l $}{
        \Begin({$ X_s \sim P(X_{s - 1}, \cdot) $} \tcp*[f]{Sample from marginal kernel}){
            $ Q \sim \sum_m \epsilon_m Q_m $ \;
            $ X' \sim Q(X_{s - 1}, \cdot) $ \;
            $ U \sim \Unif{0, 1} $ \;
            \leIf{$ U \leq h(X_{s - 1}, X') $}{$ X_s \leftarrow X' $}{$ X_s \leftarrow X_{s - 1} $}
        }
    }
    $ Y_0 \sim \pi_0 $ \;
    \While{$ X_s \neq Y_{s - l} $}{
        $ Y_{s - l} \leftarrow \textsc{Housekeeping}(X_s, Y_{s - l}) $ \;
        $ s \leftarrow s + 1 $ \;
        \Begin({$ (X_s, Y_{s - l}) \sim \bar{P}((X_{s - 1}, Y_{s - l - 1}), \cdot) $} \tcp*[f]{Sample from coupling of kernels}){
            $ \bar{Q} \sim \sum_m \epsilon_m \bar{Q}_m $ \;
            $ (X', Y') \sim \bar{Q}((X_{s - 1}, Y_{s - l - 1}), \cdot) $ \tcp*[r]{Maximal coupling at each step}
            $ U \sim \Unif{0, 1} $ \;
            \leIf{$ U \leq h(X_{s - 1}, X') $}{$ X_s \leftarrow X' $}{$ X_s \leftarrow X_{s - 1} $}
            \leIf{$ U \leq h(Y_{s - l - 1}, Y') $}{$ Y_{s - l} \leftarrow Y' $}{$ Y_{s - l} \leftarrow Y_{s - l - 1} $}
        }
    }
    $ \tau^{(l)} \leftarrow s $ \;
    \caption{Sample a meeting time $ \tau^{(l)} $ from the lag-$ l $ coupled MCMC algorithm with housekeeping and a mixture of proposal distributions.}
    \label{alg:coupled-mcmc}
\end{algorithm}

\subsection{Housekeeping}
\label{sec:housekeeping}

\begin{figure}[tb]
	\centering
     \begin{subfigure}[b]{0.32\textwidth}
         \centering
         \includegraphics[width=\textwidth]{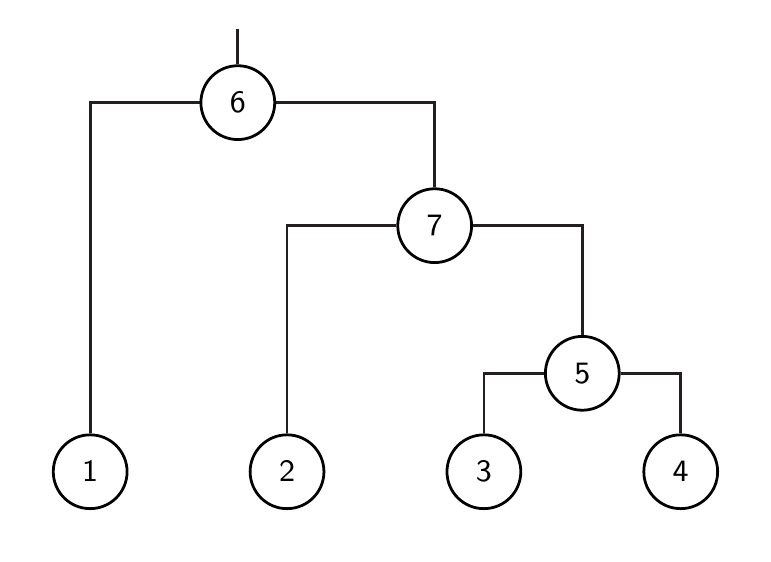}
         \caption{State $ X $}
         \label{fig:housekeeping-x}
     \end{subfigure}
     \hfill
     \begin{subfigure}[b]{0.32\textwidth}
         \centering
         \includegraphics[width=\textwidth]{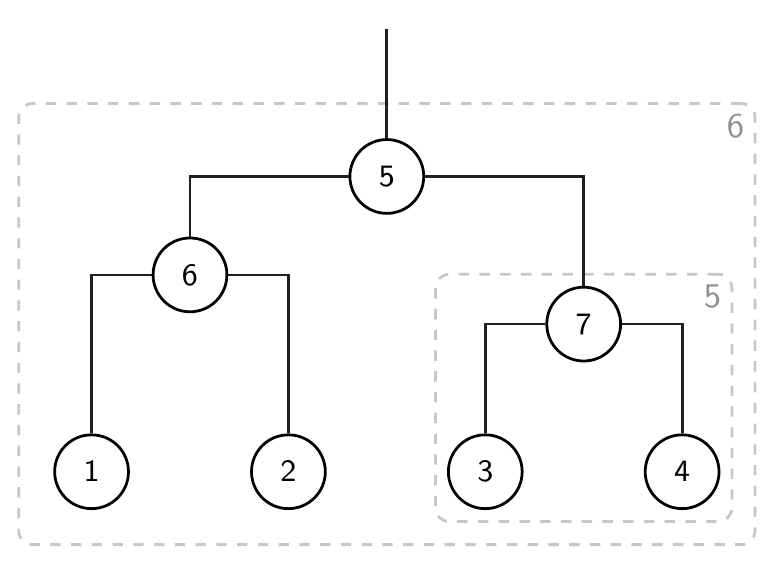}
         \caption{State $ Y $ before}
         \label{fig:housekeeping-y1}
     \end{subfigure}
     \hfill
     \begin{subfigure}[b]{0.32\textwidth}
         \centering
         \includegraphics[width=\textwidth]{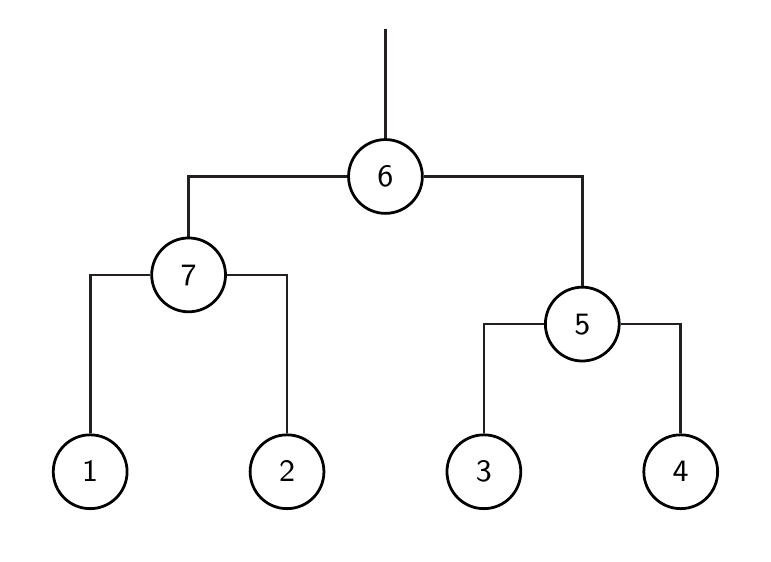}
         \caption{State $ Y $ after}
         \label{fig:housekeeping-y2}
     \end{subfigure}
    \caption{
        Housekeeping permutes node indices in $ Y $ to match $ X $ in subtrees of common clades.
        The clade with leaves $ 3 $ and $ 4 $ is common to both states so we assign its subtree root the same index in both states, likewise the root of the overall tree.
    }
    \label{fig:housekeeping}
\end{figure}

Our proposals which affect the tree or catastrophes comprise multiple steps, we typically sample a target node or branch $ i $ followed by a sequence of proposals which depend on $ i $.
As internal node indices are arbitrary, only the leaf nodes are labelled, we need to assign the same indices to components of the topology common to both states $ X $ and $ Y $.
After an update to the topology in either chain, we perform a housekeeping operation which identifies clades common to both states and assigns their subtree roots the same index, then for each common clade we use the same subset of labels for the remaining internal nodes of the subtrees, even when they are different.
\figref{fig:housekeeping} illustrates this operation.

Through a combination of housekeeping and sampling node and branch indices from maximal couplings, we tend to select and modify common tree components together.
The indices of the root and leaf nodes and the branches leading into them are always identical in both states: even if the $ X $ and $ Y $ topologies are different, we can propose moves which would bring the states closer together.
As states $ X $ and $ Y $ become closer and more components overlap, we make more similar proposals.
We illustrate this in \secref{sec:coupling-structural-moves} below.
Housekeeping is key to the success of our algorithm, without it we would often sample from independent couplings of distributions on tree components and this would be unlikely to satisfy the assumption on the tail of $ \tau^{(l)} $; indeed, we found this to be the case in our experiments.
\citet{ju20} and \citet{trippe21} perform similar housekeeping operations to relabel semantically equivalent partitions of data points.

As proposals on the topology are rarely accepted, housekeeping is only required infrequently over the course of sampling from the coupled kernel, so its contribution to the overall running time is relatively minor.
The housekeeping operation we describe requires a pair of subtrees to have identical sets of descendant leaves in order for them to treated as common components, so it is not robust to minor differences such as a single leaf node.
We leave for future work the development of more robust, efficient and informative housekeeping operations.

\subsection{Coupling proposal distributions in phylogenetic models}

Each member of our family of proposal distributions acts primarily on a component of either the topology, the node times $ T $, the set $ C $ of catastrophes on branches, or the scalar parameters of the diversification model.
A move on the topology may additionally affect node times, catastrophes or rate parameters, and a move on node times affects the location of catastrophes.
We couple these proposals by sampling from a maximal coupling at each step.
We now describe how to couple some of these proposals and provide a full description of our kernel couplings in \appref{app:coupling-proposals-phylogenetic}.

\subsubsection{Structural moves}
\label{sec:coupling-structural-moves}

Recall the rooted subtree prune-and-regraft (SPR) move described in \secref{sec:bayesian-approach} and illustrated in \figref{fig:spr} which moves the parent of subtree root $ i $ to a new time $ t_{\pa(i)}' $ along branch $ j $.
We couple this proposal by drawing subtree roots $ (i^{(X)}, i^{(Y)}) $, destination branches $ (j^{(X)}, j^{(Y)}) $ and times $ (t_{\pa(i^{(X)})}^{(X) \prime}, t_{\pa(i^{(Y)})}^{(Y) \prime}) $ from maximal couplings of their respective distributions.
\begin{enumerate}
    \item Sample a pair of subtree roots $ (i^{(X)}, i^{(Y)}) $ from a maximal coupling of $ \Unif{V \setminus \{r^{(X)}\}} $ and $ \Unif{V \setminus \{r^{(Y)}\}} $, discrete Uniform distributions on nodes beneath the root in each state.
    As $ r^{(X)} = r^{(Y)} = r $ through housekeeping, the distributions are identical so draw $ i \sim \Unif{V \setminus \{r\}} $ and set $ (i^{(X)}, i^{(Y)}) \leftarrow (i, i) $.
    \item The set of possible destination branches for the subtree with root $ i $ in state $ X $ is $ J_i^{(X)} = \{j' \in V : j' \neq i, t_{\pa(j')}^{(X)} > t_i^{(X)}\} $, the branches where we could reattach $ \pa(i) $ at a new time $ t_{\pa(i)}^{(X)\prime} > t_i^{(X)} $, and similarly define $ J_i^{(Y)} $ for state $ Y $.
    Draw destination branches $ (j^{(X)}, j^{(Y)}) $ from a maximal coupling of discrete Uniform distributions $ \Unif{J_i^{(X)}} $ and $ \Unif{J_i^{(Y)}} $.
    \item The range of valid times for $ \pa(i) $ along branch $ j $ in state $ X $ is the interval $ I_{i,j}^{(X)} = (t_i^{(X)} \vee t_j^{(X)}, t_{\pa(j)}^{(X)}) $, and likewise $ I_{i,j}^{(Y)} $ for state $ Y $.
    Sample new node times $ (t_{\pa(i)}^{(X) \prime}, t_{\pa(i)}^{(Y) \prime}) $ from a maximal coupling of continuous Uniform distributions $ \Unif{I_{i,j^{(X)}}^{(X)}} $ and $ \Unif{I_{i,j^{(Y)}}^{(Y)}} $ when $ j^{(X)} \neq r $ and $ j^{(Y)} \neq r $, $ \Exp{\theta} $ distributions truncated from below at $ t_r^{(X)} $ and $ t_r^{(Y)} $ when $ j^{(X)} = j^{(Y)} = r $, or a combination of the two otherwise.
\end{enumerate}
As in the marginal SPR proposal, this move fails in state $ X $ if we propose $ j^{(X)} = \pa(i) $ or $ j^{(X)} = \sib(i) $ as the destination branch since it corresponds to the current topology and we have a separate move to modify a single node age, and likewise in state $ Y $.
If the coupled proposal fails at an intermediate step in one chain, the other chain continues to sample according to the marginal SPR move.
We form the proposed state $ X' $ by detaching $ \pa(i) $ from its current location in state $ X $ then reattaching it along branch $ j^{(X)} $ at time $ t_{\pa(i)}^{(X) \prime} $, and similarly form $ Y' $ from $ Y $.
If catastrophes are included in the model, then we also couple the update to their number and locations on branches affected by the SPR move.

We sample destination branches $ (j^{(X)}, j^{(Y)}) $ from a maximal coupling regardless of whether $ i $ is a common subtree or not, and likewise $ (t_{\pa(i)}^{(X) \prime}, t_{\pa(i)}^{(Y) \prime}) $.
We have not investigated whether it would be beneficial to sample from alternative couplings when proposing to move different subtrees.
In any case, the coupling preserves the marginal properties of the move and ensures that proposals are identical when the target components are equal.
Under our coupling, the probability of choosing the same destination branch index in both states is
\begin{equation}
    \label{eq:spr-j}
    \PP(j^{(X)} = j^{(Y)} \given i)
        = \frac{\abs{J_i^{(X)} \cap J_i^{(Y)}}}{\abs{J_i^{(X)}} \vee \abs{J_i^{(Y)}}},
\end{equation}
where $ \abs{\cdot} $ denotes set cardinality.
\eqnref{eq:spr-j} approaches $ 1 $ as the trees get closer together.
When regrafting to a destination below the root in each state, $ \PP(t_{\pa(i)}^{(X) \prime} = t_{\pa(i)}^{(Y) \prime} \given i, j^{(X)}, j^{(Y)}) $ has a similar form to \eqnref{eq:spr-j} but with time intervals $ I_{i,j^{(X)}}^{(X)} $ and $ I_{i,j^{(Y)}}^{(Y)} $ replacing branch index sets $ J_i^{(X)} $ and $ J_i^{(Y)} $.

\begin{figure}[tbp]
	\centering
    \begin{subfigure}[t]{0.495\textwidth}
        \includegraphics[width=\textwidth]{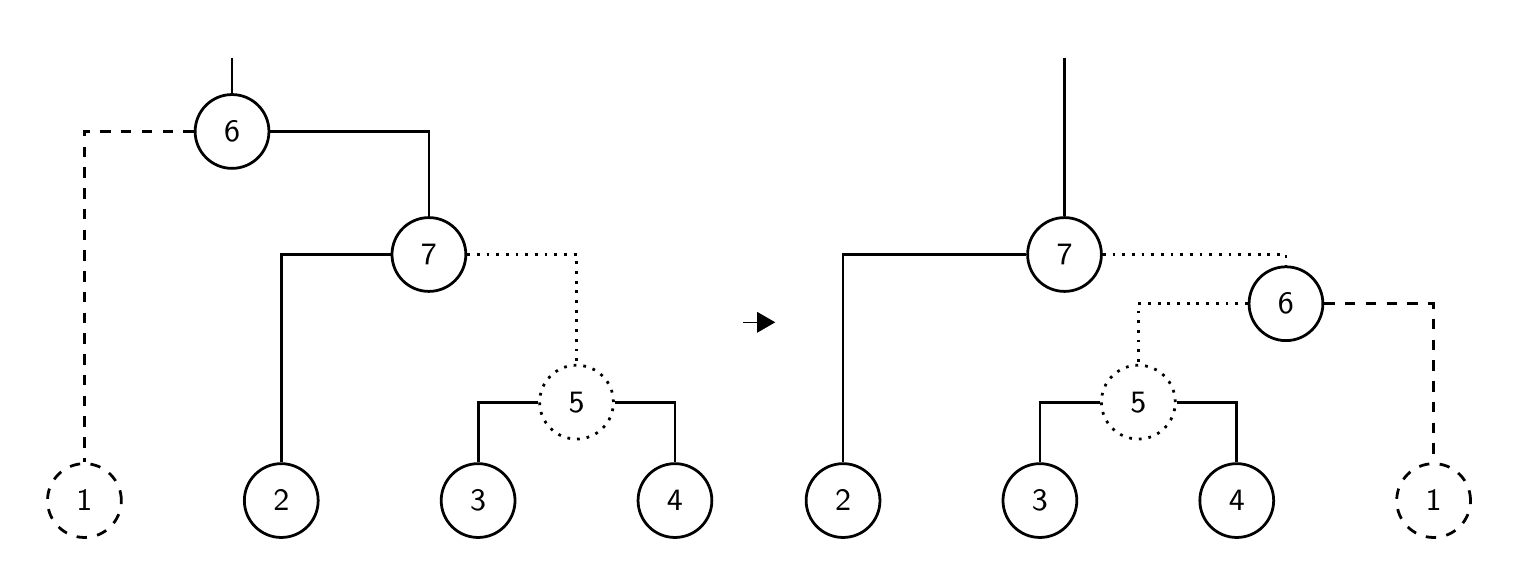}
        \caption*{$ j^{(X)} = 5 $}
    \end{subfigure}
    \hfil
    \begin{subfigure}[t]{0.495\textwidth}
        \includegraphics[width=\textwidth]{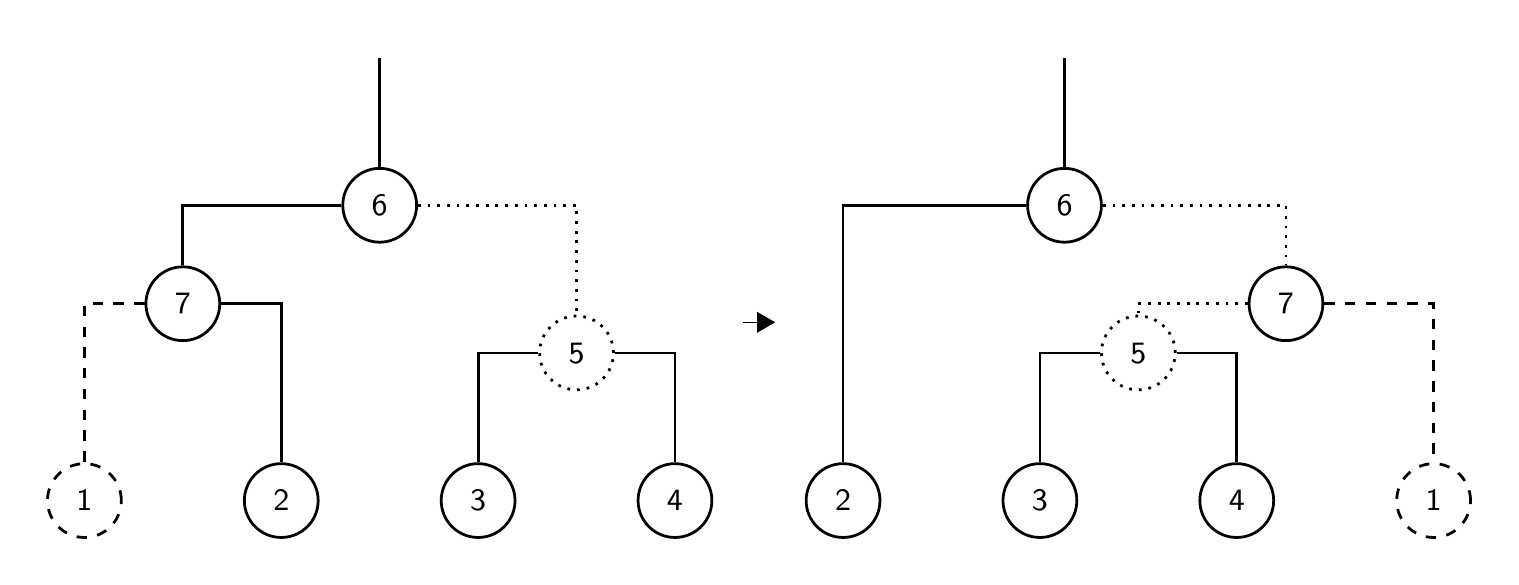}
        \caption*{$ j^{(Y)} = 5 $}
    \end{subfigure}

    \begin{subfigure}[t]{0.495\textwidth}
        \includegraphics[width=\textwidth]{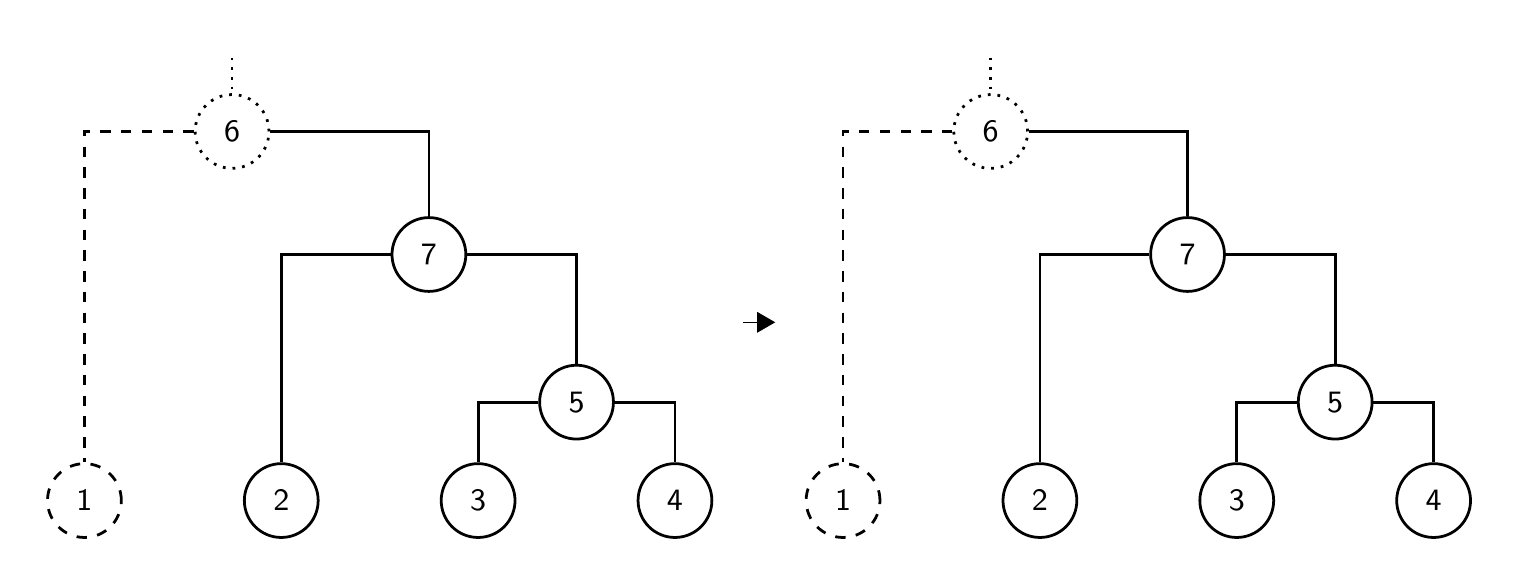}
        \caption*{$ j^{(X)} = 6 $, move fails}
    \end{subfigure}
    \hfil
    \begin{subfigure}[t]{0.495\textwidth}
        \includegraphics[width=\textwidth]{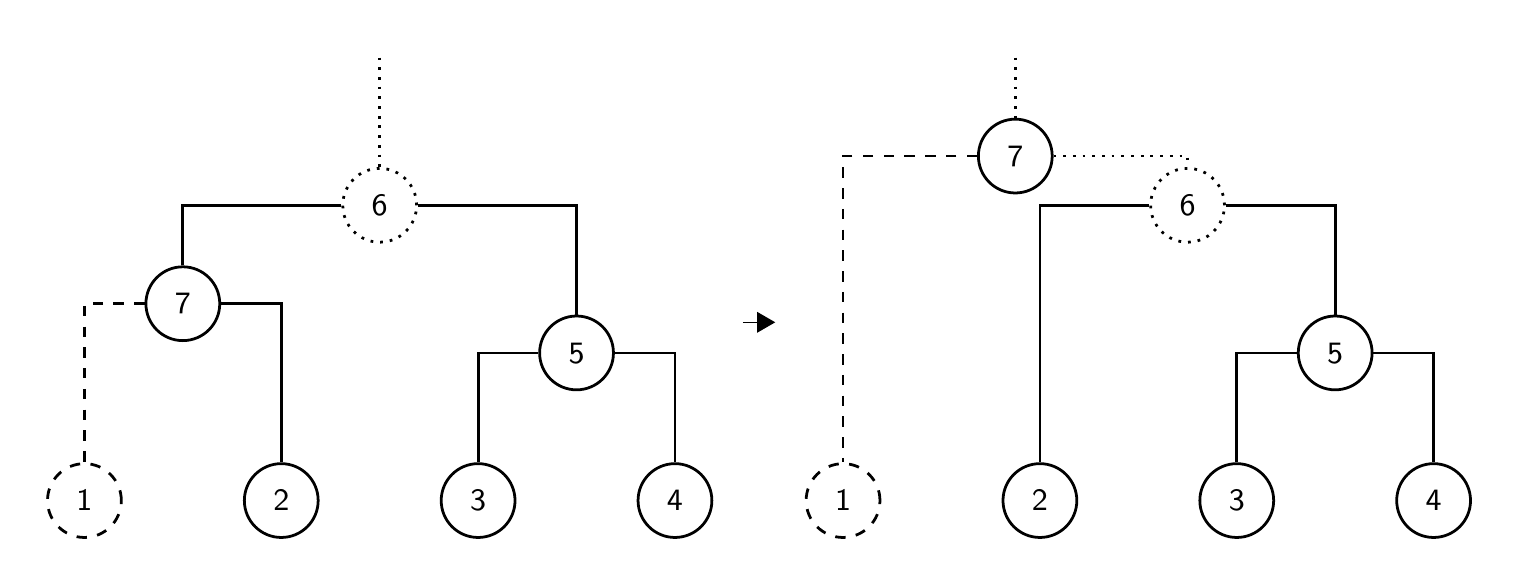}
        \caption*{$ j^{(Y)} = 6 $}
    \end{subfigure}
    \caption{
        Two possible outcomes of a coupled SPR move from $ X \rightarrow X' $ (left column) and $ Y \rightarrow Y' $ (right column) when the current states $ X $ and $ Y $ are the trees in \figref{fig:housekeeping} and we propose to move the subtrees with root $ i^{(X)} = i^{(Y)} = 1 $.
        The possible destination index sets are identical in this example so $ j^{(X)} = j^{(Y)} $ when sampling from a maximal coupling.
    }
    \label{fig:spr-coupled}
\end{figure}

Through housekeeping, common components have the same labels so tend to be selected together for SPR moves.
To illustrate this property, suppose that the current state of the chains is the pair of trees in \figref{fig:housekeeping} after housekeeping, the trees are currently one SPR move apart, and we attempt a coupled SPR move on the subtrees with root $ i = 1 $.
The potential destination indices $ J_1^{(X)} = J_1^{(Y)} = \{2, 3, 4, 5, 6, 7\} $, so from \eqnref{eq:spr-j} we always draw $ j^{(X)} = j^{(Y)} $ when sampling from a maximal coupling.
\figref{fig:spr-coupled} displays the outcomes for two pairs of destination branches, the pairs of proposed topologies are identical in both cases here; \figref{fig:spr-coupled-all} in the appendix displays the outcomes for all pairs of destination branches.

\subsubsection{Node times and model parameters}
\label{sec:coupling-nodes-parameters}

As with moves on the topology, proposals to update node times are implicitly coupled through housekeeping.
For example, we may sample a pair of internal node indices $ (i^{(X)}, i^{(Y)}) $ from a maximal coupling of $ \Unif{V \setminus L} $ and $ \Unif{V \setminus L} $, in which case $ i^{(X)} = i^{(Y)} = i $, say, then propose node times $ (t_i^{X \prime}, t_i^{Y \prime}) $ from a coupling of Uniform distributions.
Our marginal kernel includes proposals which scale groups of ancestral node times and/or parameters by a common factor $ \nu \sim \Unif{1/2, 2} $.
Scaling multiple terms by a common factor cannot in general propose a meeting as the TV distance between the proposal distributions is $ 1 $ unless the terms in one state differ from the other by a common multiplicative factor; this will only occur if those parts of the state are already equal or have just separated by a scaling move.
We attempt to couple the overall or subtree root times, since they constrain the times of nodes beneath them, then scale the remaining nodes accordingly.

In our experiments, we observed that disabling the moves which scale multiple parameters greatly reduced the meeting times of chains.
If the branch lengths and rate parameters are strongly identified in the model, then these scaling proposals will have low acceptance rates, but if they are only identifiable up to their product, then chains may fail to meet despite being similar in other respects.
For the latter, we constrain the model, such as fixing the death rate $ \mu $, to avoid such ridges in the posterior distribution.
Alternatively, we could change multiple node times through HMC proposals \citep{zhao16} coupled using the techniques developed by \citet{heng19}.

\subsubsection{Catastrophes}
\label{sec:coupling-catastrophes}

Each catastrophe $ (i, u) \in C $ comprises a branch index $ i \in V \setminus \{r\} $ and relative location $ u \in (0, 1) $ between $ t_i $ and $ t_{\pa(i)} $; that is, it occurs on branch $ i $ at time $ t_i + u (t_{\pa(i)} - t_i) $.
We do not consider catastrophes along the branch leading into the root $ r $.
Define $ C_i = \{u : (i, u) \in C\} $ and $ n_i = \abs{C_i} $ for each branch $ i $, and $ n = \sum_{i \in V \setminus \{r\}} n_i $.
Proposals for modifying $ C $ are motivated by its Poisson process prior; for example, we propose to add catastrophes at uniformly sampled locations on the tree and to delete them at random \citep{geyer94} through Reversible Jump MCMC \citep{green95}.

To select a catastrophe for removal from $ C $ in the marginal kernel, we sample a branch index $ i $ with probability $ n_i / n $ and then a location $ u \sim \Unif{C_i} $.
To couple this move, we select $ (i^{(X)}, i^{(Y)}) $ from a maximal coupling
 of the marginal distributions on branch indices, then sample $ (u^{(X)}, u^{(Y)}) $ from a maximal coupling of discrete Uniform distributions $ \Unif{C_{i^{(X)}}^{(X)}} $ and $ \Unif{C_{i^{(Y)}}^{(Y)}} $.
This construction maximises the probability of proposing to remove a catastrophe from the same branch in both states even if the locations are different.
We remove the same catastrophe from both states with probability
\begin{equation*}
    \PP(i^{(X)} = i^{(Y)}, u^{(X)} = u^{(Y)})
        = \sum_{i \in V \setminus \{r\}} \left( \frac{n_i^{(X)}}{n^{(X)}} \wedge \frac{n_i^{(Y)}}{n^{(Y)}}\right) \cdot \frac{\abs{C_i^{(X)} \cap C_i^{(Y)}}}{n_i^{(X)} \vee n_i^{(Y)}}.
\end{equation*}
We leave for future work to investigate alternative couplings when the states are different, such as proposing to add a catastrophe to $ X $ and remove one from $ Y $ when $ n^{(X)} < n^{(Y)} $.

Catastrophes have a common strength parameter $ \kappa $.
The effect of strong catastrophes are more readily identifiable in the model but have a large effect on the posterior, while weak catastrophes have a comparatively smaller effect on the likelihood but are also less identifiable in the model.
As the number of possible topologies is large and the position of a catastrophe depends on the topology, it can be difficult to diagnose issues with mixing over catastrophes from marginal chains.
By coupling chains, we can identify failures to mix over catastrophes from pairs of chains which struggle or fail to meet, such as when chains are stuck in separate modes induced by catastrophes or the catastrophes are not identifiable, and develop moves to facilitate better mixing.
The experiments in \secref{sec:synthetic-full} illustrate the power of couplings to diagnose mixing over catastrophes.

\subsection{Software}
\label{sec:software}

We have implemented our coupled algorithm in TraitLab \citep{nicholls13}, a \texttt{Matlab} \citep{matlab21} toolbox for fitting SD models available at \url{https://github.com/traitlab-mcmc/TraitLab}. Tests to validate our software implementation are described in \appref{app:software-validation}.

In our implementation with all steps taken serially, the computational cost of updating a pair of chains with a sample from our coupled kernel $ \bar{P} $ was approximately twice that for a single draw from the marginal kernel $ P $.
In addition to the occasional need to perform housekeeping on the states, there is extra computational cost from testing whether states have met.
As we only store a fraction of the samples in practice, we just check for equality between trees at those iterations.
We draw a random number of times from the coupled kernel --- in a typical experiment, we begin with $ l $ draws from $ P $ then sample $ \tau^{(l)} - l $ times from $ \bar{P} $ --- but to continue sampling after the chains have met, we need only draw from $ P $ since coupled chains remain together.
In other respects, our code is not optimised.

We profiled the code in fitting the basic model to a tree with 32 leaves on a 2016 Macbook Pro with a 3.3~GHz Intel Core i7 dual-core processor and 16~GB of memory.
On average, a complete update (proposal and accept/reject) to a single chain with the marginal transition kernel took 0.0036 seconds versus 0.0091 seconds for a pair of chains drawing proposals from the coupled kernel.
Housekeeping was only required after 3.2\% of draws from the coupled kernel, adding approximately 0.027 seconds on each occasion.
Over half the running time in this experiment was spent on likelihood calculations, so the duration of the coupled algorithm could be reduced by performing any marginal operation on the states in parallel using graphics processing units \citep{ayres19}.

\section{Experiments}
\label{sec:experiments}

We illustrate the power of our coupled MCMC approach to diagnose convergence on the Stochastic Dollo (SD) model fit to synthetic and real data sets.
With the exception of \secref{sec:synthetic-large}, we ran $ 100 $ pairs of chains targeting the posterior distribution at each lag.
Each pair of chains was initialised by short independent MCMC runs targeting the prior without catastrophes, we then ran \algoref{alg:coupled-mcmc} until the chains met.
We only stored one sample every $ 100 $ iterations and did not check whether the chains met in the interim, so there is a mild loss of resolution in the meeting times $ \tau^{(l)} $.
For lags which are sufficiently high, their estimated TV bounds \eqref{eq:tv-bound} should be indistinguishable from each other.
For the largest lag $ l $ in each experiment, we keep samples $ X_1, \dotsc, X_l $ from each pair of chains and obtain $ 100 $ independent chains targeting the posterior: we use these samples to compute the ASDSF \eqref{eq:asdsf} convergence diagnostic as a comparison.
We modified \texttt{RWTY} \citep{warren17} to evaluate the ASDSF on sliding windows of the most recent 75\% of samples as in \texttt{MrBayes}.
We only consider splits with a sampling frequency of at least 10\% in one or more chains on each window and take $ 0.01 $ as our threshold for diagnosing convergence.
Figures were created in \texttt{R} \citep{r2022} using \texttt{ggplot2} \citep{wickham16}.

\subsection{Synthetic data: only topology and internal node ages unknown}
\label{sec:synthetic-basic}

We first randomly generated trees with 8, 12 and 16 leaves and rescaled each to have a root age of $ 10^3 $.
For each tree, we simulated data from the SD model with birth rate $ \lambda = 0.1 $ and death rate $ \mu = 2.5 \times 10^{-4} $.
For the MCMC, we fixed $ \mu $ at the value used to generate the data placed an upper bound of $ 2 \times 10^3 $ on the root time.
The target of our inference was the posterior distribution on the tree topology and internal node ages $ \{t_i \in T : i \in V \setminus L\} $.

\begin{figure}[tb]
	\centering
    \begin{subfigure}[b]{\textwidth}
        \centering
        \includegraphics[width=\textwidth, trim = 0cm 0cm 0cm 1.5cm, clip]{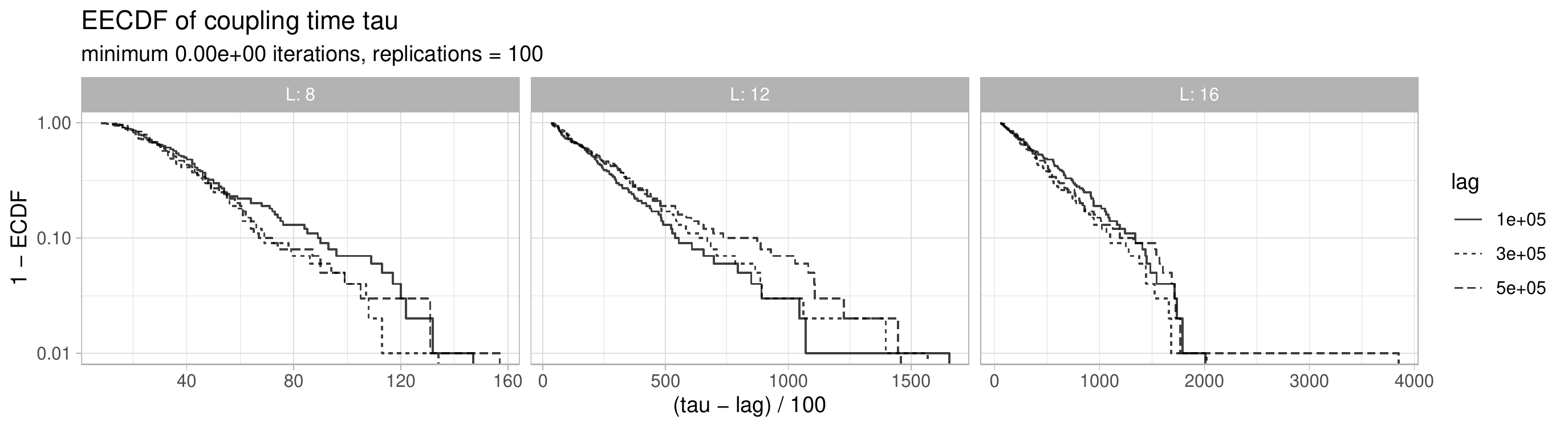}
        \caption{Tails of $ \tau^{(l)} $ decay geometrically for each lag $ l $.}
        \label{fig:simple-tau}
    \end{subfigure}

    \begin{subfigure}[b]{\textwidth}
        \centering
        \includegraphics[width=\textwidth, trim = 0cm 0cm 0cm 1.5cm, clip]{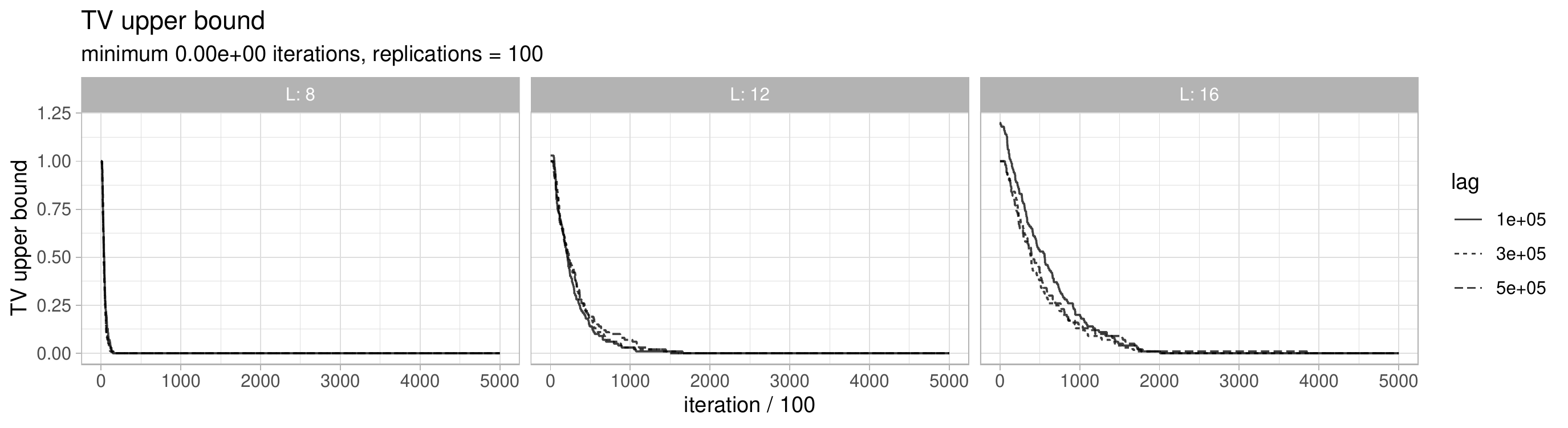}
        \caption{Estimated TV bounds decay at a similar rate for each lag.}
        \label{fig:simple-tv}
    \end{subfigure}

    \begin{subfigure}[b]{\textwidth}
        \centering
        \includegraphics[width=\textwidth, trim = 0cm 0cm 0cm 1.5cm, clip]{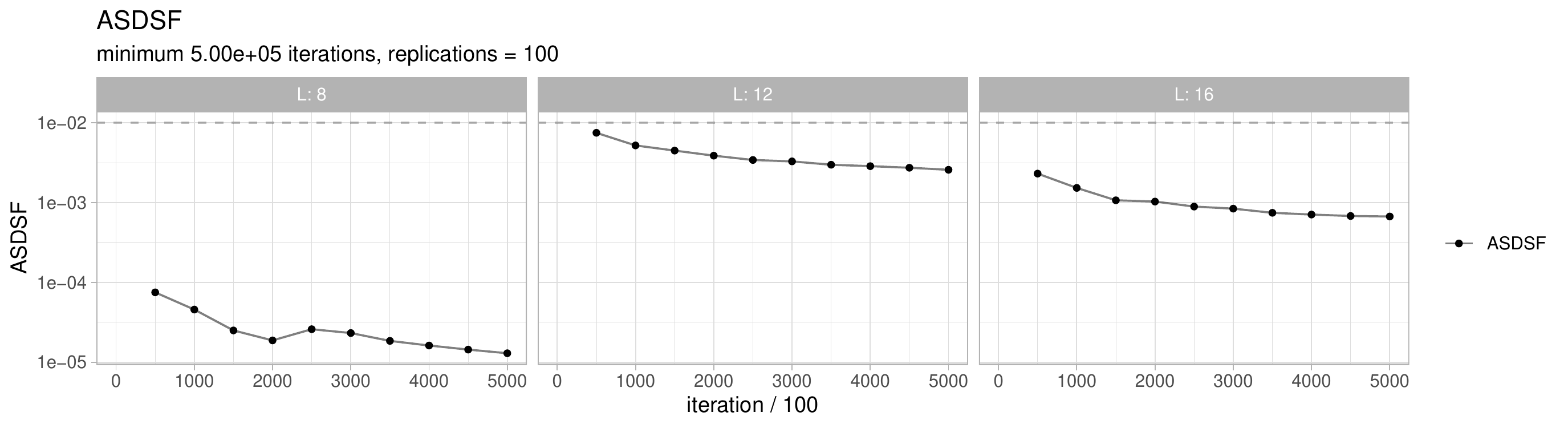}
        \caption{
            ASDSFs are below $ 0.01 $ initially and tend to decay as the window size increases.
        }
        \label{fig:simple-asdsf}
    \end{subfigure}
    \caption{
        Diagnosing convergence of MCMC chains targeting posterior distributions on tree topologies and internal node ages for synthetic data with (left to right) 8, 12 and 16 taxa.
    }
    \label{fig:simple}
\end{figure}

\figref{fig:simple} displays the results of these experiments.
We see from the empirical cumulative distribution function (ECDF) plots in \figref{fig:simple-tau} that the tails of $ \tau^{(l)} $ decay geometrically for each lag $ l $ so the assumption underlying the lagged coupling TV bound is satisfied.
For each experiment, the meeting times were much lower than the corresponding lags so the estimated bounds in \figref{fig:simple-tv} decay rapidly towards $ 0 $ and we can confidently diagnose mixing and convergence over the entire posterior distribution.
Since convergence was rapid in these examples, the ASDSF estimates in \figref{fig:simple-asdsf} are small and decay as the sample size increases.

\subsection{Synthetic data: catastrophes and missing data}
\label{sec:synthetic-full}

We now consider a situation where mixing is a concern and demonstrate how coupling can help diagnose this modelling issue.
Catastrophes allow for rate heterogeneity across branches through discrete bursts of evolutionary activity.
Mixing over catastrophes can pose an issue as each catastrophe introduces a discrete change in the likelihood.
In addition, multiple combinations of the death rate $ \mu $ and catastrophe strength $ \kappa $ can produce the same effective catastrophe duration, and multiple weak catastrophes along a branch can mimic a single stronger one or the overall trait model.
We would like to identify issues with mixing and identifiability of catastrophes.

We placed catastrophes with strength $ \kappa = 0.05 $ on three randomly selected branches on leading into leaf nodes in the same trees as \secref{sec:synthetic-basic} then generated data with the same death rate $ \mu = 2.5 \times 10^{-4} $.
Placing catastrophes higher up the tree would most likely make them more difficult to identify and result in longer meeting times.
The true presence/absence state of traits at leaf $ i \in L $ were recorded with probability $ \xi_i \sim \BetaDist{1}{1 / 3} $ and marked missing otherwise.
After marginalising the $ \Gamma(1.5, 5000) $ prior on the catastrophe rate $ \rho $ out of our model, we obtain a Negative Multinomial prior on catastrophe counts across branches.
As branches evolve independently in the absence of lateral trait transfer, we can integrate out catastrophe relative locations along branches and only consider their number on each branch.
For the MCMC, in addition to an upper bound of $ 2 \times 10^3 $ on the root time, we sampled a clade constraint at random for each tree and fixed $ \mu $ and $ \kappa $ at the values used to generate the data.
The target is the posterior distribution on the topology, internal node times in $ T $, catastrophe counts on branches $ C $, and missingness parameters $ \Xi $.

\begin{figure}[tb]
	\centering
    \begin{subfigure}[b]{\textwidth}
        \centering
        \includegraphics[width=\textwidth, trim = 0cm 0cm 0cm 1.5cm, clip]{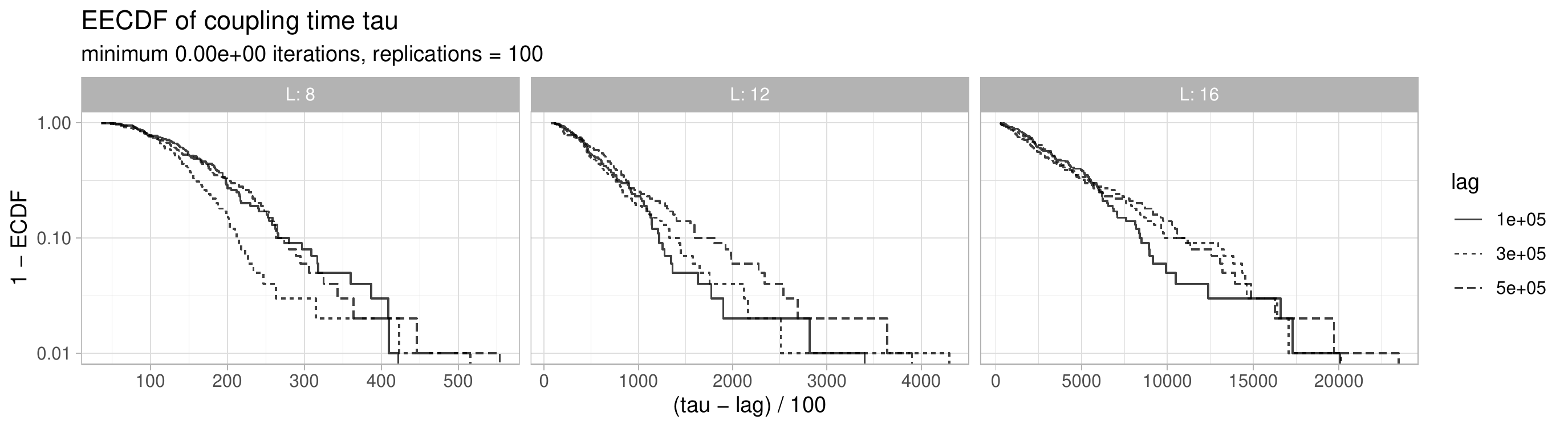}
        \caption{Tails of $ \tau^{(l)} $ decay geometrically.}
        \label{fig:weak-tau}
    \end{subfigure}

    \begin{subfigure}[b]{\textwidth}
        \centering
        \includegraphics[width=\textwidth, trim = 0cm 0cm 0cm 1.5cm, clip]{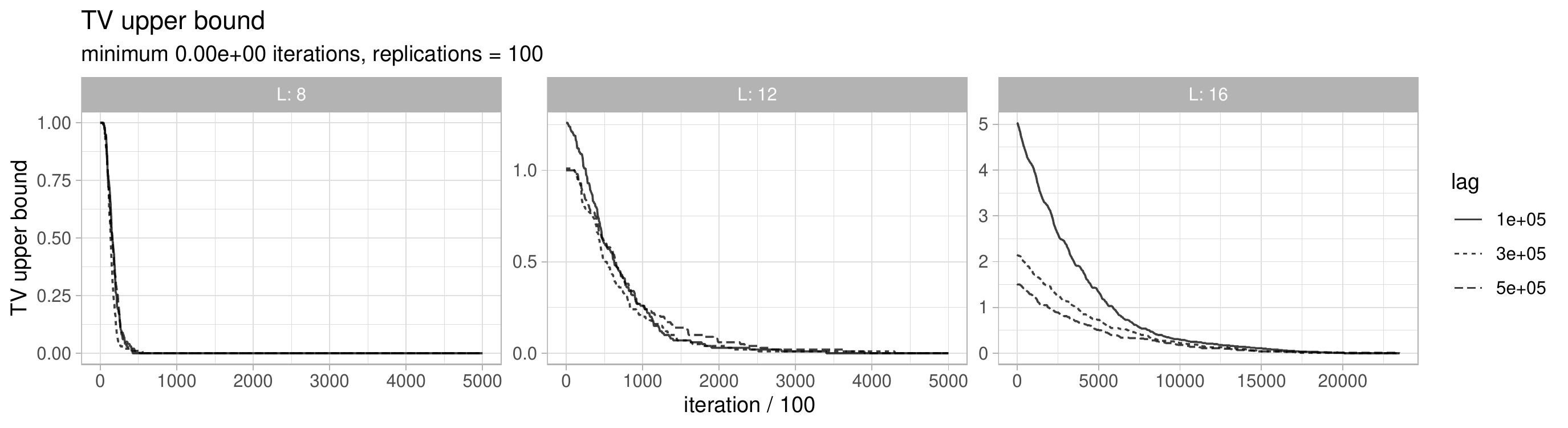}
        \caption{
            Estimated TV bounds converge but at a slower rate for smaller lags as the number of taxa increases.
        }
        \label{fig:weak-tv}
    \end{subfigure}

    \begin{subfigure}[b]{\textwidth}
        \centering
        \includegraphics[width=\textwidth, trim = 0cm 0cm 0cm 1.5cm, clip]{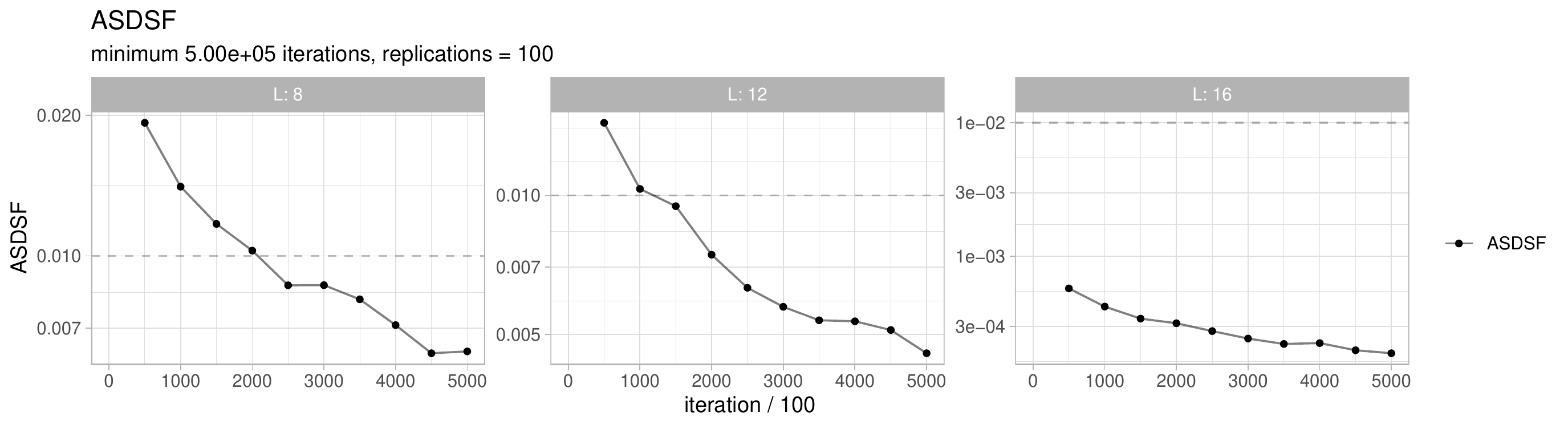}
        \caption{ASDSFs converge in each experiment.}
        \label{fig:weak-asdsf}
    \end{subfigure}
    \caption{
        Diagnosing convergence when fitting the SD model to synthetic data sets with 8, 12 and 16 taxa, missing data, three weak catastrophes and a weak prior on catastrophe counts.
    }
    \label{fig:weak}
\end{figure}

\figref{fig:weak} displays the results of our experiments with weak catastrophes.
For the data with 8 taxa, pairs of chains met rapidly and the coupling TV bound gives a significantly earlier diagnosis of convergence than ASDSF.
Both methods agree in the experiment with 12 taxa.
For the data with 16 taxa, the TV bounds are slow to converge while the ASDSF suggests a well-behaved distribution over splits.
While it is possible that the relatively slow convergence of the TV bound for 16 leaves is because our coupling is not tight enough, it is more likely to be because ASDSF is not diagnosing convergence on the other components of the model.
This experiment illustrates the power of the coupling TV bound to detect convergence quicker than ASDSF and the difficulty of diagnosing convergence as the number of parameters and latent variables increase.
The results of experiments with relatively strong catastrophes on the same trees are contained in \appref{app:experiments}.

\subsection{Eastern Polynesian lexical traits: lateral transfer}
\label{sec:eastern-polynesian}

We revisit the analysis by \citet{kelly17} of lexical trait data in 11 Eastern Polynesian languages under the SD model with lateral trait transfer.
The data was drawn from the Austronesian Basic Vocabulary Database \citep{greenhill08} and is a subset of languages previously analysed by \citet{gray09} under a number of model-based Bayesian approaches, including SD without lateral transfer, and \citet{gray10} using \texttt{Neighbor-Net} \citep{bryant04}, a likelihood-free method for constructing phylogenetic networks from splits.
The lateral trait transfer model is an ideal candidate for coupling as its likelihood computation grows exponentially in the number of taxa, so we do not want to waste resources on inefficient burn-in estimates and would like to make use of experiments run in parallel.

For these experiments, we fixed $ \kappa = 1/3 $, the centre of the range inferred by \citet{kelly17}, and allowed both $ \mu $ and $ \beta $ to vary.
We expect that also allowing $ \kappa $ to vary would increase the meeting times.
Following \citet{gray09}, \citet{kelly17} imposed a single clade constraint to fix the root age of the tree to $ [1150, 1800] $ years before the present and we do the same here.
The target of our inference is the tree topology, internal node times in $ T $, death rate $ \mu $, lateral transfer rate $ \beta $, catastrophe branches and locations $ C $, and missing data parameters $ \Xi $.

\begin{figure}[tb]
	\centering
    \begin{subfigure}[h]{0.48\textwidth}
        \centering
        \includegraphics[width=\textwidth, trim = 0cm 0cm 0cm 1.25cm, clip]{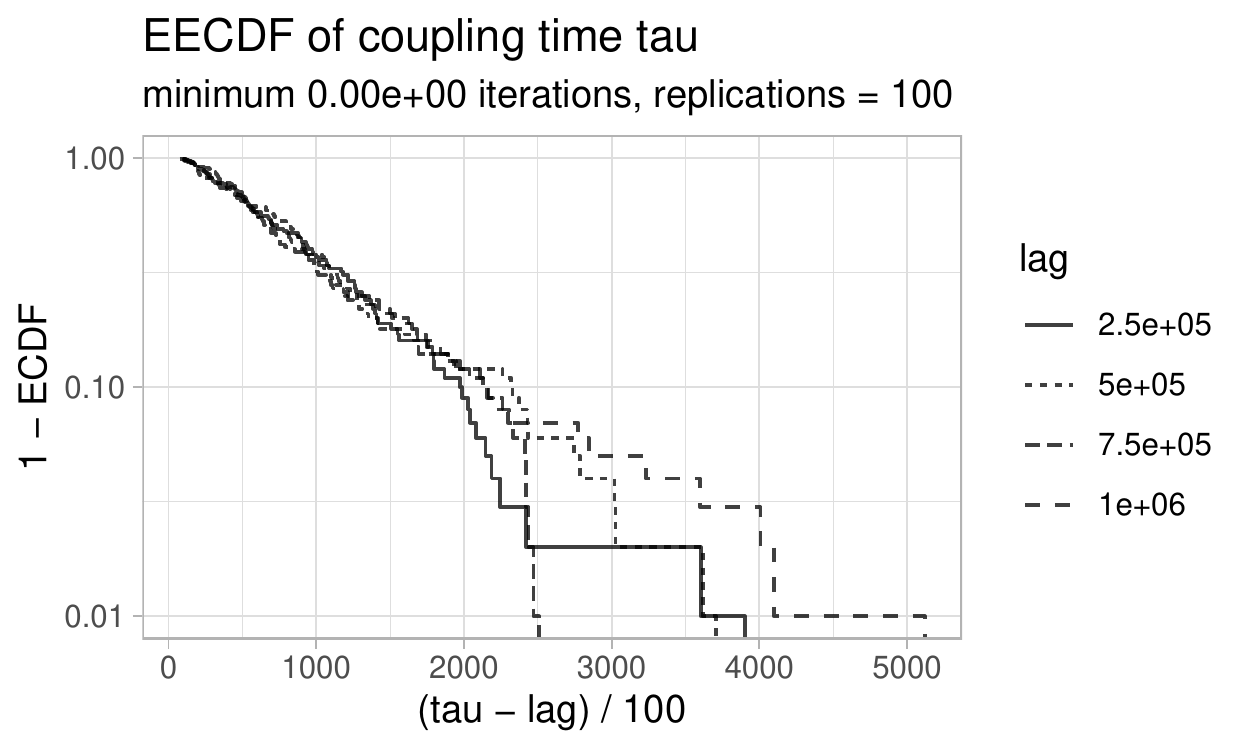}
        \caption{Tails of $ \tau^{(l)} $ decay geometrically for each lag.}
        \label{fig:poly-tau}
    \end{subfigure}
    ~
    \begin{subfigure}[h]{0.48\textwidth}
        \centering
        \includegraphics[width=\textwidth, trim = 0cm 0cm 0cm 1.25cm, clip]{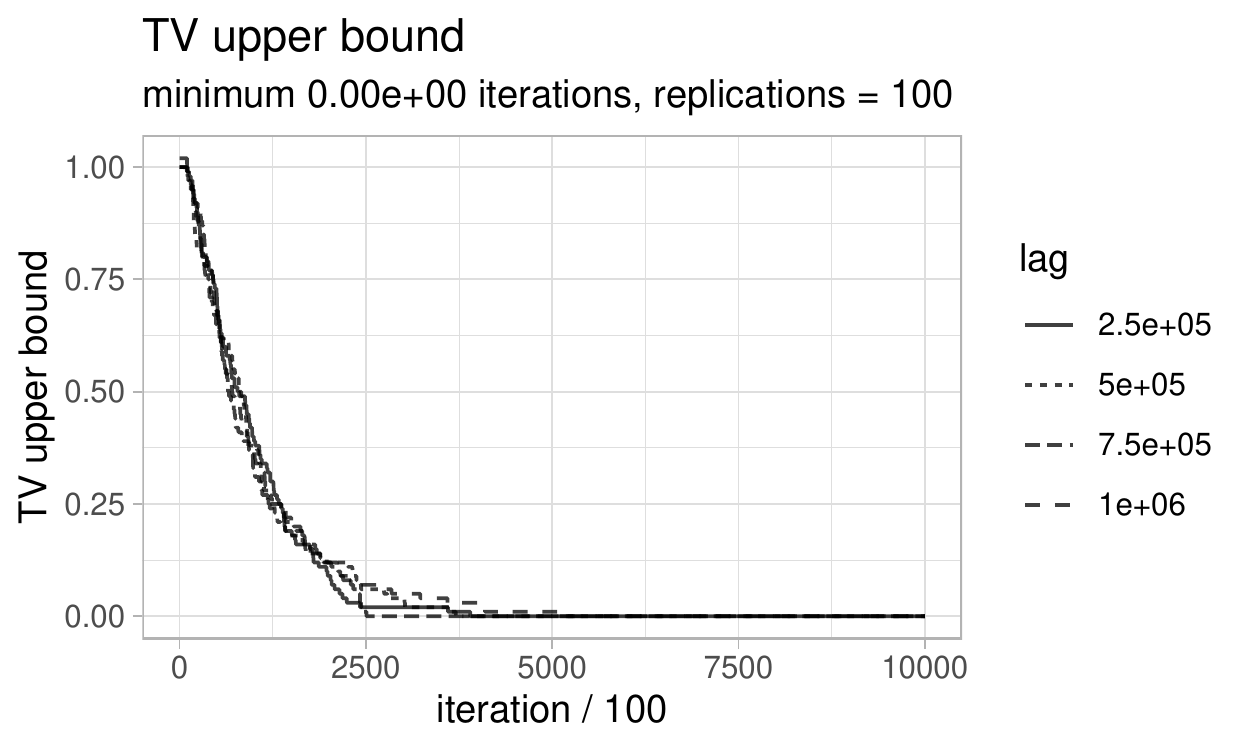}
        \caption{Estimated TV bounds are stable across lags.}
        \label{fig:poly-tv}
    \end{subfigure}

    \begin{subfigure}[h]{0.48\textwidth}
        \centering
            \includegraphics[width=\textwidth, trim = 0cm 0cm 0cm 1.25cm, clip]{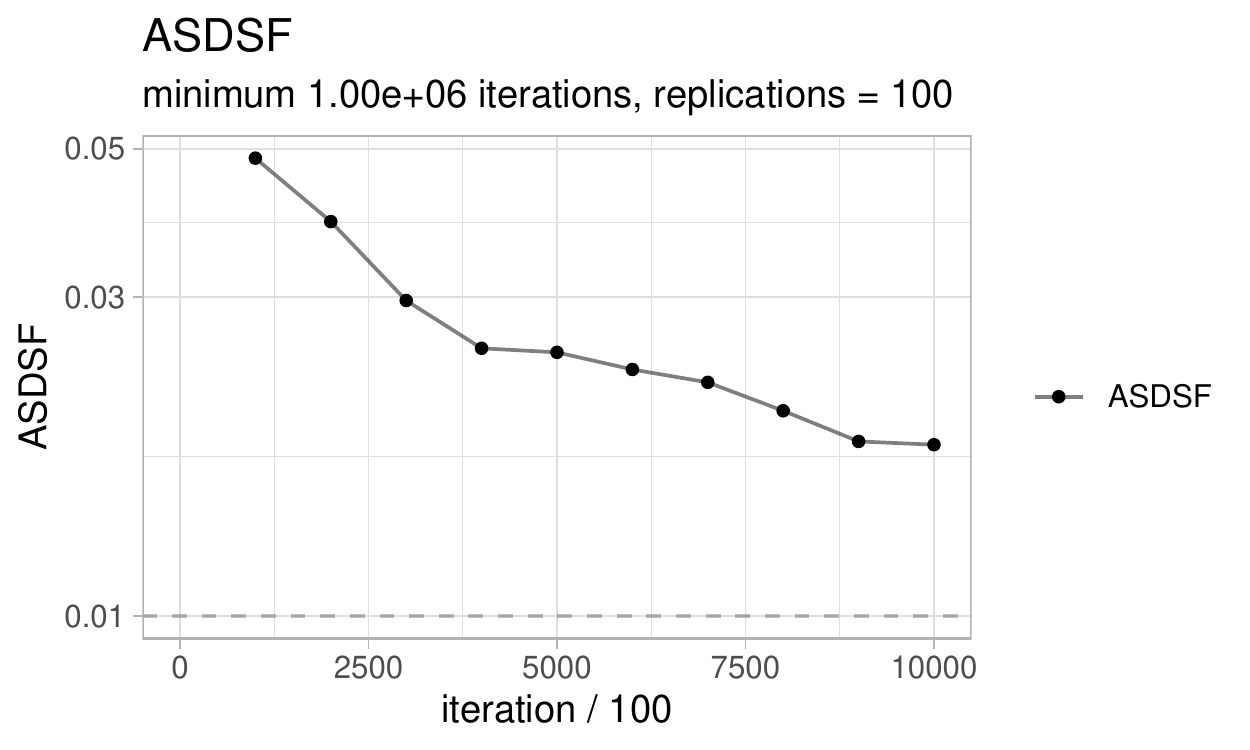}
        \caption{ASDSF decays slowly.}
        \label{fig:poly-asdsf}
    \end{subfigure}
    \caption{
        Diagnosing convergence when fitting the SD model with lateral trait transfer, catastrophes and missing data to Eastern Polynesian lexical traits in 11 taxa.
    }
    \label{fig:poly}
\end{figure}

\figref{fig:poly} displays the results of these experiments.
Estimated TV bounds are stable across lags and negligible within $ 2.5 \times 10^5 $ to $ 5 \times 10^5 $ iterations, suggesting good mixing and fast convergence.
In contrast, ASDSF fails to decay below the arbitrary threshold of $ 0.01 $ within $ 10^6 $ iterations.

\subsection{Synthetic data on larger trees}
\label{sec:synthetic-large}

\begin{figure}[tb]
	\centering
    \begin{subfigure}[b]{\textwidth}
        \centering
        \includegraphics[width=\textwidth, trim = 0cm 0cm 0cm 1.5cm, clip]{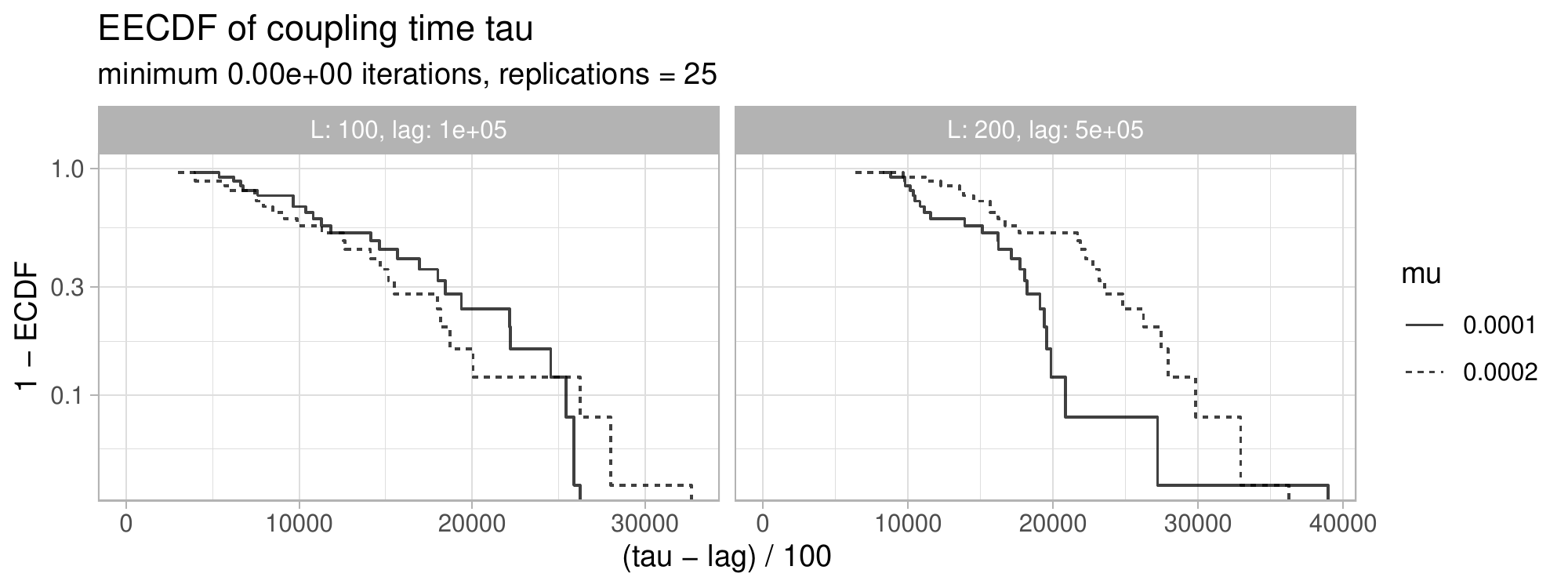}
        \caption{Tails of $ \tau^{(l)}$ decrease geometrically for each experiment.}
        \label{fig:large-tau}
    \end{subfigure}

    \begin{subfigure}[b]{\textwidth}
        \centering
        \includegraphics[width=\textwidth, trim = 0cm 0cm 0cm 1.5cm, clip]{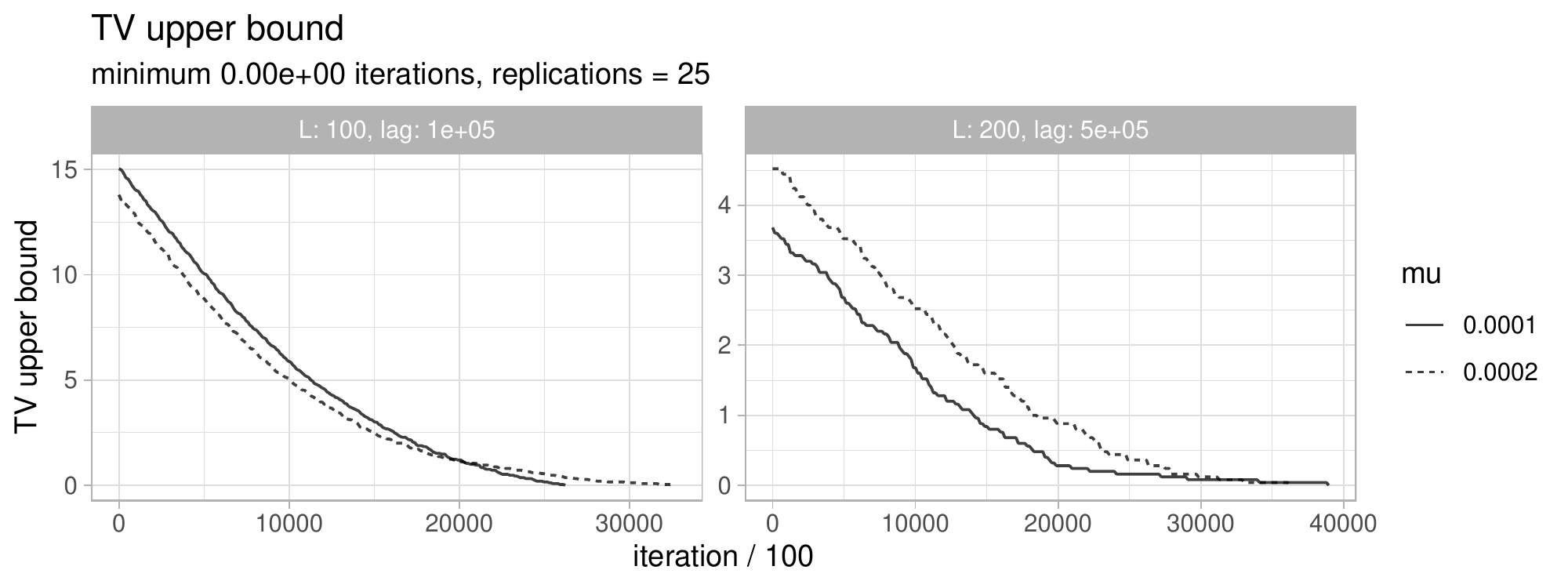}
        \caption{
            TV bounds are large initially so we should increase the lag to better diagnose convergence.
        }
        \label{fig:large-tv}
    \end{subfigure}
    \caption{
        Coupling chains targeting posterior distributions on trees with 100 and 200 leaves.
    }
    \label{fig:large}
\end{figure}

\figref{fig:large} demonstrates the ability of our coupling scheme to produce meetings of chains on larger trees.
The data for these experiments was drawn from the SD model with a death rate $ \mu = 10^{-4} $ and $ 2 \times 10^{-4} $ on trees with 100 and 200 leaves.
For each data set, we ran $ 25 $ pairs of chains coupled at a single lag ($ 10^5 $ for the trees with 100 leaves and $ 5 \times 10^5 $ for 200 leaves) until meeting.
For each experiment, we fixed the death rate $ \mu $ at the value used to simulate the data and placed an upper bound of $ 2 \times 10^3 $ on the root time.
We allowed proposals which rescaled multiple node times in the short MCMC runs targeting the prior to initialise each chain but disabled them otherwise.
This figure shows that our coupling strategy remains effective on larger trees.
Decreasing $ \mu $ increases the signal in the data so we would expect chains to meet earlier; although there is some evidence of this effect for 200 leaves, further experiments with larger lags are required to confidently assess it.
We expect that our coupling approach will extend to trees with many more taxa provided there is enough information in the data that the posterior is not too diffuse.

\section{Concluding remarks}
\label{sec:concluding-remarks}

Couplings provide a theoretical and practical framework for assessing the convergence of MCMC samplers in phylogenetic problems.
Relabelling nodes of the tree to identify common components and sampling from a maximal coupling at each step of each proposal distribution produced a sufficient coupling for chains to meet in our experiments on trees with up to 200 leaves.
Our estimated total variation bounds clearly diagnose convergence and mixing of chains across the entire model.
As well as providing a more trustworthy check of convergence, we noticed in several of our experiments that convergence is detected earlier using the coupling approach than with ASDSF, so despite only being an upper bound it is a powerful tool in practice.
The Stochastic Dollo model and associated MCMC algorithm is similar to many other phylogenetic inference schemes.
The coupling strategy we describe is generic and can be transferred to other models: the coupling method for the tree topology, node times and scalar parameters can be reused as is, while branch-dependent latent variables may be coupled using similar strategies for the node times and catastrophes.
Further work is required to develop maximal couplings of proposal operators in phylogenetic problems and obtain tighter convergence bounds.

Couplings show benefits both for practitioners using MCMC on phylogenies for applied problems and for statisticians constructing new phylogenetic models or proposal kernels.
The TV distance controls the error in estimating probabilities by $ \pi_s $ instead of $ \pi $, so can be used to quantify uncertainty in MCMC estimates of marginal distributions, such as the posterior support for a set of topologies, or bounded functions.
Each meeting time is sampled independently so we could also resample them to estimate the uncertainty in the TV bound.
In addition to estimating the coupling TV bound, samples from coupled chains may also be used to construct unbiased MCMC estimators which can be averaged across pairs of chains.
For inference, we remind the practitioner that chains should not be interrupted as soon as they have met but run for a sufficient number of iterations to obtain an adequate sample from the posterior.
Overall, since MCMC on phylogenies is typically run for a large number of iterations and is very time consuming, we hope that couplings will allow practitioners to use the output more efficiently and to make use of multiple independent chains run in parallel.

Note that the coupling diagnostic will only be as reliable as the distribution of the meeting times.
As described in \secref{sec:introduction}, modes in the posterior can manifest as plateaus in the TV bound as chains initialised in the same mode may meet quickly but may take many iterations to meet when started in separate modes.
We advocate initialising pairs of chains far apart and experimenting with a range of lags so that chains are more likely to explore the posterior before meeting.
We used samples from a pair of independent MCMC chains targeting the prior to initialise each experiment; \citet{atkins2019extremal} show that the SPR distance between a pair of rooted binary trees chosen at random is close to maximal.
We could also attempt to introduce negative correlation to the initialisation scheme to further increase the distance between initial states \citep{ryder20}.
Sampling from the marginal kernel is faster than sampling from a coupling, so the computational cost of increasing the lag is relatively minor but can greatly improve estimates of the TV bound, as illustrated by our experiments in \secref{sec:experiments}.
We used 100 pairs of chains at each lag to compare our coupling bounds to ASDSF but in practice we could choose the number of chains and sequence of lags from a small number of pilot runs.

The coupling approach is also of use earlier in the pipeline as a tool to help identify certain modelling issues and weaknesses in the MCMC kernel.
In our experiments with small catastrophes (low value of $\kappa$), some parameters are poorly identified and this leads to high multimodality in the posterior which the MCMC sampler has trouble exploring, such as when long branches are replaced by shorter branches with catastrophes.
Our initial attempts to couple the chains failed in this situation, which made the issue fully apparent and allowed us to improve the marginal kernels.
In other experiments, we occasionally observed a pair of chains to be in very similar states but failing to meet, and found this was because we needed to simultaneously update two highly correlated components of the state such as the topology and catastrophes.
If the mixture of kernels does not allow for a proposal which updates the two components jointly, then such transitions hardly ever occur and the mixing time deteriorates.
We can also use couplings to detect ancillary or non-identifiable parameters.
For example, if there is no evidence of lateral trait transfer in the data, then the $ \beta $ parameter samples will get arbitrarily small but may fail to meet in many pairs of chains because proposal distributions in the two states do not overlap.
If a particular local kernel is poorly designed or its mixture weight too low in the overall kernel, then the target components of the state may mix slowly so a chain may appear to have converged when in fact it has only done so on a subset of the posterior.
Initialising chains far apart and increasing the lag helps avoid the illusion of convergence from slow mixing components.
\secref{sec:coupling-nodes-parameters} describes a difficulty with coupling proposals which rescale multiple parameters by a common factor.
Diagnosing why pairs of chains fail to meet makes these issues more obvious and helps us construct more efficient kernels.

In this paper, we have focused on coupling the Metropolis--Rosenbluth--Teller--Hastings algorithm.
The framework described by \citet{jacob20} is generic and may be used to couple many existing marginal MCMC algorithms, such as parallel tempering, so the transition kernel couplings we describe could also be utilised to couple these algorithms in phylogenetic problems.
\citet{biswas19} compare different choices of HMC tuning parameters using the lagged coupling TV bound as a proxy for the true convergence rate, a similar approach could be taken for tuning mixture weights in phylogenetic proposal kernels or comparing sampling schemes.
\citet{biswas20} demonstrate promising performance of couplings for Gibbs sampling in high-dimensional regression, we are hopeful that couplings can also be used to diagnose MCMC behaviour in large-scale phylogenetic analyses.
In situations where other MCMC methods prove more efficient at exploring the posterior, such as sequential Monte Carlo \citep{wang19} or Hamiltonian Monte Carlo \citep{ji21}, it could be useful to adapt these coupling methods following ideas in \citet{jacob20smoothing} or \citet{heng19}.

\begin{acks}[Acknowledgments]
    We are grateful to Pierre E. Jacob, Geoff K. Nicholls and Alexandre Bouchard-C\^{o}t\'{e} for helpful discussion and feedback.
    We thank the three anonymous referees, Associate Editor and Editor for insightful comments and suggestions.
\end{acks}

\begin{funding}
    LJK was supported by the French government under management of Agence Nationale de la Recherche as part of the ABSint (reference ANR-18-CE40-0034) and PRAIRIE (reference ANR-19-P3IA-0001) programmes.
\end{funding}

\bibliographystyle{imsart-nameyear}
\bibliography{references}


\appendix

\section{Stochastic Dollo model}
\label{app:sd-model}

A phylogenetic tree $ g = (V, E, T) $ has edge set $ E $, vertex set $ V $ and node times $ T $.
The neighbours of a node $ i \in V $ are its parent $ \pa(i) $, offspring $ \off(i) = \{j \in V : \pa(j) = i\} $ and sibling $ \sib(i) = \{j \in V : j \neq i, \pa(j) = \pa(i)\} $.
For example, in \figref{fig:sd-example}: $ \pa(3) = 10 $, $ \sib(3) = 2 $ and $ \off(3) = \emptyset $.
Denote the set of node times $ T = \{t_i : i \in V\} $.
We refer to the edge from $ \pa(i) $ into node $ i $ as branch $ i $, we denote its length $ \Delta_i = t_{\pa(i)} - t_i $ and $ \Delta = \sum_{i \in V \setminus \{r\}} $ the length of the tree beneath the root.
We denote $ r $ the root of the tree, $ L \subset V $ the set of leaf indices and $ A = V \setminus L $ the set of $ \abs{L} - 1 $ ancestral nodes.
There is an infinitely long branch leading into the root but we do not include its parent node in $ V $.

Leaf node indices are identical and constant across states $ X $ and $ Y $.
Through our housekeeping operation, the root node indices are identical across states $ X $ and $ Y $ at each iteration, but are not constant across iterations of the MCMC algorithm.
Each subtree implied by a clade has a common root index across $ X $ and $ Y $ after housekeeping, and nodes within a common clade are indexed from the same subset of $ V $ in both states.

The Stochastic Dollo (SD) model posits a birth-death process of traits along the dated tree $ g = (V, E, T) $: new traits arise in each species according to a Poisson process with rate $ \lambda $ and evolve along the tree towards the leaves, instances of each trait are copied into offspring lineages at a speciation event and die independently at rate $ \mu $.
We record the binary patterns of trait presence or absence across the leaves.
For an ordering of the leaf nodes, let $ N_p $ denote the number of traits displaying binary pattern $ p \in \cP = \{0, 1\}^L \setminus \{(0, \dotsc, 0)\} $, we cannot observe a trait absent at all of the leaves so ignore the pattern $  (0, \dotsc, 0) $.
A branch of infinite length leads into the root node so the process is in equilibrium with a $ \Pois{\lambda / \mu} $ number of traits just before the first branching event.
With this initial condition at the root, we compute the expected frequency $ z_p = \EE[N_p \given g, \lambda, \mu] $ of each binary pattern $ p \in \cP $ by integrating over the possible unobserved trait events on the tree \citep{nicholls08}.
As we have a Poisson process of trait births with independent thinning, our observation model is
\begin{equation}
    \label{eq:sd-likelihood}
    N_p \given g, \lambda, \mu \sim \Pois{z_p}, \quad p \in \cP.
\end{equation}

As the traits evolve independently, $ z_p = \lambda \tilde{z}_p $ where $ \tilde{z}_p = \EE[N_p \given g, \lambda = 1, \mu] $.
In building our Bayesian model, we place a $ \Gamma(a, b) $ prior on $ \lambda $ and integrate it out of the posterior to obtain a Negative Multinomial distribution on the pattern frequencies,
\begin{equation}
    \label{eq:sd-integrated-likelihood}
    (N_p : p \in \cP) \given g, \mu \sim \NM{
        a
    }{
        \frac{b}{b + \sum_{q \in \cP} \tilde{z}_q}
    }{
        \left(\frac{\tilde{z}_p}{b + \sum_{q \in \cP} \tilde{z}_q} : p \in \cP\right)
    }.
\end{equation}

We incorporate the following extensions to this observation model.

\paragraph*{Rate heterogeneity through catastrophes}
We allow for discrete bursts of evolutionary activity in the form of catastrophes which occur according to a Poisson process of rate $ \rho $ along the branches of the tree \citep{ryder11}.
At a catastrophe, each trait present on the branch is killed with probability $ \kappa $ and a $ \Pois{\lambda \kappa / \mu} $ number of new traits are born.
This is equivalent to instantaneously advancing the trait process on the branch by $ -\log(1 - \kappa) / \mu $ units of time.
We do not consider catastrophes on the branch leading into the root.
With this definition, the effects of catastrophes are reversible with respect to the underlying trait process so the catastrophe set $ C $ only records the number of catastrophes on each branch as we can integrate their locations out of the likelihood.

\paragraph*{Missing-at-random data}
We assume that data are missing at random.
The true status of a trait at leaf $ i \in L $ is observed with probability $ \xi_i $ independently of other traits and recorded as missing otherwise \citep{ryder11}.
We denote $ \Xi = \{\xi_i : i \in L\} $.

\paragraph*{Lateral trait transfer}
Lateral trait transfer is a form of reticulate evolutionary activity whereby species can acquire traits outside of ancestral relationships.
Following \citet{kelly17}, each instance of a trait transfers a copy of itself to other contemporary species at rate $ \beta $.
As in the pure birth-death SD model, a catastrophe advances the trait process along the branch by $ {-}\log(1 - \kappa) / \mu $ units of time relative to other branches, during which traits may transfer in from other lineages but not out.
A catastrophe is not reversible in this setting as its effect depends on the number of other branches and the traits present on them, so we include the location of catastrophes along their branches in the catastrophe set $ C $.
Each catastrophe $ c = (i, u) \in C $ has a branch index $ i \in V \setminus \{r\} $ and relative location $ u \in (0, 1) $ along the branch between $ i $ and $ \pa(i) $, so catastrophe $ c $ occurs at time $ t_i + u (t_{\pa(i)} - t_i) $.

\smallskip
\figref{fig:sd-example} illustrates a trait history drawn from the SD model.
The extensions described above do not change the Poisson distribution of the data in \eqnref{eq:sd-likelihood} or Negative Multinomial in \eqnref{eq:sd-integrated-likelihood}, but do require a more computationally expensive integral to compute the expected pattern frequencies $ (z_p)_{p \in \cP} $ \citep{ryder11,kelly17}.

We place diffuse priors on all of the parameters.
Our tree prior is Uniform across topologies and approximately Uniform on the time of the root node \citep{nicholls11}.
Clade constraints represent known prior information and restrict the space of possible topologies and node times.
We can integrate the catastrophe rate $ \rho $ out of our posterior with respect to a Gamma prior to obtain a Negative Multinomial distribution of catastrophe counts on branches.
Conditional on their number, the prior distribution of catastrophe locations on a branch is Uniform.
Multiple small catastrophes on a branch can produce a similar effect to a single larger catastrophe, so we generally opt to lower-bound or fix $ \kappa $ at a reasonable value.
Under the full model, the target of our inference is $ X = (g, \mu, \beta, \Xi, C, \kappa) $, where the tree $ g = (V, E, T) $.

\section{Sampling from maximal couplings}
\label{app:sample-maximal-couplings}

For random variables $ X \sim p $ and $ Y \sim q $ on a common space $ \cX $, we consider two approaches to sampling $ (X, Y) $ from a maximal coupling with independent residuals.
We abuse notation and also use $ p $ and $ q $ to denote probability mass functions or densities.
For the majority of our couplings, we use the rejection sampling algorithm described in \secref{sec:couplings-generic}.
This algorithm is simple to apply in practice, we can write a generic function which takes as arguments functions to sample from and evaluate $ p $ and $ q $.

In certain settings, we can sample directly from a maximal coupling \citep[Section~5.4]{jacob20}.
\begin{enumerate}
    \item With probability $ 1 - d_{\mathrm{TV}}(p, q) $,
    \[
        X \sim \frac{p \wedge q}{1 - d_{\mathrm{TV}}(p, q)}
        \quad \text{and} \quad
        Y \leftarrow X.
    \]
    \item Otherwise,
    \[
        X \sim \frac{p - p \wedge q}{d_{\mathrm{TV}}(p, q)}
        \quad \text{and} \quad
        Y \sim \frac{q - p \wedge q}{d_{\mathrm{TV}}(p, q)}.
    \]
\end{enumerate}
This approach was tractable for the MH accept/reject step in \secref{sec:couplings-generic}, rescaling parameters by Uniform random variables and sampling root times in SPR moves from a coupling of truncated Exponential distributions such as in the following example.

Let $ \Exp{\theta; c} $ denote an $ \Exp{\theta} $ random variable restricted to $ [c, \infty) $, and $ \Exp{\theta; c, d} $ when restricted to $ [c, d) $.
We would like to sample $ X \sim p = \Exp{\theta; a_p} $ and $ Y \sim q = \Exp{\theta; a_q} $. With probability $ 1 - d_{\mathrm{TV}}(p, q) = \exp(-\theta \abs{a_p - a_q}) $, we sample $ (X, Y) $ from the overlap between $ p $ and $ q $,
\[
    X \sim \Exp{\theta; a_p \vee a_q} \quad \text{and} \quad Y \leftarrow X.
\]
Otherwise, we draw $ (X, Y) $ independently from their residual distributions: if $ a_p < a_q $, then
\[
    X \sim \Exp{\theta; a_p, a_q} \quad \text{and} \quad Y \sim \Exp{\theta; a_q},
\]
and similarly when $ a_q < a_p $.

\section{Coupling proposals for phylogenetic models}
\label{app:coupling-proposals-phylogenetic}

Our Metropolis--Rosenbluth--Teller--Hastings (MH) proposal kernel is a mixture of local kernels $ (Q_m)_m $ with fixed weights $ (\epsilon_m)_m $.
\tabref{tab:proposals} lists the various proposal kernels in our MCMC algorithm.
At each iteration of the marginal algorithm, we draw $ Q \sim \sum_m \epsilon_m Q_m $ and use it to propose an update.
At each iteration of the coupled algorithm, we draw $ \bar{Q} \sim \sum_m \epsilon_m \bar{Q}_m $, where each $ \bar{Q}_m $ is a coupling of $ Q_m $ with itself, and sample a pair of proposals $ (X', Y') $.
As described in \secref{sec:coupled-mcmc-phylogenetics}, we construct our coupling $ \bar{Q}_m((X, Y), \cdot) $ of local proposal kernels $ Q_m(X, \cdot) $ and $ Q_m(Y, \cdot) $ by sampling from a maximal coupling at each step of the marginal kernels.
We now give a detailed description of each marginal move and how we draw a sample from a coupling at each step.
We deal with constraints on the model by rejecting invalid proposals.
A number of proposals add a Jacobian term to the Hastings ratio for updating the chains, such as those which rescale parameters by a common factor $ \nu \sim \Unif{1/2, 2} $, but these do not require any adjustments to our algorithm to sample from a maximal coupling of proposal steps or MH accept/reject decisions.

\begin{table}[t]
    \centering
    \caption{MCMC moves act primarily on either the tree topology, node times, catastrophes, or the parameters of the trait diversification or observation processes.}
    \label{tab:proposals}
    \begin{tabular}{@{}lccl@{}}
        \toprule
        Primary target & Section & Move & Description \\ \midrule
        \multirow{4}{*}{Topology} & \multirow{4}{*}{\ref{app:moves-topology}}
            & 1 & Exchange parents of neighbouring node pair \\
            & & 2 & Exchange parents of randomly chosen node pair \\
            & & 3 & SPR onto neighbouring branch \\
            & & 4 & SPR onto randomly chosen branch \\ \midrule
        \multirow{5}{*}{Node times} & \multirow{5}{*}{\ref{app:moves-times}}
            & 5 & Resample internal node time\\
            & & 6 & Resample leaf time \\
            & & 7 & Rescale tree \\
            & & 8 & Rescale subtree \\
            & & 9 & Rescale tree above clade bounds \\ \midrule
        \multirow{5}{*}{Catastrophes} & \multirow{5}{*}{\ref{app:moves-catastrophes}}
            & 10 & Add catastrophe to an edge \\
            & & 11 & Delete catastrophe from an edge \\
            & & 12 & Move catastrophe to neighbouring edge \\
            & & 13 & Resample all catastrophes on branch \\
            & & 14 & Resample catastrophe location on branch \\ \midrule
        \multirow{5}{*}{Model parameters} & \multirow{5}{*}{\ref{app:moves-parameters}}
            & 15 & Rescale death rate $ \mu $ \\
            & & 16 & Rescale transfer rate $ \beta $ \\
            & & 17 & Rescale catastrophe strength $ \kappa $ \\
            & & 18 & Rescale one missing data parameter $ \xi_i \in \Xi $ \\
            & & 19 & Rescale all missing data parameters $ \Xi $ \\
        \bottomrule
    \end{tabular}
\end{table}

\subsection{Moves 1--4: tree topology}
\label{app:moves-topology}

\subsubsection{Moves 1 \& 2: subtree swap}

We switch the parents of a randomly chosen pair of nodes $ i $ and $ j $,
\begin{align*}
    \pa(i)' &\leftarrow \pa(j), \\
    \pa(j)' &\leftarrow \pa(i).
\end{align*}
A narrow move (1 in \tabref{tab:proposals}) exchanges the parents of neighbouring nodes, we select $ i $ and $ j $ as follows:
\begin{enumerate}
    \item sample node $ i \sim \Unif{\{i' \in V : \pa(i') \neq r\}} $;
    \item set $ j \leftarrow \sib[\pa(i)] $.
\end{enumerate}
If $ t_j \geq t_{\pa(i)} $ or any calibration constraints are violated, then the move fails.
For a wide move (2 in \tabref{tab:proposals}), we draw $ (i, j) $ uniformly from the set of pairs of nodes which are not neighbours and do not violate any calibration constraints.

In both the narrow and wide cases, we can easily sample from a maximal coupling of the corresponding discrete Uniform distributions.

\subsubsection{Moves 3 \& 4: subtree prune-and-regraft}

As illustrated in \figref{fig:spr} and \secref{sec:coupling-structural-moves}, we randomly choose a subtree with root $ i $, detach its parent $ \pa(i) $ from the tree and reattach it at new time $ t_{p}' $ on branch $ j $.
To form the proposed state, we set
\begin{align*}
    \pa[\pa(i)]' &\leftarrow \pa(j), \\
    \pa[\sib(i)]' &\leftarrow \pa(j), \\
    \pa(j)' &\leftarrow  \pa(i).
\end{align*}
To simplify notation, for the rest of this section we denote $ p = \pa(i) $ and $ q = \pa(j) $.

In a narrow SPR move (3 in \tabref{tab:proposals}), we select $ i $, $ j $ and $ t_p' $ as follows:
\begin{enumerate}
    \item sample $ i \sim \Unif{V \setminus \{r\}} $;
    \item set $ j \leftarrow \sib(p) $;
    \item sample a new time $ t_p' \sim \Unif{t_i \vee t_j, t_q} $.
\end{enumerate}
The move fails after the first step if $ p = r $ as $ \sib(p) = \emptyset $.
We sample from maximal couplings at each step of these moves.
If the proposal fails for one state at an intermediate step, then we proceed as in the marginal move for the remainder of the move in the other state.

\begin{figure}[tbp]
	\centering
    \begin{subfigure}[t]{0.495\textwidth}
        \centering
        \begin{tikzpicture}
            \node (xc) [anchor=south east] at (0, 0) {
                \includegraphics[width=0.4\textwidth]{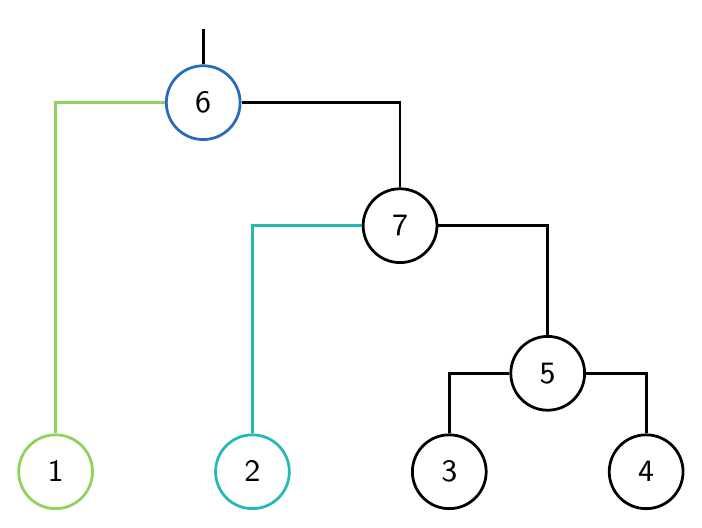}
            };
            \node (xp) [anchor=south west]  at (0.33, 0) {
                \includegraphics[width=0.4\textwidth]{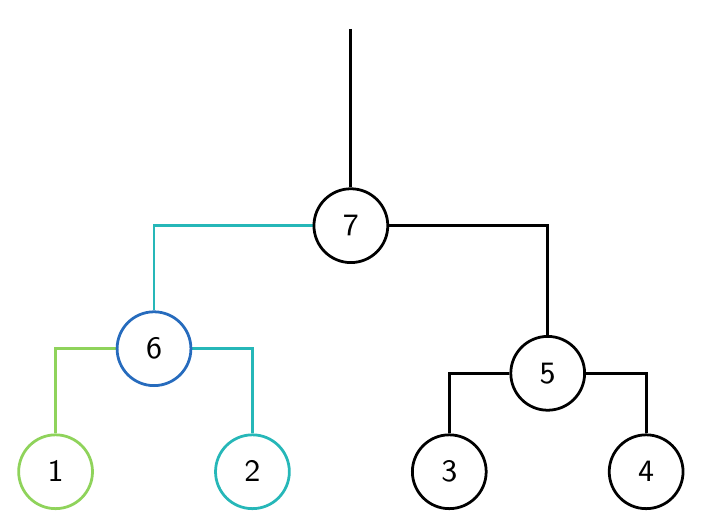}
            };
            \draw [->, >= triangle 60] (0, 1) -- (0.33, 1);
        \end{tikzpicture}
        \caption*{$ j^{(X)} = 2 $}
    \end{subfigure}
    \hfil
    \begin{subfigure}[t]{0.495\textwidth}
        \centering
        \begin{tikzpicture}
            \node (xc) [anchor=south east] at (0, 0) {
                \includegraphics[width=0.4\textwidth]{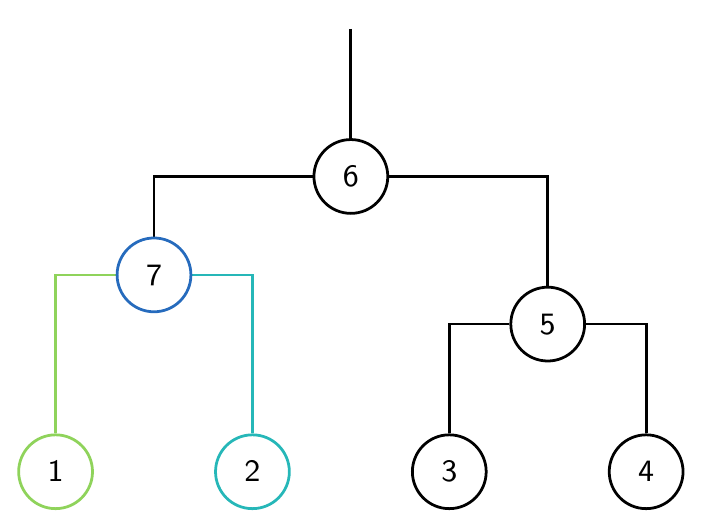}
            };
            \node (xp) [anchor=south west]  at (0.33, 0) {
                \includegraphics[width=0.4\textwidth]{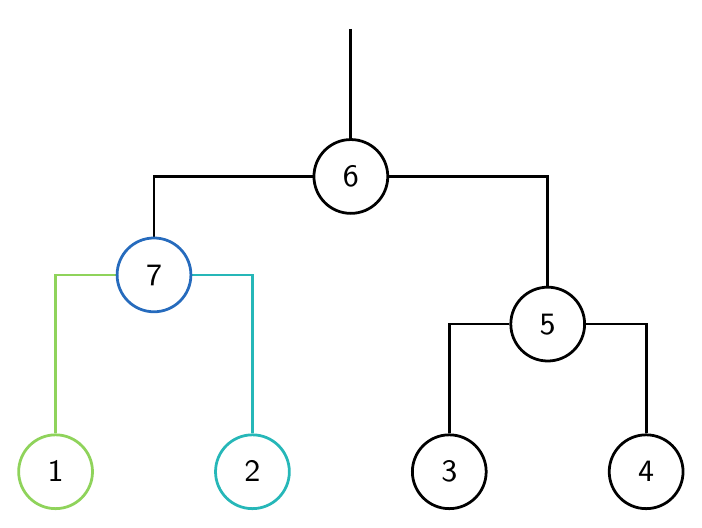}
            };
            \draw [->, >= triangle 60] (0, 1) -- (0.33, 1);
        \end{tikzpicture}
        \caption*{$ j^{(Y)} = 2 $ (move fails)}
    \end{subfigure}

    \begin{subfigure}[t]{0.495\textwidth}
        \centering
        \begin{tikzpicture}
            \node (xc) [anchor=south east] at (0, 0) {
                \includegraphics[width=0.4\textwidth]{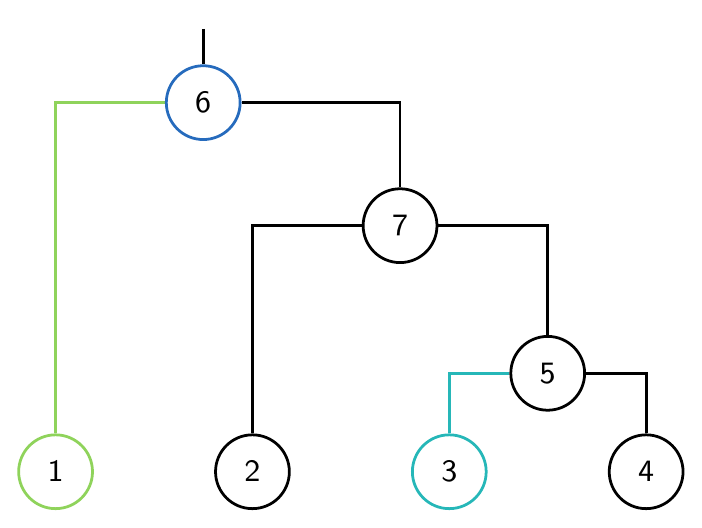}
            };
            \node (xp) [anchor=south west]  at (0.33, 0) {
                \includegraphics[width=0.4\textwidth]{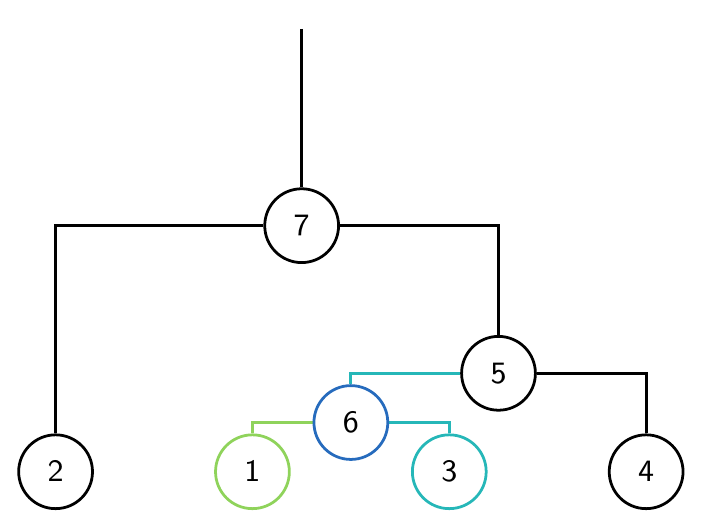}
            };
            \draw [->, >= triangle 60] (0, 1) -- (0.33, 1);
        \end{tikzpicture}
        \caption*{$ j^{(X)} = 3 $}
    \end{subfigure}
    \hfil
    \begin{subfigure}[t]{0.495\textwidth}
        \centering
        \begin{tikzpicture}
            \node (xc) [anchor=south east] at (0, 0) {
                \includegraphics[width=0.4\textwidth]{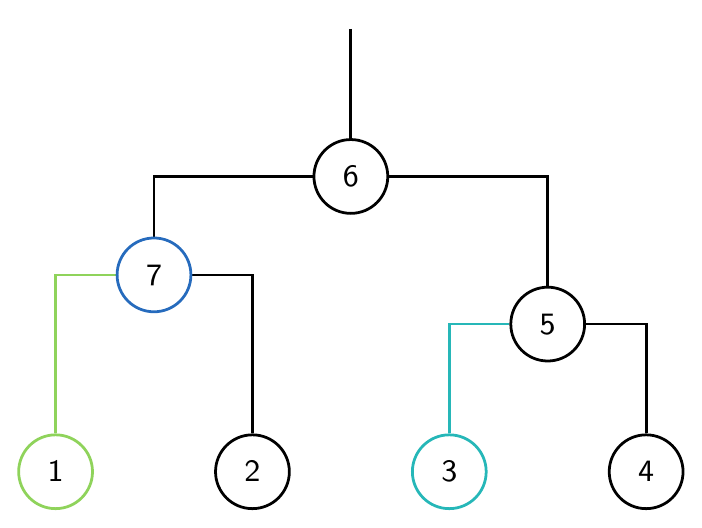}
            };
            \node (xp) [anchor=south west]  at (0.33, 0) {
                \includegraphics[width=0.4\textwidth]{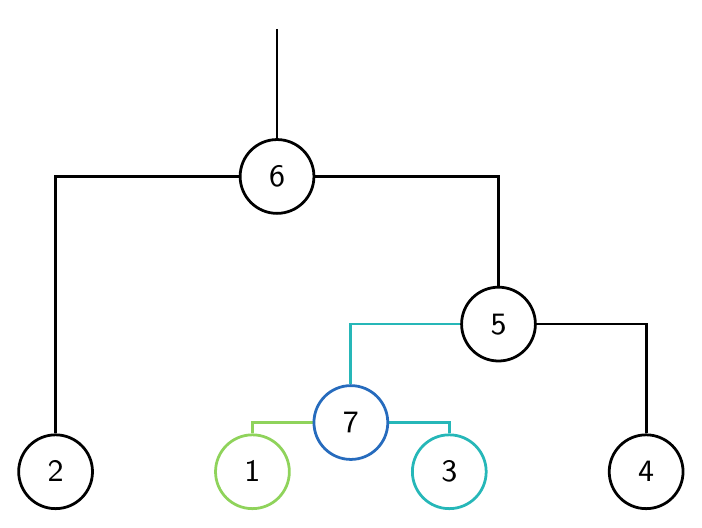}
            };
            \draw [->, >= triangle 60] (0, 1) -- (0.33, 1);
        \end{tikzpicture}
        \caption*{$ j^{(Y)} = 3 $}
    \end{subfigure}

    \begin{subfigure}[t]{0.495\textwidth}
        \centering
        \begin{tikzpicture}
            \node (xc) [anchor=south east] at (0, 0) {
                \includegraphics[width=0.4\textwidth]{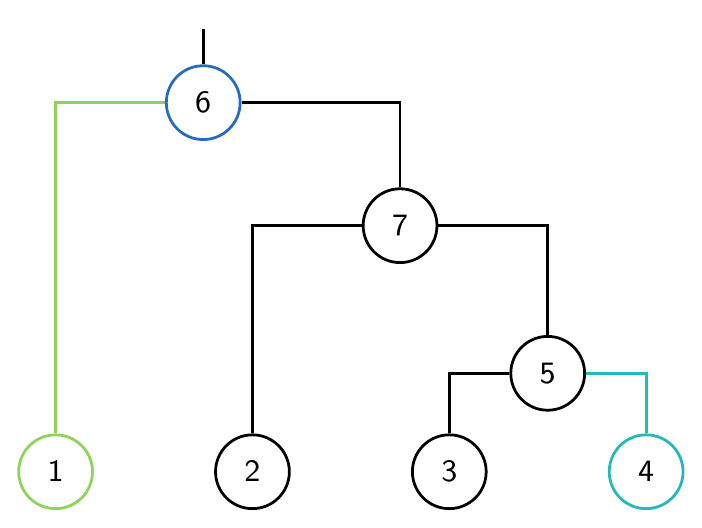}
            };
            \node (xp) [anchor=south west]  at (0.33, 0) {
                \includegraphics[width=0.4\textwidth]{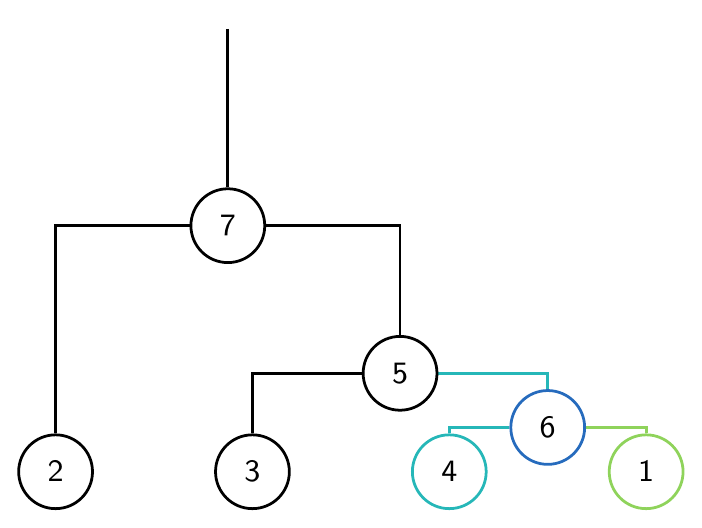}
            };
            \draw [->, >= triangle 60] (0, 1) -- (0.33, 1);
        \end{tikzpicture}
        \caption*{$ j^{(X)} = 4 $}
    \end{subfigure}
    \hfil
    \begin{subfigure}[t]{0.495\textwidth}
        \centering
        \begin{tikzpicture}
            \node (xc) [anchor=south east] at (0, 0) {
                \includegraphics[width=0.4\textwidth]{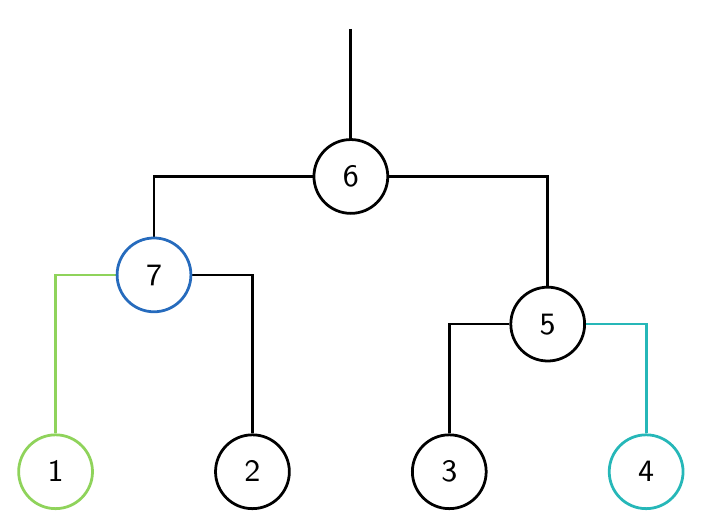}
            };
            \node (xp) [anchor=south west]  at (0.33, 0) {
                \includegraphics[width=0.4\textwidth]{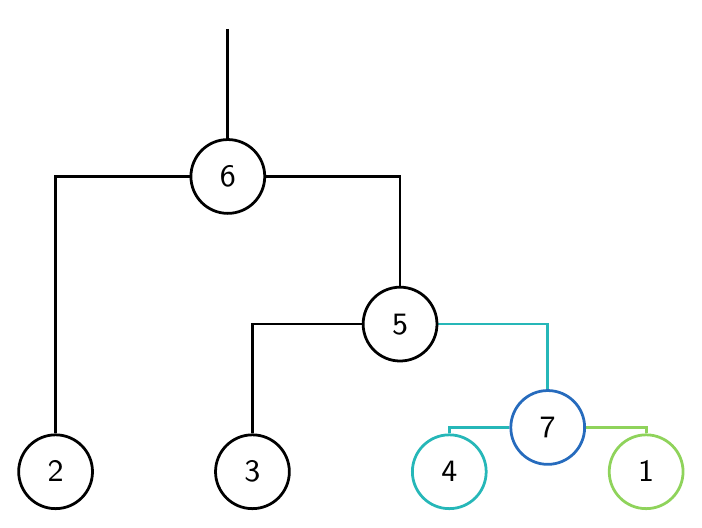}
            };
            \draw [->, >= triangle 60] (0, 1) -- (0.33, 1);
        \end{tikzpicture}
        \caption*{$ j^{(Y)} = 4 $}
    \end{subfigure}

    \begin{subfigure}[t]{0.495\textwidth}
        \centering
        \begin{tikzpicture}
            \node (xc) [anchor=south east] at (0, 0) {
                \includegraphics[width=0.4\textwidth]{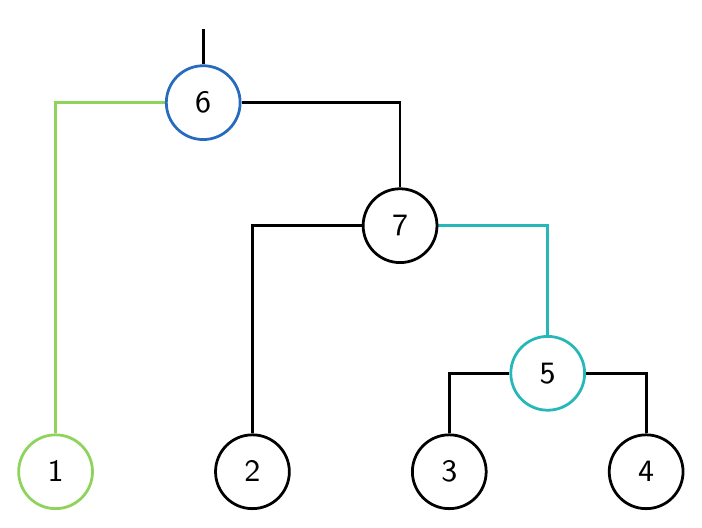}
            };
            \node (xp) [anchor=south west]  at (0.33, 0) {
                \includegraphics[width=0.4\textwidth]{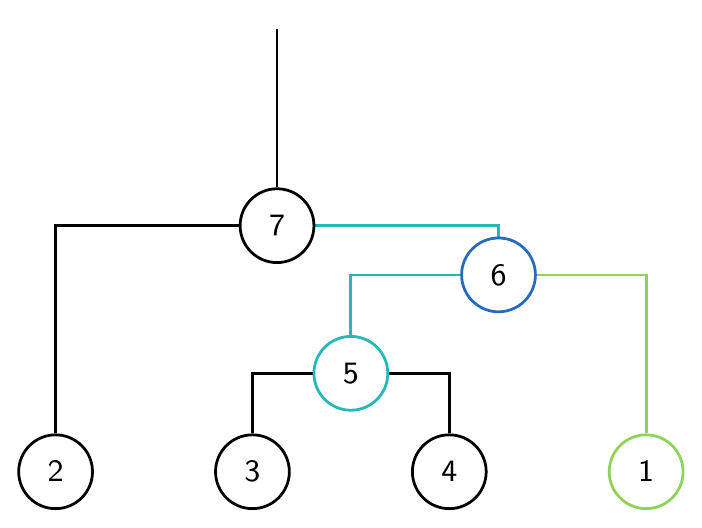}
            };
            \draw [->, >= triangle 60] (0, 1) -- (0.33, 1);
        \end{tikzpicture}
        \caption*{$ j^{(X)} = 5 $}
    \end{subfigure}
    \hfil
    \begin{subfigure}[t]{0.495\textwidth}
        \centering
        \begin{tikzpicture}
            \node (xc) [anchor=south east] at (0, 0) {
                \includegraphics[width=0.4\textwidth]{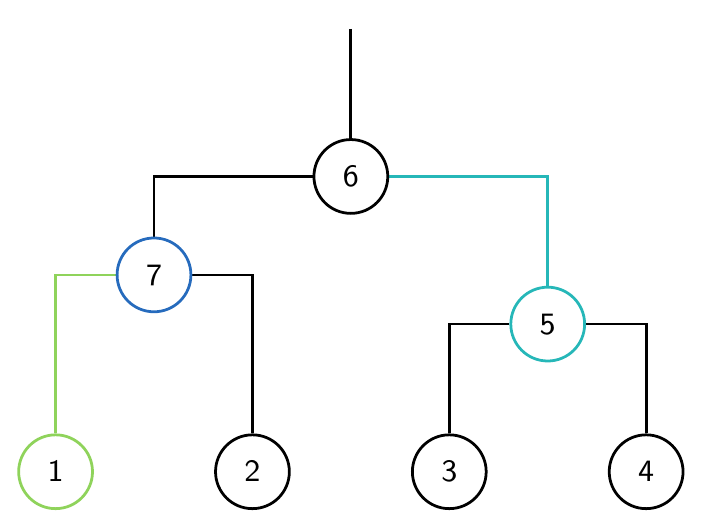}
            };
            \node (xp) [anchor=south west]  at (0.33, 0) {
                \includegraphics[width=0.4\textwidth]{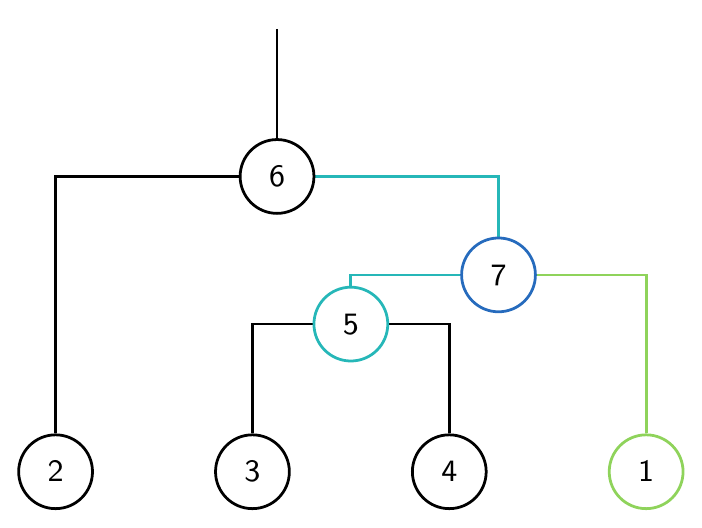}
            };
            \draw [->, >= triangle 60] (0, 1) -- (0.33, 1);
        \end{tikzpicture}
        \caption*{$ j^{(Y)} = 5 $}
    \end{subfigure}

    \begin{subfigure}[t]{0.495\textwidth}
        \centering
        \begin{tikzpicture}
            \node (xc) [anchor=south east] at (0, 0) {
                \includegraphics[width=0.4\textwidth]{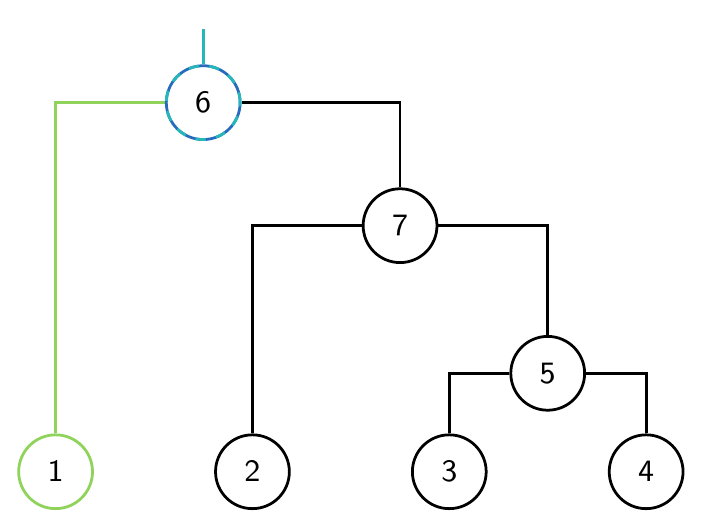}
            };
            \node (xp) [anchor=south west]  at (0.33, 0) {
                \includegraphics[width=0.4\textwidth]{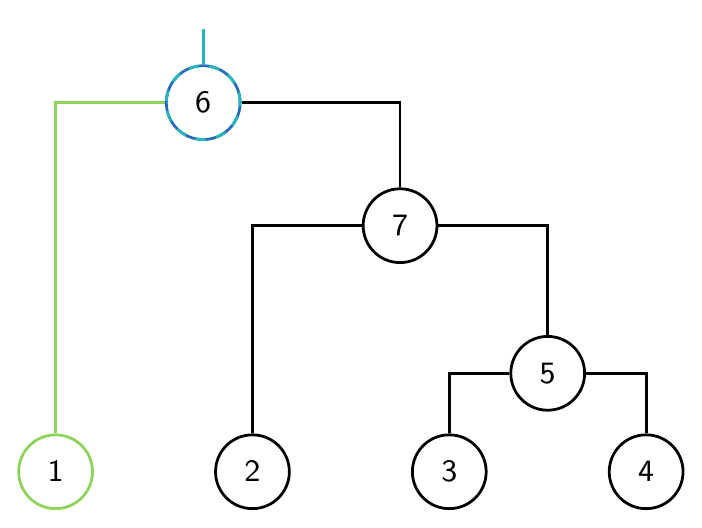}
            };
            \draw [->, >= triangle 60] (0, 1) -- (0.33, 1);
        \end{tikzpicture}
        \caption*{$ j^{(X)} = 6 $ (move fails)}
    \end{subfigure}
    \hfil
    \begin{subfigure}[t]{0.495\textwidth}
        \centering
        \begin{tikzpicture}
            \node (xc) [anchor=south east] at (0, 0) {
                \includegraphics[width=0.4\textwidth]{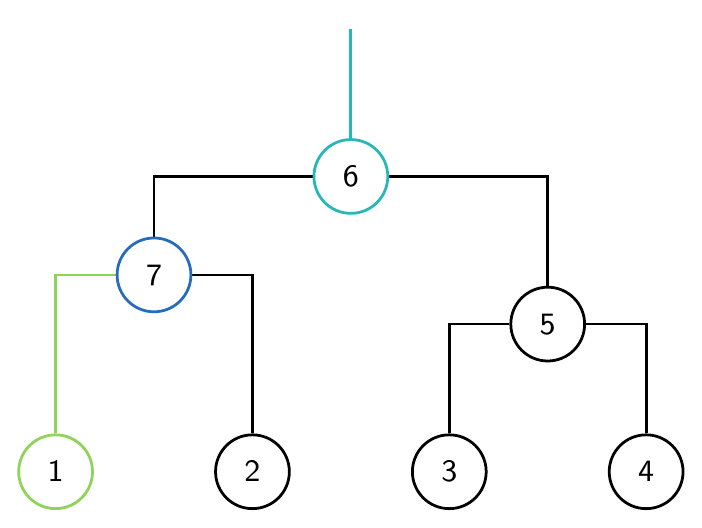}
            };
            \node (xp) [anchor=south west]  at (0.33, 0) {
                \includegraphics[width=0.4\textwidth]{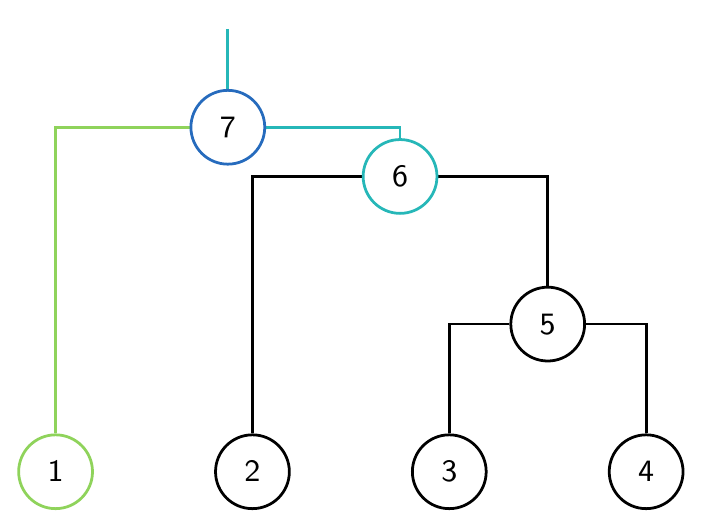}
            };
            \draw [->, >= triangle 60] (0, 1) -- (0.33, 1);
        \end{tikzpicture}
        \caption*{$ j^{(Y)} = 6 $}
    \end{subfigure}

    \begin{subfigure}[t]{0.495\textwidth}
        \centering
        \begin{tikzpicture}
            \node (xc) [anchor=south east] at (0, 0) {
                \includegraphics[width=0.4\textwidth]{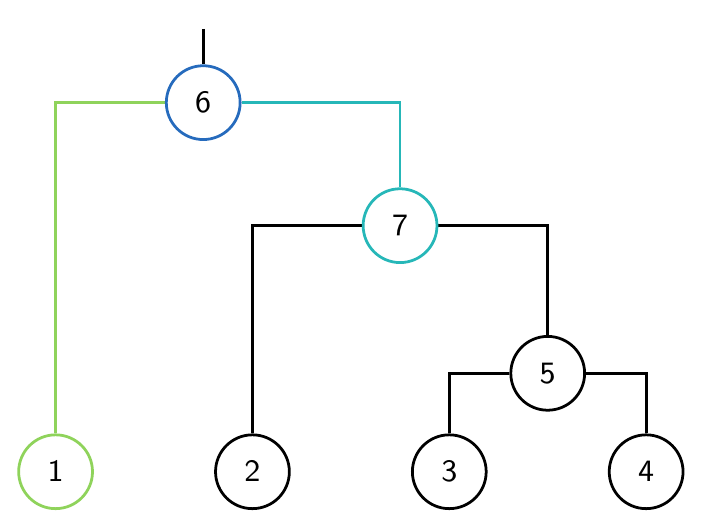}
            };
            \node (xp) [anchor=south west]  at (0.33, 0) {
                \includegraphics[width=0.4\textwidth]{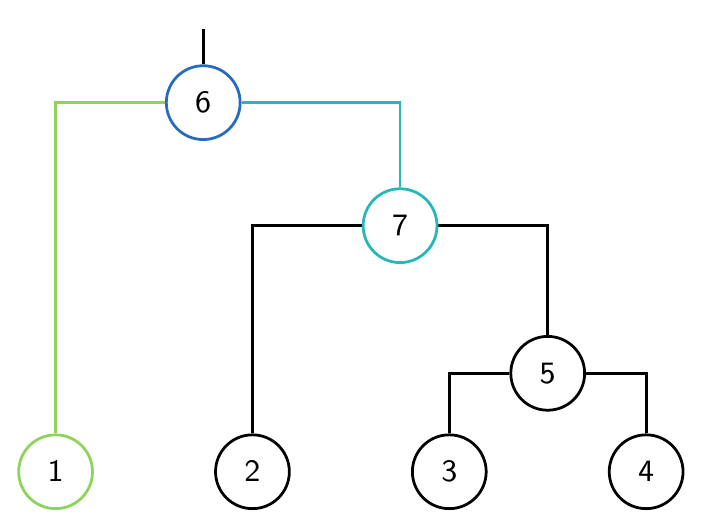}
            };
            \draw [->, >= triangle 60] (0, 1) -- (0.33, 1);
        \end{tikzpicture}
        \caption*{$ j^{(X)} = 7 $ (move fails)}
    \end{subfigure}
    \hfil
    \begin{subfigure}[t]{0.495\textwidth}
        \centering
        \begin{tikzpicture}
            \node (xc) [anchor=south east] at (0, 0) {
                \includegraphics[width=0.4\textwidth]{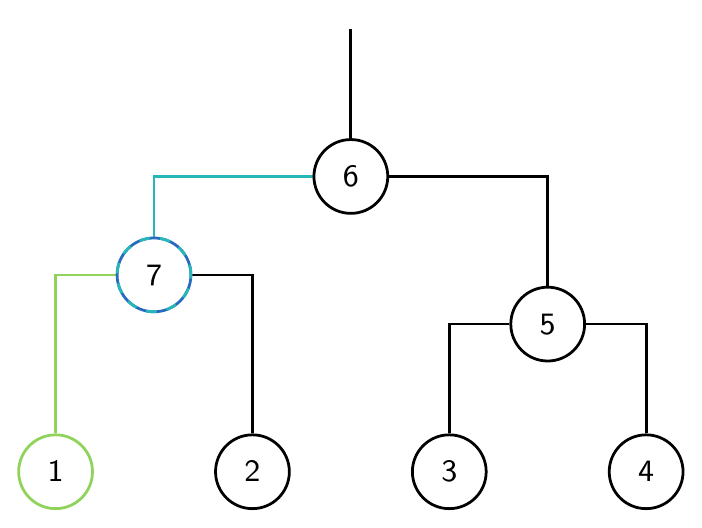}
            };
            \node (xp) [anchor=south west]  at (0.33, 0) {
                \includegraphics[width=0.4\textwidth]{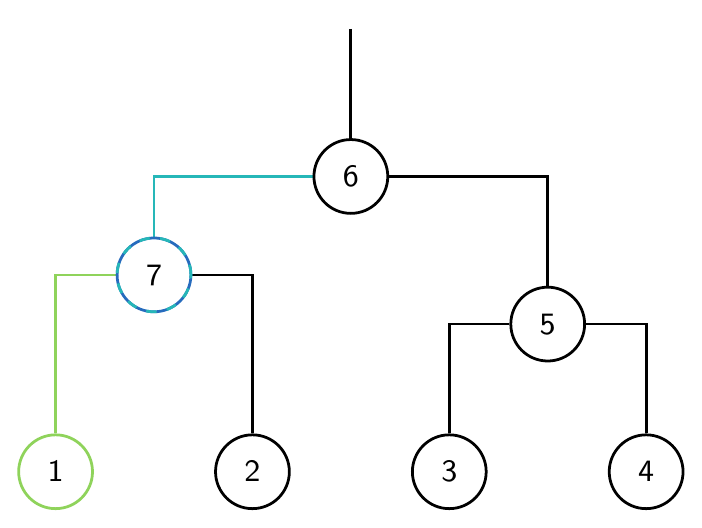}
            };
            \draw [->, >= triangle 60] (0, 1) -- (0.33, 1);
        \end{tikzpicture}
        \caption*{$ j^{(Y)} = 7 $ (move fails)}
    \end{subfigure}
    \caption{
        An SPR move detaches the \textcolor{NavyBlue}{parent} of a subtree \textcolor{YellowGreen}{root} $ i $ and reattaches it along the \textcolor{BlueGreen}{destination} branch $ j $.
        We couple this proposal by sampling subtree roots $ (i^{(X)}, i^{(Y)}) $ and destination branches $ (j^{(X)}, j^{(Y)}) $ from maximal couplings of their respective distributions.
        Each row displays the current and proposed states, $ X \rightarrow X' $ (left) and $ Y \rightarrow Y' $ (right), for $ i^{(X)} = i^{(Y)} = 1 $ and different $ (j^{(X)}, j^{(Y)}) $.
        The possible destination index sets are identical for both $ X $ and $ Y $ in this example so $ j^{(X)} = j^{(Y)} $ when drawn from a maximal coupling.
        The outcome topologies are identical except when the destination branch is $ 7 $.
    }
    \label{fig:spr-coupled-all}
\end{figure}
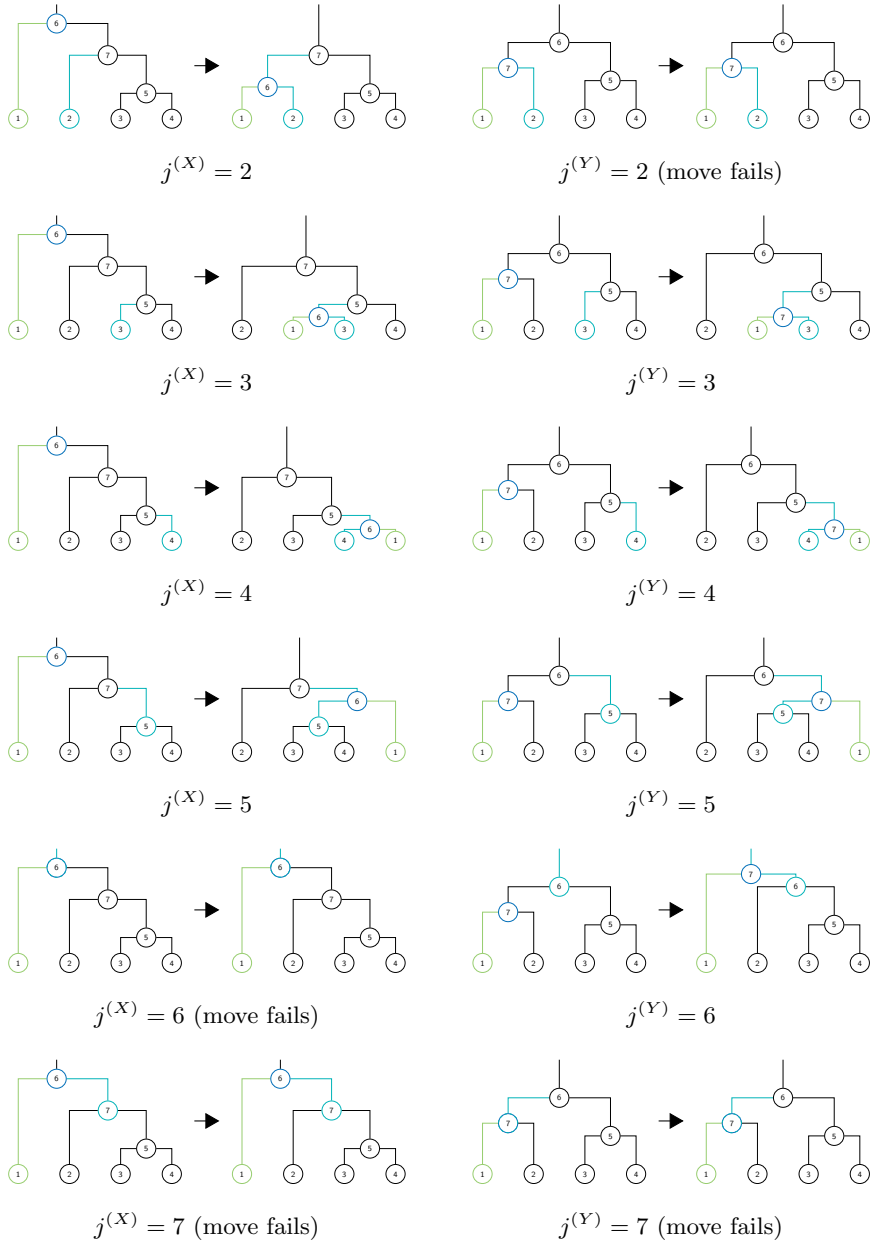

We describe a coupling of wide SPR moves (4 in \tabref{tab:proposals}) in \secref{sec:coupling-structural-moves}.
\figref{fig:spr-coupled-all} displays the possible outcome topologies in a coupled wide SPR move on a pair of subtrees.
When calibration constraints are imposed, we sample $ j $ uniformly from branches under the same set of constraints.

If catastrophes are included in the model then we also update them on any branches affected by an SPR move.
Let $ n_i $ denote the number of catastrophes currently on branch $ i $.
As the length of branch $ i $ changes in the proposal, we rescale the times of catastrophes on $ i $ accordingly.
This introduces a Jacobian term in the Hastings ratio which is cancelled out by terms in the corresponding ratio of prior distributions \citep{kelly17}.
We move any catastrophes currently on branch $ p $ onto $ h = \sib(i) $ in the new state, and $ p $ acquires a Binomial sample of the catastrophes currently on the destination branch $ j $,
\begin{align*}
    n_h' &\leftarrow n_h + n_p, \\
    n_j' &\leftarrow \Binom{n_j}{\frac{t_p' - t_j}{t_q - t_j}}, \\
    n_p' &\leftarrow n_j - n_j',
\end{align*}
where $ q = \pa(j) $.
If instead $ p $ becomes the root, then it loses its catastrophes and $ n_j' $ is sampled from the prior in the new state, likewise $ n_h' $ if $ h $ is the new root.

We sample $ n_j' $ in the proposed states for each chain from a maximal coupling of their respective distributions, the catastrophe updates on the remaining branches are deterministic given $ n_j' $.
This is different to what was previously implemented by \citet{ryder11} and \citet{kelly17} for the marginal move --- a situation where we were able to diagnose a mixing issue in the marginal kernel thanks to couplings.
We also refresh the locations of catastrophes on branches $ h $, $ j $ and $ p $ as part of the SPR move.
We sample from a maximal coupling of the distributions of locations along branches given their respective counts, this procedure is described in detail in \appref{app:moves-catastrophes} below.

\subsection{Moves 5--9: resampling or rescaling node times}
\label{app:moves-times}

We either resample a single node time or rescale a set of internal node ages by a common factor $ \eta \sim \Unif{1/2, 2} $.
We cannot sample from a maximal coupling here unless $ t_i^{(X)} / t_i^{(Y)} $ is a constant in $ (1/4, 4) $ for each $ i \in A' $, which only occurs when the node ages have already met or have just separated.
Instead, we sample from a maximal coupling of the implied proposals on the eldest node age in $ A' $ in each state then rescale the remaining node times accordingly.

\subsubsection{Move 5: resample internal node time}

We select an internal node $ i $ at random and sample a new node time between its eldest child and parent as follows:
\begin{enumerate}
    \item sample $ i \sim \Unif{A} $;
    \item let $ j = \argmax \{t_{j'} : j' \in \off(i)\} $;
    \item if $ i = r $, then sample $ t_i' \sim \Unif{\frac{t_i + t_j}{2}, 2 t_i -  t_j} $, otherwise $ t_i' \sim \Unif{t_j, t_{\pa(i)}} $.
\end{enumerate}
As the ancestral node labels are identical in both states, we always have $ i^{(X)} = i^{(Y)} $ when sampling from a maximal coupling.
Through housekeeping, if we select the root index in $ X $ then we do the same in $ Y $.
We then sample $ (t_i^{(X)\prime}, t_i^{(Y)\prime}) $ from a maximal coupling of their respective distributions using the algorithm in \secref{sec:couplings-generic} or the direct approach in \appref{app:sample-maximal-couplings}.

We also resample catastrophe counts and locations on the branches connected to $ i $.
Let $ j $ and $ k $ denote the offspring of $ i $.
If $ i \neq r $, then we sample new catastrophe counts as
\[
    (n_i', n_j', n_k')
        \sim \Munom{
            n_i + n_j + n_k
        }{
            \left(
                \frac{\Delta_i'}{\Delta_{ijk}'},
                \frac{\Delta_j'}{\Delta_{ijk}'},
                \frac{\Delta_k'}{\Delta_{ijk}'}
            \right)
        },
\]
where $ \Delta_i' $ is length of branch $ i $ in the proposed state, and $ \Delta_{ijk}' = \Delta_i' + \Delta_j' + \Delta_k' $.
If $ i $ is the root, then we only resample catastrophe counts on $ j $ and $ k $.
We sample from a maximal coupling of Multinomial distributions if the total catastrophe count $ n_i + n_j + n_k $ is equal in both states, otherwise we sample the counts independently.
\appref{app:moves-catastrophes} describes how we couple the sampling of catastrophe positions.

\subsubsection{Move 6: resample leaf time}

For certain taxa, we may only know upper and lower bounds for when they were recorded rather than the exact sampling times.
In this move, we select a leaf at random and propose a new time sampled uniformly at random along its range.
If the new leaf time is greater than its parent age, then the move fails.
When sampling from a maximal coupling here, we always choose the same node in both states, and as the proposed time is independent of the current state and the ranges fixed, we always sample the same node time in both.
This is one of only two kernels in our mixture which can be maximally coupled with a common random numbers coupling.

\subsubsection{Move 7: rescale tree}

We rescale the ages of internal nodes by $ \eta \sim \Unif{1/2, 2} $.
Let $ t_0 $ denote the age of the youngest leaf node, typically $ 0 $.
In the proposed state,
\begin{equation}
    \label{eq:rescale-times}
    t_i' \leftarrow t_0 + \nu (t_i - t_0), \quad i \in A.
\end{equation}
If the death rate $ \mu $ is allowed to vary then we also propose to update it to $ \mu' \leftarrow \mu / \nu $, likewise $ \beta $.
The move fails if in the proposed state a leaf is older than its parent or a node violates its clade time constraints.

We sample from a maximal coupling of the root time distributions and scale the remaining internal nodes in each state accordingly; that is, we sample from a maximal coupling of continuous Uniform distributions of the form
\begin{equation}
    \label{eq:rescale-root}
    t_r' \sim \UNIF{t_0 + \frac{t_r - t_0}{2}, 2 t_r - t_0},
\end{equation}
in each state, and use $ \nu = (t_r' - t_0) / (t_r - t_0) $ to update the remaining node times according to \eqnref{eq:rescale-times}.

\subsubsection{Move 8: rescale sub tree}

We sample a subtree root $ i $ with probability proportional to the number of leaves beneath it, then rescale the node times in the subtree by $ \nu \sim \Unif{1/2, 2} $.
For this move, the update is identical to \eqnref{eq:rescale-times} except we only consider the subtree nodes and $ t_0 $ is the age of the youngest leaf of the subtree.
This move fails if the new subtree root age is older than its parent or any leaf node is older than its parent.

To couple this move, we sample the subtree root $ i $ in each state from a maximal coupling of the corresponding distributions in a similar fashion to \eqnref{eq:rescale-root}, then update the remaining subtree node times accordingly.

\subsubsection{Move 9: rescale tree above clade bounds}

Similar to moves~7 and 8, we rescale the ages of nodes above clade bounds by $ \nu \sim \Unif{1/2, 2} $.
We proceed in an identical fashion to move~7 and sample new root times from a maximal coupling of their respective distributions, as described in \eqnref{eq:rescale-root}, then rescale the times of nodes above clade bounds according to \eqnref{eq:rescale-times}.

\subsection{Moves 10--14: catastrophes}
\label{app:moves-catastrophes}

Catastrophes are latent variables which we cannot integrate analytically out of our posterior.
As described in \appref{app:sd-model}, we only consider catastrophes on branches beneath the root.
In the absence of lateral transfer, $ C $ only records the number of catastrophes on each branch and not their locations.

\subsubsection{Move 10: add one catastrophe}

We propose to add a catastrophe uniformly at random across the tree as follows:
\begin{enumerate}
    \item select a branch $ i $ with probability $ \Delta_i / \Delta $;
    \item sample a relative location $ u \sim \Unif{0, 1} $.
\end{enumerate}
The proposed catastrophe set is $ C' \leftarrow C \cup \{(i, u)\} $.
When coupling this move, we first sample the target branch in each state from a maximal coupling of the distributions on branch indices, so the event we observe $ i^{(X)} = i^{(Y)} = i $ occurs with probability $ (\Delta_i^{(X)} / \Delta^{(X)}) \wedge (\Delta_i^{(Y)} / \Delta^{(Y)}) $.
As the relative locations are Uniform, we propose the same $ u \sim \Unif{0, 1} $ to both states, $ (u^{(X)}, u^{(Y)}) \leftarrow (u, u) $.

\subsubsection{Move 11: delete one catastrophe}

We describe our coupling of this move in \secref{sec:coupling-catastrophes}.
This move fails if there are no catastrophes on the current tree.

\subsubsection{Move 12: move one catastrophe to a neighbouring edge}

Let $ n_i $ denote the number of catastrophes on a branch $ i $ below the root and $ n $ the total on the tree.
We sample a catastrophe at random to move to a neighbouring branch as follows:
\begin{enumerate}
    \item select a branch $ i \in V \setminus \{r\} $ with probability $ n_i / n $;
    \item select a catastrophe location $ u $ uniformly at random from the $ n_i $ catastrophes located on $ i $;
    \item select a destination $ j \sim \Unif{\{\pa(i), \off(i), \sib[\pa(i)]\} \setminus \{r\}\}} $;
    \item sample a new relative location $ u' \sim \Unif{0, 1} $ on the destination branch.
\end{enumerate}
The proposed catastrophe set is $ C' \leftarrow C \cup \{(j, u')\} \setminus \{(i, u)\} $.
As in move~11, this move fails if there are no catastrophes on the tree in the current state.

When coupling this move, we sample the branch $ i $ and catastrophe location $ u $ identically to move~11.
We then sample the destination branch from a maximal coupling of the discrete Uniform distributions on the neighbouring branch indices in each state, and propose a common location $ u' \in \Unif{0, 1} $ as in move~10.

\subsubsection{Move 13: resample catastrophes on a single branch}

When $ \rho $ is fixed, the prior distribution on the number of catastrophes on branch $ i $ is $ \Pois{\rho \Delta_i} $, and when $ \rho $ is unknown and integrated out with respect to its Gamma prior, the prior distribution on catastrophe counts on a single branch is Negative Binomial and Negative Multinomial on the tree.
The prior distribution over relative locations along branches conditional on their counts remains Uniform in both cases.
We propose to update the catastrophes on a randomly chosen branch $ i $ with a sample from the prior:
\begin{enumerate}
    \item select a branch $ i $ with probability $ \Delta_i / \Delta $;
    \item sample new catastrophe count $ n_i' $ from the prior on branch $ i $;
    \item sample new relative locations $ u_1' \sim \Unif{0, 1}, \dotsc, u_{n_i'}' \sim \Unif{0, 1} $.
\end{enumerate}
We form the proposed state $ C' $ by removing the catastrophes currently on $ i $ in $ C $ and adding $ (i, u_1'), \dotsc, (i, u_{n_i'}') $.
To couple this move, we first sample the target branches $ i^{(X)} $ and $ i^{(Y)} $ from maximal couplings of their respective distributions, in which case
\[
    \PP(i^{(X)} = i^{(Y)})
        = \sum_{i \in V \setminus \{r\}} \frac{\Delta_i^{(X)}}{\Delta^{(X)}} \wedge\frac{\Delta_i^{(Y)}}{\Delta^{(Y)}}.
\]
We then draw $ (n_i^{(X) \prime}, n_i^{(Y) \prime}) $ from a maximal coupling of Poisson distributions when $ \rho $ is fixed or Negative Binomial distributions when $ \rho $ is marginalised out.
Finally, we sample new catastrophe locations from a maximal coupling: in the proposed states, $ n_{i^{(X)}}^{(X) \prime} \wedge n_{i^{(Y)}}^{Y \prime} $ catastrophes will have common locations with the remainder sampled independently.

This procedure for coupling proposed catastrophe locations is equivalent to simulating a dominating catastrophe process and maximally coupling the random thinning operation to obtain the proposed catastrophes for $ X $ and $ Y $.
For example, if the prior on catastrophes is a Poisson process with rate $ \rho $, then we could simulate a Poisson process with rate $ \rho (\Delta_{i^{(X)}} \vee \Delta_{i^{(Y)}}) $ on $ [0, 1] $, and include each point as a relative location for a catastrophe on $ i^{(X)} $ in $ X $ with probability $ \Delta_{i^{(X)}} / (\Delta_{i^{(X)}} \vee \Delta_{i^{(Y)}}) $, and likewise for $ Y $ \citep{kingman92}.
If we sample from a maximal coupling to include each candidate point, then $ n_{i^{(X)}}^{(X) \prime} \sim \Pois{\rho \Delta_{i^{(X)}}} $ and $ n_{i^{(Y)}}^{(Y) '} \sim \Pois{\rho \Delta_{i^{(Y)}}} $, with branches $ i^{(X)} $ and $ i^{(Y)} $ having $ n_{i^{(X)}}^{(X) \prime} \wedge n_{i^{(Y)}}^{(Y) \prime} $ relative locations in common.

\subsubsection{Move 14: resample catastrophe location on branch}

We only attempt this move when fitting the lateral transfer model and it fails if there are no catastrophes on the tree.
We sample a catastrophe $ (i, u) \in C $ at random and propose a new location $ u' $ along its branch $ i $ as follows:
\begin{enumerate}
    \item select branch $ i $ with probability $ n_i / n $;
    \item select a location $ u $ uniformly at random from those on $ i $;
    \item sample a new relative location $ u' \sim \Unif{0, 1} $.
\end{enumerate}
The proposed catastrophe set is $ C' \leftarrow (C \setminus \{(i, u)\}) \cup \{(i, u')\} $.
As with moves~11 and 12, we sample the target branches from a maximal coupling of their distributions, we sample a catastrophe identified by its location from a maximal coupling of discrete Uniform distributions, and we propose the same relative location $ u' $ to both states $ X $ and $ Y $.

\subsection{Moves 15--19: Stochastic Dollo model parameters}
\label{app:moves-parameters}

For all of these proposals, we sample $ \eta \sim \Unif{1/2, 2} $ and use it to rescale one or more parameters.
The move fails if any proposed value is outside its valid range.

\subsubsection{Move 15: rescale death rate}

We propose a new death rate $ \mu' \leftarrow \nu \mu $.
When coupling this move, we sample from a maximal coupling of $ \Unif{\mu / 2, 2 \mu} $ for $ \mu $ in each state.
If $ \mu^{(X)} \leq \mu^{(Y)} $, then
\[
    \PP(\mu^{(X) \prime} = \mu^{(Y) \prime} \given \mu^{(X)}, \mu^{(Y)})
        = \left\{
            \begin{array}{ll}
                0, \quad & 2 \mu^{(X)} < \mu^{(Y)} / 2, \\
            \frac{2}{3}
            \frac{
                2 \mu^{(X)} - \mu^{(Y)} / 2
            }{
                \mu^{(Y)}
            }, \quad & \text{otherwise},
            \end{array}
        \right.
\]
and likewise when $ \mu^{(Y)} < \mu^{(X)} $.

\subsubsection{Move 16: rescale transfer rate}

The proposal for $ \beta $ is identical to that for $ \mu $ in move~15.

\subsubsection{Move 17: rescale catastrophe strength}

To avoid issues with identifiability, we often fix $ \kappa $ or else enforce a lower bound in order to avoid weak catastrophes.
As with the other scalar parameters, we propose $ \kappa' \leftarrow \nu \kappa $ in a marginal move so the coupling is identical to those for $ \mu $ and $ \beta $ in moves~15 and 16 except the proposal fails if $ \kappa' $ violates its bounds.

\subsubsection{Move 18: rescale one missing data parameter}

We expect that most $ \xi_i $ terms will be close to $ 1 $ (high probability of observing the true state) so instead scale $ 1 - \xi_i $ by $ \eta $.
We propose to update a single $ \xi_i $ to $ \xi_i' $ as follows:
\begin{enumerate}
    \item select a leaf $ i \sim \Unif{L} $;
    \item set $ \xi_i' \leftarrow 1 - \nu (1 - \xi_i) $.
\end{enumerate}
The move fails if $ \xi_i' \not\in [0, 1] $.
To couple this proposal, we select the same leaf $ i $ for both $ X $ and $ Y $, and sample $ (\xi_i^{(X)\prime}, \xi_i^{(Y)\prime}) $ from a maximal coupling of the corresponding Uniform distributions.

\subsubsection{Move 19: rescale all missing data parameters}

We propose to update all of the missingness parameters $ \Xi $ via a common $ \nu $:
\[
    \xi_i' \leftarrow 1 - \nu (1 - \xi_i), \quad i \in L.
\]
As above, the move fails if any $ \xi_i' \not\in [0, 1] $.
As with rescaling multiple node times by a common factor (moves~7--9), we cannot sample from a maximal coupling here.
We could attempt to couple one $ \xi_i $ and rescale the others accordingly but have not done so.
Using the same $ \nu $ variate in updating $ \Xi^{(X)} $ and $ \Xi^{(Y)} $ is equivalent to a common random number coupling.

\section{Experiments}
\label{app:experiments}

To complement the experiments with relatively weak catastrophes  in \secref{sec:synthetic-full} (strength $ \kappa = 0.05 $, $ \Gamma(1.5, 5000) $ prior on the catastrophe rate $ \rho $), we now consider data generated with two larger catastrophes with strength $ \kappa = 1 / 3 $ and place a more stringent $ \Gamma(1.5, 10^5) $ prior on $ \rho $.
The catastrophes were placed on randomly selected branches leading into leaf nodes before generating the data with the same death rate $ \mu = 2.5 \times 10^{-4} $ as M5.1.
The true presence/absence state of each trait at leaf $ i \in L $ was recorded with probability $ \xi_i \sim \BetaDist{1}{1/3} $ and marked missing otherwise.
For the MCMC, we applied a single clade constraint selected at random for each tree and set an upper bound of $ 2 \times 10^3 $ on the root time, $ \mu $ and $ \kappa $ were fixed at the values used to generate the data.
We are not modelling lateral transfer so only consider the number of catastrophes along each branch, the target is the posterior distribution on the tree topology, internal node times in $ T $, the number of catastrophes on each branch $ C $, and the missing data parameters for each leaf $ \Xi $.

\begin{figure}[tb]
	\centering
    \begin{subfigure}[b]{\textwidth}
        \centering
        \includegraphics[width=\textwidth, trim = 0cm 0cm 0cm 1.5cm, clip]{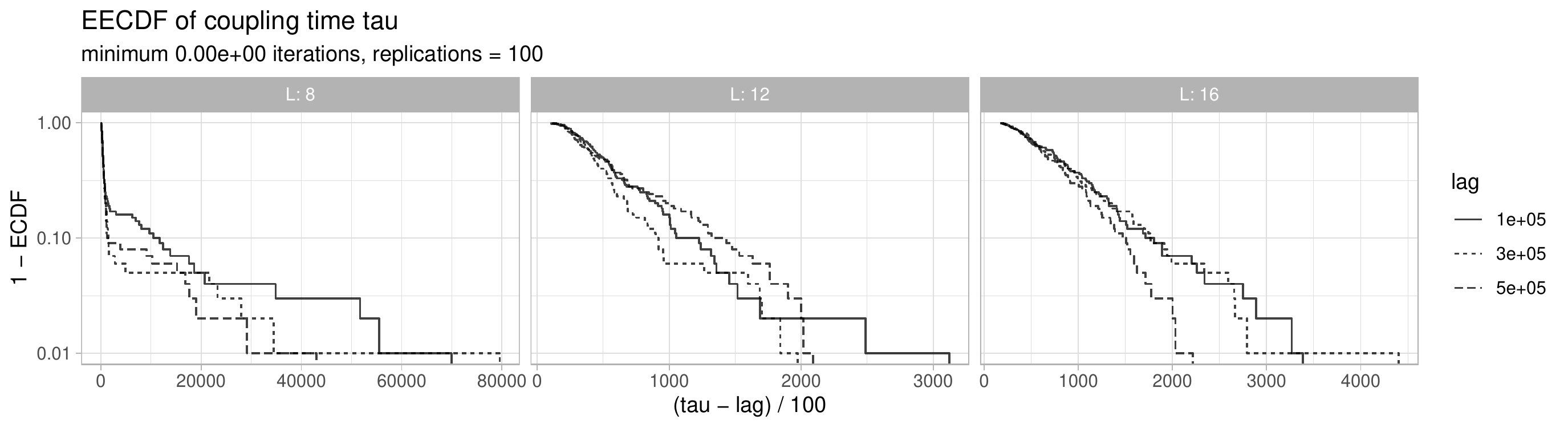}
        \caption{
            Tails of $ \tau^{(l)} $ decay geometrically for each lag $ l $.
        }
        \label{fig:strong-tau}
    \end{subfigure}

    \begin{subfigure}[b]{\textwidth}
        \centering
        \includegraphics[width=\textwidth, trim = 0cm 0cm 0cm 1.5cm, clip]{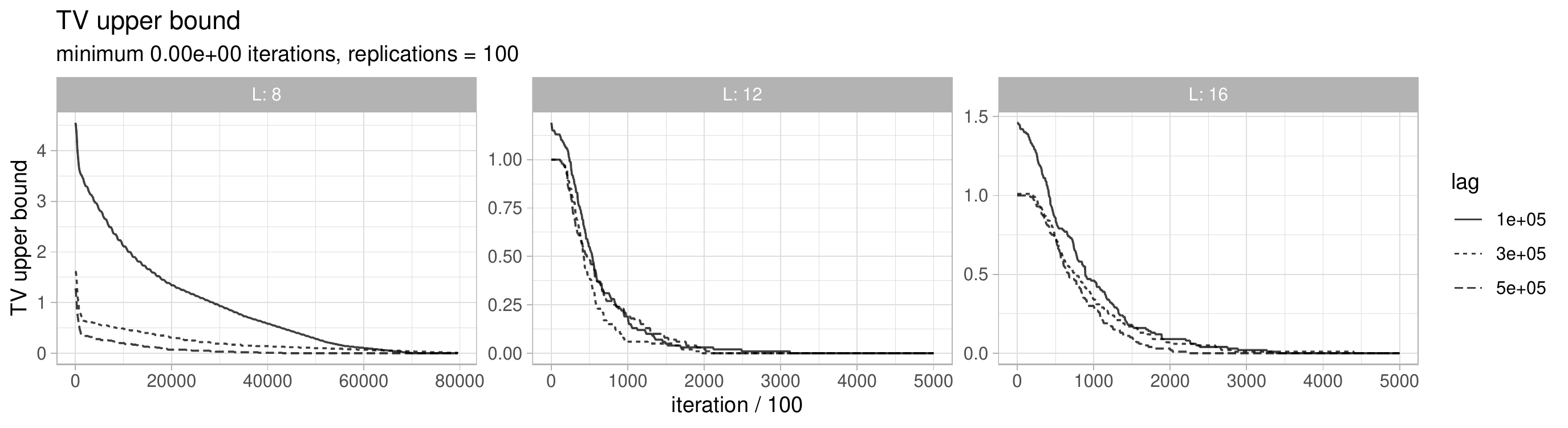}
        \caption{
            Estimated TV bounds eventually converge at similar rates for each lag.
        }
        \label{fig:strong-tv}
    \end{subfigure}

    \begin{subfigure}[b]{\textwidth}
        \centering
        \includegraphics[width=\textwidth, trim = 0cm 0cm 0cm 1.5cm, clip]{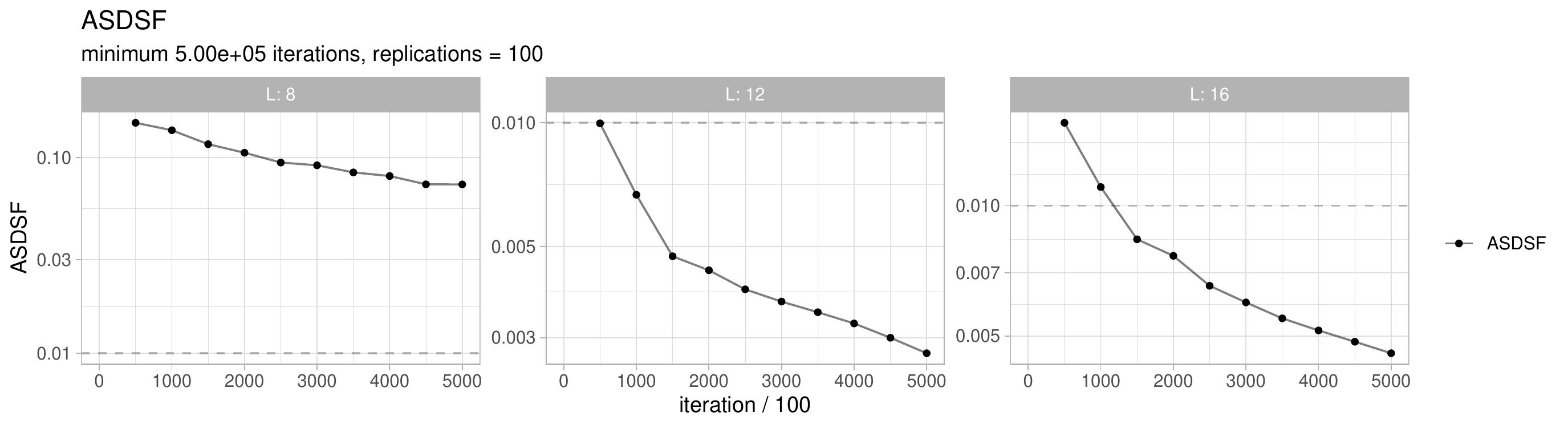}
        \caption{ASDSFs decay monotonically but fail to reach $ 0.01 $ for the trees with 8 taxa.}
        \label{fig:strong-asdsf}
    \end{subfigure}
    \caption{
        Diagnosing convergence on synthetic data sets with missing data, two strong catastrophes and a stringent prior on their number.
    }
    \label{fig:strong}
\end{figure}

\figref{fig:strong} displays the results of our experiments running $ 100 $ pairs of chains at each lag.
For 8 taxa, we see in \figref{fig:strong-tau} that many chains met quickly but the rest took significantly longer, particularly for the smallest lag $ l = 10^5 $, with corresponding effects on the estimated bounds in \figref{fig:strong-tv}.
The slow decay of ASDSF estimates on the first $ 5 \times 10^5 $ iterations suggests it will require many more iterations to reach the convergence threshold of $ 0.01 $.
The coupling TV bound and ASDSF display similar behaviour for the experiments with 12 and 16 taxa, suggesting that the catastrophes caused fewer mixing issues in these problems.

\section{Software validation}
\label{app:software-validation}

\begin{figure}[tp]
    \centering
    \includegraphics[width=\textwidth]{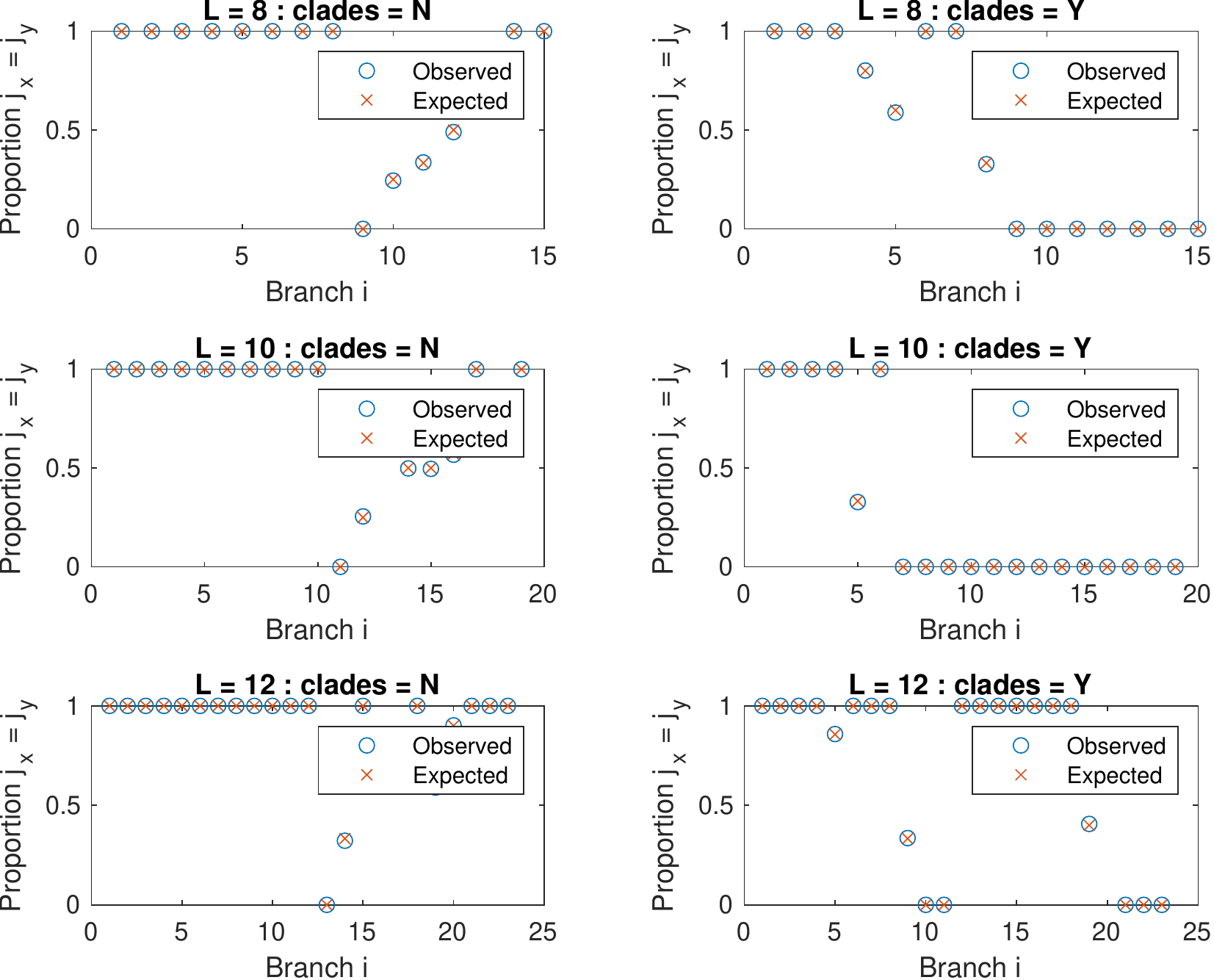}
    \caption{
        Checking that our maximal coupling to sample SPR destinations branch indices is correct.
        For each experiment, we sample a pair of trees with $ L $ leaves, perform housekeeping, and possibly add a random selection of clades.
        For each node index $ i \in V \setminus \{r\} $, we sampled $ 10^4 $ destination index pairs $ (j^{(X)}, j^{(Y)}) $ from a maximal coupling of $ \Unif{J_i^{(X)}} $ and $ \Unif{J_i^{(Y)}} $, discrete Uniform distributions on the sets of valid destination branches $ J_i^{(X)} $ and $ J_i^{(Y)} $ defined in \secref{sec:coupling-structural-moves}.
        The proportions of samples where $ j^{(X)} = j^{(Y)} $ are the \emph{Observed} terms in the figures.
        The proportion of matching samples we would expect to see under a maximal coupling is given by \eqnref{eq:spr-j} and are the \emph{Expected} terms in the figure.
        In each case, we see that the observed and expected proportions match.
    }
    \label{fig:test-spr-j}
\end{figure}

Our algorithm is implemented in \texttt{TraitLab} \citep{nicholls13}, a \texttt{Matlab} toolbox for fitting Stochastic Dollo models.
We validated our software implementation through a variety of tests.
For each maximal coupling of a step of a proposal distribution, we wrote unit tests to:
\begin{itemize}
    \item check that identical pairs of states produce identical proposals;
    \item compare the distribution of coupled draws with their marginal counterparts;
    \item compare the proportion of identical samples with what we expect theoretically from a maximal coupling.
\end{itemize}
For example, \figref{fig:test-spr-j} compares the proportion of matching destination branches in a coupled SPR move with what we would expect under a maximal coupling.
We also wrote unit tests to compare the distribution of samples from the overall couplings of the local transition kernels against their marginal counterparts.
The unit tests are included with the software.
Finally, we ran multiple pairs of chains for many iterations after they met to ensure that chains did not decouple through an undetected bug and to verify that the distribution of samples from coupled chains matched those from marginal chains.

\end{document}